\documentclass[twocolumn,showpacs,preprintnumbers,amsmath,amssymb,nofootinbib,epsfig]{revtex4-2}

\usepackage{graphicx}
\usepackage{dcolumn}
\usepackage{bm}
\usepackage{epsfig}

\usepackage{float}
\usepackage{amsmath}
\usepackage{epsfig,floatflt}
\usepackage{subfigure}
\usepackage[usenames]{color}
\usepackage{rotating}
\usepackage{comment} 
\usepackage{hyperref}

\newcommand{\mii}{Mg\,\textsc{ii}}
\newcommand{\civ}{C\,\textsc{iv}}

\usepackage[T1]{fontenc}
\usepackage{scalerel}
\usepackage{tikz}
\usetikzlibrary{svg.path}

\definecolor{orcidlogocol}{HTML}{A6CE39}
\tikzset{
  orcidlogo/.pic={
    \fill[orcidlogocol] svg{M256,128c0,70.7-57.3,128-128,128C57.3,256,0,198.7,0,128C0,57.3,57.3,0,128,0C198.7,0,256,57.3,256,128z};
    \fill[white] svg{M86.3,186.2H70.9V79.1h15.4v48.4V186.2z}
                 svg{M108.9,79.1h41.6c39.6,0,57,28.3,57,53.6c0,27.5-21.5,53.6-56.8,53.6h-41.8V79.1z M124.3,172.4h24.5c34.9,0,42.9-26.5,42.9-39.7c0-21.5-13.7-39.7-43.7-39.7h-23.7V172.4z}
                 svg{M88.7,56.8c0,5.5-4.5,10.1-10.1,10.1c-5.6,0-10.1-4.6-10.1-10.1c0-5.6,4.5-10.1,10.1-10.1C84.2,46.7,88.7,51.3,88.7,56.8z};
  }
}

\newcommand\orcidicon[1]{\href{https://orcid.org/#1}{\mbox{\scalerel*{
\begin{tikzpicture}[yscale=-1,transform shape]
\pic{orcidlogo};
\end{tikzpicture}
}{|}}}}

\begin{document}


\title{Is the $w_0w_a$CDM cosmological parameterization evidence for dark energy dynamics partially caused by the excess smoothing of Planck PR4 CMB anisotropy data?}

\author{Javier de Cruz P\'erez$^{\orcidicon{0000-0001-8603-5447}}$}%
\email{jdecruz@ugr.es}
\affiliation{Departamento de Física Te\'orica y del Cosmos, Universidad de Granada, E-18071, Granada, Spain}%

\author{Chan-Gyung Park$^{\orcidicon{0000-0002-3076-2781}}$}%
\email{park.chan.gyung@gmail.com}
\affiliation{Division of Science Education and Institute of Fusion Science, Jeonbuk National University, Jeonju 54896, Republic of Korea}%

\author{Bharat Ratra$^{\orcidicon{0000-0002-7307-0726}}$}%
\email{ratra@phys.ksu.edu}
\affiliation{Department of Physics, Kansas State University, 116 Cardwell Hall, Manhattan, KS 66506, USA}%

\date{\today}

\begin{abstract}

We study the performance of the spatially-flat $\Lambda$CDM model and the spatially-flat dynamical dark energy parameterizations $w_0$CDM and $w_0w_a$CDM, in which the dark energy equation of state parameter is either constant ($w=w_0$) or redshift-dependent [$w(z)=w_0+w_a z/(1+z)$], without and with a varying CMB lensing consistency parameter $A_L$, using combinations of Planck PR4 cosmic microwave background (CMB) anisotropy data (PR4 and lensing), and a compilation of non-CMB observations composed of baryon acoustic oscillation (BAO) measurements that do not include DESI BAO data, Pantheon+ type Ia supernova observations, Hubble parameter $H(z)$ measurements, and growth rate $f\sigma_8$ data points. We also compare results from earlier Planck PR3 data with those obtained using PR4 data in order to assess the stability of cosmological constraints when moving from PR3 to PR4. For the largest data combinations, PR3/PR4+lensing+non-CMB, the cosmological parameters inferred from PR3 and PR4 data are consistent, almost all differing by $1\sigma$ or less. When $A_L$ is allowed to vary, PR4-based data show a weaker preference for anomalous $A_L>1$ values than do PR3-based data. For the $\Lambda$CDM$+A_L$ model using PR3/PR4+lensing+non-CMB data the inferred value changes from PR3-based $A_L=1.087 \pm 0.035$ ($2.5\sigma$ above unity) to the PR4-based $A_L=1.053 \pm 0.034$ ($1.6\sigma$ above unity), with similar reductions occurring in the $w_0$CDM$+A_L$ and $w_0 w_a$CDM$+A_L$ parameterizations. This indicates that the significance of the CMB lensing anomaly is reduced when PR4 data are used, although mild evidence for $A_L > 1$ remains when CMB and non-CMB data are combined. For the most restrictive PR4+lensing+non-CMB dataset and the $w_0 w_a$CDM parameterization with $A_L=1$, we find $w_0 = -0.863\pm0.060$ (quintessence-like and $2.3\sigma$ away from $w_0=-1$) and $w_0+w_a=-1.37^{+0.19}_{-0.17}$ (phantom-like and $1.9\sigma$ away from $w_0+w_a=-1$), suggesting that the current observations favor dynamical dark energy over a cosmological constant at about $1.8\sigma$, slightly smaller than the $2\sigma$ preference we obtained using PR3-based data. When we use the $w_0w_a$CDM$+A_L$ parameterization with these data we find $w_0=-0.877\pm 0.060$ (quintessence-like and $2.1\sigma$ away from $w_0=-1$) and $w_0 + w_a =-1.29_{-0.17}^{+0.20}$ (phantom-like and $1.5\sigma$ away from $w_0+w_a=-1$), corresponding to a preference for dynamical dark energy over a cosmological constant of about $1.5\sigma$ (similar to what we found earlier from PR3-based data) and with $A_L = 1.042 \pm 0.037$ exceeding unity at $1.1\sigma$ (smaller than the $2.0\sigma$ significance we found earlier from PR3-based data). The reduction in evidence for a deviation from the $\Lambda$ point at $w_0 = -1$ and $w_a = 0$ is because the mean $w_0$ and $w_a$ values move closer to $w_0 = -1$ and $w_a = 0$ when $A_L$ is allowed to vary and not because the error bars become larger when $A_L$ is allowed to vary.  These results indicate that while these PR4 data mildly favor a time-evolving dark energy component, part of this preference may be associated with possible residual excess smoothing present in the Planck PR4 CMB anisotropy spectra.
 
\end{abstract}

\maketitle

\maxdeadcycles=200

\section{Introduction}
\label{sec:Introduction} 

On macroscopic scales general relativity is the current best theory of gravity and so currently provides the framework for the standard model of cosmology. This spatially-flat $\Lambda$CDM model \cite{Peebles:1984ge} has flat spatial hypersurfaces and is characterized by six cosmological parameters. Photons, neutrinos, ordinary baryonic matter, cold dark matter (CDM), and a cosmological constant $\Lambda$ contribute to the flat $\Lambda$CDM model cosmological energy budget. The $\Lambda$ contribution dominates now and is responsible for the observed currently accelerated cosmological expansion.

The $\Lambda$CDM model provides an excellent description of a wide range of cosmological observations, including cosmic microwave background (CMB) anisotropies, baryon acoustic oscillations (BAO) imprinted on large-scale structure, type Ia supernova (SNIa) luminosity distances, and Hubble parameter measurements. Joint analyzes of CMB data from the Planck satellite together with other non-CMB measurements allow for very precise determinations of the six cosmological parameters of this model. Nevertheless, despite its success in passing most observational tests, the $\Lambda$CDM model still faces several conceptual and observational challenges, with some recent measurements questioning whether some of the model's predictions remain fully consistent with current data \cite{Hu:2023jqc, CosmoVerseNetwork:2025alb}. 

Among the contributors to the cosmological energy budget, $\Lambda$ stands out as a parameter, unlike the photons, neutrinos, and ordinary baryonic matter, as well as possibly the CDM, which are spacetime fields. This was one motivation for replacing $\Lambda$ by a dynamical dark energy scalar field, \cite{Peebles:1987ek, Ratra:1987rm}, with the currently-dominating evolving (and spatially inhomogeneous) scalar field, $\phi$, energy density, comprising of a potential energy part and a kinetic energy part, powering the observed currently accelerated cosmological expansion.\footnote{For discussions of observational constraints on dynamical dark energy scalar field models see \cite{Sola:2016hnq, Ooba:2017lng, Ooba:2018dzf, Park:2018fxx, Park:2018tgj, SolaPeracaula:2018wwm, Park:2019emi, Khadka:2020whe, Cao:2022ugh, Berghaus:2024kra, Shlivko:2024llw, Bhattacharya:2024hep, VanRaamsdonk:2024sdp, DESI:2025fii, Gialamas:2025pwv, Park:2025fbl, Zhang:2025lam, Wang:2026sqy} and references therein.} While such $\phi$CDM models are physically consistent and complete, dynamical dark energy fluid parameterizations, that are physically incomplete, have attracted more attention, perhaps partially because of their relative simplicity.

XCDM or $w_0$CDM is the simplest dynamical dark energy fluid parameterization, with a constant equation of state parameter, $w_0 = p/\rho$, where $p$ and $\rho$ are the pressure and energy density of the fluid, and here $w_0$ need not equal the $\Lambda$ value of $w_0 = -1$, but must be sufficiently negative to provide accelerated cosmological expansion.

Extensions of the $w_0$CDM parameterization have also been studied. A popular two-parameter extension is the $w_0w_a$CDM parameterization, \cite{Chevallier:2000qy, Linder:2002et}, where the equation of state parameter, $w(z) = w_0 + w_a (1-a) = w_0 + w_az/(1+z) $, now also retains the next term in a Taylor expansion of $w(z)$ in terms of $(1 - a)$, where $a$ is the scale factor, $z$ is the redshift, and $w_a$ is an additional parameter. Part of the motivation in considering $w_0w_a$CDM was the interest in seeing what happens if $w(z)$ was not a constant but instead varied with redshift. Parameterizations with other combinations of terms in the $(1-a)$ Taylor expansion of $w(z)$ have also been studied \cite{Park:2025azv}. However, as discussed next, the $w_0w_a$CDM and other such $w(z)$CDM fluid  parameterizations are also physically incomplete.

The speed of sound squared in a fluid is given by $c_s^2 = \partial p/\partial\rho$, in the rest frame and considering adiabatic perturbations, so in the $w_0$CDM parameterization with negative $p$, $c_s$ is imaginary and the parameterization is inconsistent with the observations. Consequently, a modified $w_0$CDM parameterization is defined by setting $c_s = 1$ (in the fluid rest frame) in the equations that govern the spatial inhomogeneities. While this makes the parameterization physically consistent, it introduces an additional free parameter, $c_s$, that has been arbitrarily set to unity in order for the perturbed density and pressure to be interpreted as those of a scalar field. A similar issue arises in the $w_0w_a$CDM and other $w(z)$CDM parameterizations. On the other hand, $\phi$CDM models, \cite{Peebles:1987ek, Ratra:1987rm}, do not have to deal with this incompleteness/inconsistency issue and are guaranteed to result in a physically sensible $c_s$ that does not have to be arbitrarily defined. In any scalar field $\phi$CDM model with a canonical kinetic term (where the scalar field Lagrangian density takes the form $\mathcal{L}(\phi,X)=X - V(\phi)$, with $X$ being the kinetic term and $V(\phi)$ the potential term), under the conditions mentioned before, i.e., in the rest frame of the scalar field and considering adiabatic perturbations, the speed of sound squared is $c^2_s=1$. This means that perturbations associated to the scalar field do not grow efficiently, and the clustering is very small. On the other hand, the scalar field can indirectly suppress the growth of spatial inhomogeneities by affecting the accelerated expansion of the universe.

In addition to the theoretical motivation for considering a dynamical dark energy $\phi$CDM model, in the last few years there has also been observational evidence that favors dynamical dark energy over a $\Lambda$, at the $\sim 2\sigma$ level \cite{Cao:2023eja, Rubin:2023jdq, deCruzPerez:2024abc, DESI:2024mwx, Park:2024jns}, and now at the $\sim 3\sigma$ level when including the DESI collaboration DR2 BAO data \cite{DESI:2025zgx}. For other discussions of the DESI results see \cite{Bansal:2025ipo, Chakraborty:2025syu, Borghetto:2025jrk, Ishiyama:2025bbd, Moghtaderi:2025cns, Urena-Lopez:2025rad, Paliathanasis:2025dcr, Shah:2025ayl, Shlivko:2025fgv, Wolf:2025jed, Mirpoorian:2025rfp, RoyChoudhury:2025dhe, Liu:2025mub, Ye:2025ark, Cheng:2025lod, Wang:2025vfb, Mukherjee:2025ytj, vanderWesthuizen:2025iam, Cai:2025mas, Gonzalez-Fuentes:2025lei, Barua:2025ypw, Ozulker:2025ehg, Liu:2025myr, Mishra:2025goj, Qiang:2025cxp, Lee:2025pzo, Chaudhary:2025pcc, Wang:2025znm, RoyChoudhury:2025iis, Dong:2025ukl, Cheng:2025yue, deCruzPerez:2025dni, Gokcen:2026pkq, Avsajanishvili:2026fqi} and references therein.

The DESI collaboration uses the $w_0w_a$CDM parameterization to determine whether dark energy is dynamical. In \cite{Park:2024jns}, we analyze a combination of Planck PR3 CMB data and a compilation of non-CMB observations that include BAO, Pantheon+ SNIa, Hubble parameter, and $f\sigma_8$ growth factor measurements, but do not include the recent DESI BAO measurements. With the spatially-flat $w_0w_a$CDM parameterization, our data compilation provides slightly more restrictive constraints, giving $w_0=-0.850\pm 0.059$, $w_a=-0.59^{+0.26}_{-0.22}$, and $w_0+w_a=-1.44^{+0.20}_{-0.17}$, than the DESI collaboration DESI(DR1)+CMB+PantheonPlus dataset results, \cite{DESI:2024mwx}, $w_0 = -0.827 \pm 0.063$ and $w_a = -0.75^{+0.29}_{-0.25}$, with both analyses indicating a $\sim 2\sigma$ preference for dynamical dark energy over a $\Lambda$. Since we did not use the DESI BAO measurements, our results provide independent confirmation of the results of \cite{DESI:2024mwx}, and also show that the evidence for dark energy dynamics does not depend on using DESI BAO measurements but is instead a more general feature of current cosmological observations. In \cite{Park:2024jns}, we also showed that this $\sim 2\sigma$ preference for dynamical dark energy over a $\Lambda$ also does not depend on Pantheon+ SNIa data, also see \cite{Shlivko:2026jxa}. Depending on the data compilation used, dark energy dynamics is favored at $\sim 3\sigma$ when including DESI DR2 BAO data \cite{DESI:2025zgx}.

In the $w_0w_a$CDM parameterization, $w(z) \sim w_0$ at low $z\sim 0$ while $w(z) = w_0 + w_a$ at $z \gg 1$, so the two-parameter $w_0w_a$CDM parameterization behaves like single-parameter $w_0$CDM and $(w_0+w_a)$CDM parameterizations at low and high $z$, respectively. From our results listed in the previous paragraph, the low-$z$ $w_0$CDM parametrization is quintessence-like while the high-$z$ $(w_0+w_a)$CDM parameterization is phantom-like. These are consistent with the single-parameter $w_0$CDM results \cite{deCruzPerez:2024abc}, where low-$z$ non-CMB data favor quintessence-like $(w > -1)$ dark energy dynamics while high-$z$ CMB data favor phantom-like $(w < -1)$ dark energy dynamics and the phantom-divide crossing seen in the two-parameter $w_0w_a$CDM parameterization follows from what is seen in the one-parameter $w_0$CDM parameterization. There are models that do not have a phantom-divide crossing that are reasonably consistent with these data, see \cite{Park:2025fbl, Wang:2026sqy}.

In the $w_0$CDM parameterization, the non-CMB data and the CMB data constraints differ by more than $3\sigma$, \cite{deCruzPerez:2024abc}. To try to more carefully understand this we also used our data compilation to constrain the $w_0$CDM$+A_L$ parameterization, where $A_L$ is the CMB weak lensing consistency parameter\footnote{The CMB weak lensing consistency amplitude parameter $A_L$ is a phenomenological parameter used in CMB analyses to test the consistency of gravitational lensing effects in the data. It is defined as a rescaling factor of the CMB lensing potential power spectrum $C_{\ell}^{\phi\phi} \rightarrow A_L C_{\ell}^{\phi\phi}$, where $\phi({\hat n})$ is the lensing potential and $C_{\ell}^{\phi\phi}$ is its angular power spectrum.  $A_L=1$ is the expected value in the best-fit cosmological model and $A_L > 1$ implies more lensing than predicted in the best-fit cosmological model, implying excessive smoothing of CMB acoustic peaks.} \cite{Calabreseetal2008} that is now allowed to vary and be determined from these data. It is known that Planck PR3 data favors $A_L > 1$ at $> 2\sigma$ significance, qualitatively consistent with the observed excess smoothing of some of the Planck data relative to what is predicted in the best-fit cosmological model. In the $w_0$CDM$+A_L$ parameterization and our data compilation we find that allowing $A_L$ to vary and be determined by these data alters the CMB data constraints enough to reduce the inconsistency with the non-CMB data constraints to less than our $3\sigma$ threshold, with the dark energy now being quintessence-like, $w_0 = -0.968 \pm 0.24$, but with $A_L = 1.101 \pm 0.037$ (for PR3+lensing+non-CMB data), $2.7\sigma$ greater than unity, \cite{deCruzPerez:2024abc}. 

While non-CMB data and CMB data constraints in the $w_0w_a$CDM parameterization are inconsistent to less than our $3\sigma$ threshold, \cite{Park:2024jns}, it is useful to study these data constraints in the $w_0w_a$CDM$+A_L$ parameterization and investigate the relationship between the evidence for dark energy dynamics and the value of the CMB lensing consistency parameter $A_L$. In \cite{Park:2024pew}, for our PR3 based data compilation, we find that when $A_L$ is allowed to vary in the $w_0w_a$CDM$+A_L$ parameterization, the evidence for dark energy dynamics over a $\Lambda$ decreases to $\sim 1.5\sigma$ (compared to the $\sim 2\sigma$ evidence for the $w_0w_a$CDM parameterization case) and that $A_L > 1$ is favored at $\sim 2\sigma$, indicating that these data favor more weak lensing of the CMB than that predicted by the best-fit model.\footnote{It is interesting that in the $\phi$CDM$(+A_L)$ model, where dark energy dynamics can only be quintessence-like, the opposite happens in that the evidence for dark energy dynamics increases (not decreases) from $1.3\sigma$ for $\phi$CDM to $1.7\sigma$ for $\phi$CDM$+A_L$ (with $A_L$ greater than unity at $2.8\sigma$) for our PR3+lensing+non-CMB data compilation \cite{Park:2025fbl}.} This suggests that at least part of the support for dark energy dynamics in the $w_0w_a$CDM parameterization comes from the observed excess smoothing of some of the Planck CMB anisotropy data. For related analyses and results, see \cite{RoyChoudhury:2024wri, Park:2025azv, RoyChoudhury:2025dhe, RoyChoudhury:2025iis}.

From our PR3 CMB data (without CMB lensing data) analysis in the flat $\Lambda$CDM$+A_L$ model we find $A_L = 1.181 \pm 0.067$, $2.7\sigma$ larger than unity, \cite{deCruzPerez:2024abc}. The newer Planck NPIPE pipeline PR4 data gives $A_L = 1.039 \pm 0.052$, only $0.75\sigma$ larger than unity, \cite{Tristram:2023haj}.\footnote{We note that smaller angular scale CMB experiments give $\Lambda$CDM$+A_L$ model $A_L$ values consistent with unity, with $A_L = 1.007 \pm 0.057$ for ACT \cite{AtacamaCosmologyTelescope:2025blo} and $A_L = 0.972^{+0.079}_{-0.089}$ for SPT-3G D1 \cite{SPT-3G:2025bzu}.} In this paper we repeat our analyses of the $\Lambda$CDM$(+A_L)$ models and the $w_0$CDM$(+A_L)$ and $w_0w_a$CDM$(+A_L)$ parameterizations, but now using PR4 CMB data, \cite{Tristram:2023haj}, instead of PR3 CMB data, to determine what the reduced PR4 evidence for $A_L$ does to evidence for dark energy dynamics. In our analyses here we retain the non-CMB data compilation of \cite{deCruzPerez:2024abc}, that does not include DESI BAO data, as we are also interested in determining how significantly the cosmological constraints change when we replace PR3 data by PR4 data, but hold everything else fixed.

In this paper we compare our Planck PR4-based data results with those we obtained using earlier Planck PR3-based data in order to assess the robustness of cosmological constraints. For the largest data combinations, PR3/PR4+lensing+non-CMB, and with $A_L = 1$, the measured cosmological parameters from the PR3-based analysis and the PR4-based analysis show consistency at the $1\sigma$ level or better, while when $A_L$ is allowed to vary and also be determined from these data the biggest differences are in the physical baryonic matter density parameter $\Omega_b h^2$, but are still small at $1.2\sigma$ or less. When the CMB lensing consistency parameter $A_L$ is allowed to vary PR4-based data show a reduced preference for anomalous $A_L > 1$ values compared to PR3-based data, indicating that the significance of the CMB lensing anomaly is smaller in the PR4 datasets, as expected from the analyses of \cite{Tristram:2023haj}.

For the most restrictive PR4+lensing+non-CMB data set, the $w_0 w_a$CDM parameterization with $A_L=1$ yields quintessence-like $w_0=-0.863\pm 0.060$ and phantom-like $w_0 + w_a =-1.37_{-0.17}^{+0.19}$ corresponding to a preference for dynamical dark energy over a cosmological constant of about $1.8\sigma$, slightly smaller than the $2\sigma$ preference inferred using PR3 data \cite{Park:2024jns}. When we consider the $w_0w_a$CDM$+A_L$ parameterization we find for these data a quintessence-like $w_0=-0.877\pm 0.060$ and a phantom-like $w_0 + w_a =-1.29_{-0.17}^{+0.20}$ corresponding to a preference for dynamical dark energy over a cosmological constant of about $1.5\sigma$ and with $A_L = 1.042 \pm 0.037$ exceeding unity at $1.1\sigma$. We emphasize that the reduction in evidence for a deviation from the $\Lambda$ point at $w_0 = -1$ and $w_a = 0$ comes about because the mean $w_0$ and $w_a$ values move closer to $w_0 = -1$ and $w_a = 0$ when $A_L$ is allowed to vary and not because the error bars become larger when $A_L$ is allowed to vary. While the PR3 analysis favored an $A_L$ larger than unity at $2.0\sigma$ significance, it favored a similar (reduced) $1.5\sigma$ deviation towards dark energy dynamics away from $\Lambda$, \cite{Park:2024pew}. These results indicate that while these observations mildly favor a time-evolving dark energy component, part of this preference may be associated with the residual excess smoothing present in the Planck CMB anisotropy spectra, even when we use Planck PR4 data, in agreement with our earlier conclusions \cite{Park:2024pew}, also see \cite{RoyChoudhury:2024wri, Park:2025azv, RoyChoudhury:2025dhe, RoyChoudhury:2025iis}.

It is important to bear in mind that our results are not that statistically significant and that $w_0w_a$CDM$(+A_L)$ is not a physically consistent cosmological model but rather just a redshift-dependent parameterization of a dynamical dark energy equation of state that is somewhat arbitrarily modified by setting $c_s^2$ to unity. However, the small effect we have discovered is non-negligible and should be better understood.

A brief description of the structure of our article follows. In Sec.\ \ref{sec:Data} we provide general details of the different datasets we use to constrain the cosmological parameters and also to test the models under study. A brief summary of the main features of the analysis can be found in Sec.\ \ref{sec:Methods}. In Sec.\ \ref{sec:ResultsandDiscussion} our main results are presented and discussed and finally in Sec.\ \ref{sec:Conclusion} we deliver our conclusions.

\section{Data}
\label{sec:Data}

In this section we list and briefly describe the CMB datasets we use as well as list the non-CMB datasets we use but refer the reader to Sec.\ II of \cite{deCruzPerez:2024abc} for more details. When the covariance matrices are available, we use them in all of our analyses. 
\newline
\newline 
{\bf PR4.} We consider a combination of PR3 Planck data (Planck 2018) \cite{Planck:2018vyg} and reanalyzed PR4 data \cite{Tristram:2023haj} for the temperature, polarization and the cross spectra, obtained using the new NPIPE pipline. From PR3 we utilize the TT power spectra at low-$\ell$ ($2\leq \ell \leq 30$) and from PR4 we use the \texttt{LoLLiPoP} likelihood, containing the EE power spectra at low-$\ell$ ($2\leq \ell \leq 30$), and the \texttt{HiLLiPoP} likelihood, which provides the TT power spectra at high-$\ell$ ($30\leq \ell \leq 2500$), as well as the TE and EE power spectra at high-$\ell$ ($30\leq \ell \leq 2000$). In \texttt{Cobaya} the corresponding likelihoods are named \texttt{planck\_2018\_lowl.TT}, \texttt{planck\_2020\_lollipop.lowlE}, and \texttt{planck\_2020\_hillipop.TTTEEE}. Although PR3 data are also utilized, hereafter we jointly denote these data as PR4 to shorten the notation. Our PR4 dataset is identical to the TTTEEE dataset of \cite{Tristram:2023haj} in \texttt{Cobaya}. 
\newline 
\newline
{\bf (PR4) lensing.} From the analysis of the Planck PR4 CMB maps \cite{Carron:2022eyg} we use the extracted lensing potential power spectrum. In \texttt{Cobaya} the corresponding likelihood is referred to as \texttt{PlanckPR4Lensing}. When combined with PR4 data above, our PR4+lensing dataset is identical to the TTTEEE+lensing dataset of \cite{Tristram:2023haj} in \texttt{Cobaya}. 
\newline
\newline 
{\bf Non-CMB.} This combination of data is the one denoted as non-CMB (new) data in reference \cite{deCruzPerez:2024abc}. Non-CMB data are comprised of 

\begin{itemize}
\item 16 BAO data points from isotropic and anisotropic analyses, spanning $0.122 \le z \le 2.334$, listed in Table I of \cite{deCruzPerez:2024abc}. We do not use DESI 2024 or DR2 BAO data, \cite{DESI:2024mwx, DESI:2025zgx}.
\item From the Pantheon+ compilation \cite{Brout:2022vxf} a subset of 1590 SNIa data points, obtained after removing those SNIa at $z < 0.01$, so that the impact of the dependence on the modeling of peculiar velocities is reduced. The range covered by these data is $0.01016 \le z \le 2.26137$.
\item From the differential age technique a compilation of 32 Hubble parameter [$H(z)$]  measurements, with redshifts $0.070 \le z \le 1.965$. These values are listed in Table 1 of \cite{Cao:2023eja} and also in Table II of \cite{deCruzPerez:2024abc}. 
\item 9 uncorrelated growth rate ($f\sigma_8$) data points covering $0.013 \le z \le 1.36$. The complete list is provided in Table III of \cite{deCruzPerez:2024abc}.
\end{itemize}
We constrain the parameter spaces of the $\Lambda$CDM(+$A_L$) models and the $w_0$CDM(+$A_L$) and $w_0w_a$CDM(+$A_L$) parameterizations, using five different combinations of data, namely: non-CMB, PR4, PR4+lensing, PR4+non-CMB, and PR4+lensing+non-CMB.

\section{Methods}
\label{sec:Methods} 

Here we present a brief summary of the methods we use in our study. To constrain the parameter space of the different models under study, by testing them against the combinations of observational data described in the previous section, we jointly use the \texttt{CLASS} \cite{Blas:2011rf} and \texttt{Cobaya} \cite{Torrado:2020dgo} codes. \texttt{CLASS} is an Einstein-Boltzmann system solver whose main purpose is to compute the evolution of the universe at the background and perturbation level in a given cosmological model and so compute observable quantities as a function of the set of cosmological parameters in that model. Then \texttt{Cobaya}, through the use of the Markov chain Monte Carlo (MCMC) algorithm, uses the predicted observables to extract from the utilized datasets an estimate of the cosmological parameter values for the model under study. As a convergence criterion, for the Gelman and Rubin $R$ estimator, we consider $R-1 <0.01$ in most cases (exceptions are discussed below).  Once the converged chains are obtained, we utilize the \texttt{GetDist} code \cite{Lewis:2019xzd} to extract average values, confidence intervals, and likelihood distributions of the cosmological model parameters. 

The six primary parameters we choose for the standard spatially-flat $\Lambda$CDM model are the current value of the physical baryonic matter and cold dark matter density parameters, $\Omega_b h^2$ and $\Omega_c h^2$, respectively ($h$ is the Hubble constant $H_0$ in units of 100 km s$^{-1}$ Mpc$^{-1}$), the present value of the Hubble parameter $H_0$, the reionization optical depth $\tau$, the primordial scalar-type perturbation power spectral index $n_s$, and the power spectrum amplitude $A_s$. For all the parameters considered we use flat priors, non-zero within $0.005 \le \Omega_b h^2 \le 0.1$, $0.001 \le \Omega_c h^2 \le 0.99$, $20\leq H_0[\text{km/s/Mpc}]\leq 100$, $0.01 \le \tau \le 0.8$, $0.8 \le n_s \le 1.2$, and $1.61 \le \ln(10^{10} A_s) \le 3.91$. 

We study two spatially-flat dynamical dark energy parameterizations, where the dark energy fluid is assumed to be perfect and the equation of state parameter is $w\neq -1$. ($w$ is the ratio of the dark energy fluid pressure and energy density.) The simplest parameterization is known as $w_0$CDM (or $w$CDM or XCDM) and has a constant equation of state parameter but is free to vary from $w_0=-1$.\footnote{The simplest version of this parameterization has an imaginary speed of sound, resulting in observationally inconsistent rapidly growing spatial inhomogeneities, and must be arbitrarily modified to deal with this issue.} The second one is the $w_0w_a$CDM parameterization with a time-evolving equation of state parameter $w(z) = w_0 + w_a z/(1+z)$, \cite{Chevallier:2000qy, Linder:2002et}.\footnote{This parameterization is also physically inconsistent.} For the varying dark energy equation of state parameters we adopt flat priors again, being non-zero over $-3.0 \le w_0 \le 0.2$ and $-3 < w_a < 2$. 

For the $\Lambda$CDM model and the two dynamical dark energy parameterizations we additionally also consider the variation of the CMB weak lensing consistency parameter $A_L$, \cite{Calabreseetal2008} (with these cases indicated by adding $+A_L$ to the model name), with a flat prior non-zero over $0 \le A_L \le 10$. 

Since the non-CMB data we use are incapable of constraining the values of $\tau$ and $n_s$, in the corresponding analyses we fix the values of these parameters to the values obtained from PR4 data. We also provide constraints on three derived parameters, namely: the angular size of the sound horizon evaluated at recombination $\theta_{\rm rec}$, the current value of the non-relativistic matter density parameter $\Omega_m$, and the amplitude of matter fluctuations $\sigma_8$. These values are obtained from the values of the primary parameters of the cosmological model. Finally, for the $w_0w_a$CDM(+$A_L$) parameterization, we also compute the value of the sum of the dark energy equation of state parameters $w_0+w_a$, to which $w(z)$ asymptotes at high $z$. 

A comment about one of the derived parameters used in this work, $100\theta_\textrm{rec}$, is in order. In this paper we use \texttt{CLASS} to get the cosmological parameters constraints, and there it is possible to use two slightly different quantities related to the angular scale of the sound horizon. First we have \texttt{theta\_s\_100} that represents the angular scale of the sound horizon at recombination. The second one is \texttt{theta\_star\_100}, which is the same but computed at photon decoupling. In this work, we use the first one, and we denote it by $100\theta_\textrm{rec}$. This quantity differs from the $100 \theta_\textrm{MC}$ used in our previous works \cite{Park:2024jns, Park:2024pew, deCruzPerez:2022hfr, deCruzPerez:2024abc, Park:2025azv, Park:2025fbl} and in the Planck collaboration papers \cite{Planck:2018lbu}. In \texttt{CosmoMC}, $\theta_\textrm{MC}$ is not computed from the exact recombination or decoupling history; instead, it relies on the well-known analytic fitting formulae to approximate the decoupling redshift $z_*$, and thus $\theta_\textrm{MC}$ is only an approximate representation of the true angular sound horizon scale. Minor differences are expected when comparing  $100\theta_\textrm{rec}$ with the $100\theta_{*}$ used in \cite{Tristram:2023haj}; the latter corresponds to the same physical quantity as $\theta_\textrm{rec}$ but is computed using \texttt{CAMB}. Consequently there are small but non-negligible deviations between $100 \theta_*$, $100\theta_\textrm{rec}$, and $100\theta_\textrm{MC}$.

In our analyses we use \texttt{GetDist} to compute marginalized one-dimensional constraints. While \texttt{Cobaya} does not provide separate information for the $1\sigma$ and $2\sigma$ limits, \texttt{GetDist} determines the $1\sigma$ and $2\sigma$ limits directly from the smoothed marginalized posteriors. It is important to note that \texttt{Cobaya} smooths the likelihood near hard prior boundaries, causing the marginalized likelihood to approach zero at these boundaries. This behavior differs from our previous analyses (e.g., \cite{Park:2024pew}), where the likelihood could remain large at the edge of the allowed prior range. For certain parameters and data combinations, most notably $H_0$, $w_0$, $w_a$, $\Omega_m$, and $\sigma_8$ in the PR4 and PR4+lensing analyses, strong parameter degeneracies cause the one-dimensional marginalized distributions to be sharply truncated by the imposed priors (e.g., $H_0 < 100~ \textrm{km}\textrm{s}^{-1} \textrm{Mpc}^{-1}$ and $w_a > -3$).

We note that the priors imposed on the primary cosmological parameters induce  a cutoff-like behavior in the posterior distributions of derived parameters. For example, there are sharp lower cutoffs in $\Omega_m$ and upper cutoffs in $\sigma_8$ for $w_0$CDM($+A_L$) and $w_0 w_a$CDM($+A_L$) parameterizations with PR4(+lensing) data. In particular, strong prior bounds on parameters like the Hubble constant propagate through parameter degeneracies and produce apparent truncations that are not driven by observational data. Since this effect is secondary and prior-induced, quoting $2\sigma$ limits for these derived parameters does not have the usual statistical interpretation and is therefore not reported here. Similar prior cutoff effects were observed in our previous analyses, where $2\sigma$ constraints on these parameters were likewise not reported.

For all the flat models under study, the primordial scalar-type energy density perturbation power spectrum has the form
\begin{equation}
    P_\delta (k) = A_s \left( \frac{k}{k_0} \right)^{n_s},
\label{eq:powden-flat}
\end{equation}
where $k$ is wavenumber and $n_s$ and $A_s$ are the spectral index and the amplitude of the spectrum at pivot scale $k_0=0.05~\textrm{Mpc}^{-1}$. This power spectrum is generated by quantum fluctuations during an early epoch of power-law inflation in a spatially-flat inflation model powered by a scalar field inflaton potential energy density that is an exponential function of the inflaton \cite{Lucchin:1984yf, Ratra:1989uv, Ratra:1989uz}.

In order to compare the performance of the models when it comes to fitting the different datasets considered in this work, we utilize the deviance information criterion difference ($\Delta$DIC) between the deviance information criterion (DIC) values for the model under study and for the flat $\Lambda$CDM model. See Sec.\ III of \cite{deCruzPerez:2024abc}, and references therein, for an extended description of this statistical estimator. According to Jeffreys' scale, when $-2 \leq \Delta\textrm{DIC}<0$ there is {\it weak} evidence in favor of the model under study. For $-6 \leq \Delta\textrm{DIC} < -2$ there is {\it positive} evidence, for $-10\leq\Delta\textrm{DIC} < -6$ there is {\it strong} evidence, and when $\Delta\textrm{DIC} < -10$ we can claim {\it very strong} evidence in favor of the model under study relative to the $\Lambda$CDM model. Conversely, if $\Delta\textrm{DIC}$ values are positive, the $\Lambda$CDM model is favored over the model under study. 

We want to determine the consistency of two sets of cosmological parameter constraints, obtained using two different data sets, in a given model. In earlier work we used two different statistical estimators for this purpose, \cite{Park:2024jns, Park:2024pew, deCruzPerez:2022hfr, deCruzPerez:2024abc, Park:2025azv, Park:2025fbl}.  The first of them, $\log_{10}\mathcal{I}$, is based on DIC values (see \cite{Joudaki:2016mvz} and Sec.\ III of \cite{deCruzPerez:2024abc}). While positive values ($\log_{10}\mathcal{I}>0$) indicate consistency between the two data sets, negative values ($\log_{10}\mathcal{I}<0$), on the other hand, indicate inconsistency. According to Jeffreys' scale, the degree of concordance or discordance between two data sets is classified as {\it substantial} if $\lvert \log_{10}\mathcal{I} \rvert >0.5$, {\it strong} if $\lvert \log_{10}\mathcal{I} \rvert >1$, and {\it decisive} if $\lvert \log_{10}\mathcal{I} \rvert >2$ \cite{Joudaki:2016mvz}. The second estimator considered is known as the tension probability $p$ and the related Gaussian approximation "sigma value" $\sigma$ (see \cite{Handley:2019pqx, Handley:2019wlz, Handley:2019tkm} and also Sec.\ III of \cite{deCruzPerez:2024abc}). Roughly speaking, a value of $p=0.05$ corresponds to $2\sigma$ and $p=0.003$ corresponds to a 3$\sigma$ Gaussian standard deviation. Here we emphasize that the characteristics of PR4 likelihood data differ from those of PR3 (or P18) data. For PR4 data, the $\chi^2$ minimum often tends to increase as the number of model parameters increases. This contrasts with what happens with PR3 data, where adding parameters generally leads to a lower $\chi^2$ minimum. The $\chi^2$ values of each element in the MCMC chains obtained from PR4 data has different characteristics compared to those obtained from previous PR3 data. This leads to problems where the second consistency check estimator used in our previous papers are not properly measured using PR4 data. Therefore, in this study, we utilize the DIC along with the $\log_{10} \mathcal{I}$ statistics calculated from the DICs, to measure consistency between data sets.


\begin{table*}
	\caption{Mean and 68\% confidence limits of $\Lambda$CDM model parameters
        from non-CMB, PR4, PR4+lensing, PR4+non-CMB, and PR4+lensing+non-CMB data.
        $H_0$ has units of km s$^{-1}$ Mpc$^{-1}$. We also include the values of $\chi^2_{\text{min}}$ and DIC.
}
\begin{ruledtabular}
\begin{tabular}{lccccc}
  Parameter                       &  Non-CMB                     & PR4                         &  PR4+lensing               &  PR4+non-CMB            & PR4+lensing+non-CMB     \\[+1mm]
\hline \\[-1mm]
  $\Omega_b h^2$                  & $0.0258^{+0.0036}_{-0.0031}$ & $0.02225 \pm 0.00013$       & $0.02223 \pm 0.00014$      &  $0.02231 \pm 0.00012$  &  $0.02231 \pm 0.00012$  \\[+1mm]
  $\Omega_c h^2$                  & $0.1211\pm 0.0081$           & $0.1188 \pm 0.0012$         & $0.1191 \pm 0.0011$        &  $0.11780 \pm 0.00083$  &  $0.11804 \pm 0.00082$  \\[+1mm]
  $H_0$                           & $70.6^{+2.6}_{-2.2}$         & $67.65 \pm 0.56$            & $67.54 \pm 0.52$           &  $68.11 \pm 0.38$       &  $68.01 \pm 0.37$       \\[+1mm]
  $\tau$                          & $0.0577$                     & $0.0577 \pm 0.0064$         & $0.0588 \pm 0.0062$        &  $0.0583 \pm 0.0062$    &  $0.0598 \pm 0.0060$    \\[+1mm]
  $n_s$                           & $0.9675$                     & $0.9675 \pm 0.0042$         & $0.9670 \pm 0.0040$        &  $0.9698 \pm 0.0035$    &  $0.9693 \pm 0.0035$    \\[+1mm]
  $\ln(10^{10} A_s)$              & $3.01\pm 0.12$               & $3.039 \pm 0.014$           & $3.045 \pm 0.012$          &  $3.037 \pm 0.014$      &  $3.046 \pm 0.012$      \\[+1mm]
\hline \\[-1mm]
  $100\theta_\textrm{rec}$        & $1.0429\pm 0.0091$           & $1.04182 \pm 0.00025$       & $1.04179 \pm 0.00025$      &  $1.04189 \pm 0.00024$  &  $1.04187 \pm 0.00024$  \\[+1mm]
  $\Omega_m$                      & $0.296\pm 0.011$             & $0.3097 \pm 0.0076$         & $0.3112 \pm 0.0071$        &  $0.3035 \pm 0.0050$    &  $0.3048 \pm 0.0049$    \\[+1mm]
  $\sigma_8$                      & $0.787\pm 0.026$             & $0.8059 \pm 0.0067$         & $0.8090 \pm 0.0051$        &  $0.8024 \pm 0.0061$    &  $0.8065 \pm 0.0049$    \\[+1mm]
\hline\\[-1mm]
  $\chi_{\textrm{min}}^2$         & $1460.61$                    & $30548.16$                  & $30558.58$                        & $32010.06$              &  $32021.05$      \\[+1mm]
  $\textrm{DIC}$                  & $1468.20$                    & $30601.92$                  & $30609.10$                 & $32066.16$              &  $32074.05$             \\[+1mm]
\end{tabular}
\\[+1mm]
\begin{flushleft}
\end{flushleft}
\end{ruledtabular}
\label{tab:results_flat_LCDM}
\end{table*}


\begin{table*}
	\caption{Mean and 68\% confidence limits of $\Lambda$CDM$+A_L$ model parameters
        from non-CMB, PR4, PR4+lensing, PR4+non-CMB, and PR4+lensing+non-CMB data.
        $H_0$ has units of km s$^{-1}$ Mpc$^{-1}$. We also include the values of $\chi^2_{\text{min}}$, DIC, and $\Delta$DIC the difference with respect to the $\Lambda$CDM model.
}
\begin{ruledtabular}
\begin{tabular}{lccccc}
  Parameter                       &  Non-CMB                     & PR4                         &  PR4+lensing                 &  PR4+non-CMB            & PR4+lensing+non-CMB     \\[+1mm]
\hline \\[-1mm]
  $\Omega_b h^2$                  & $0.0258^{+0.0036}_{-0.0031}$ & $0.02228^{+0.00017}_{-0.00015}$ & $0.02229^{+0.00016}_{-0.00014}$ &  $0.02236 \pm 0.00012$ &  $0.02236 \pm 0.00012$  \\[+1mm]
  $\Omega_c h^2$                  & $0.1211\pm 0.0081$           & $0.1184 \pm 0.0014$         & $0.1183^{+0.0013}_{-0.0015}$ &  $0.11753 \pm 0.00086$  &  $0.11750 \pm 0.00086$  \\[+1mm]
  $H_0$                           & $70.6^{+2.6}_{-2.2}$         & $67.85 \pm 0.65$            & $67.92^{+0.66}_{-0.59}$      &  $68.25 \pm 0.40$       &  $68.26 \pm 0.39$       \\[+1mm]
  $\tau$                          & $0.0577$                     & $0.0572 \pm 0.0063$         & $0.0573 \pm 0.0063$          &  $0.0568 \pm 0.0062$    &  $0.0568 \pm 0.0062$    \\[+1mm]
  $n_s$                           & $0.9675$                     & $0.9686 \pm 0.0046$         & $0.9690^{+0.0047}_{-0.0042}$ &  $0.9710 \pm 0.0036$    &  $0.9709 \pm 0.0035$    \\[+1mm]
  $\ln(10^{10} A_s)$              & $3.01\pm 0.12$               & $3.036 \pm 0.015$           & $3.037 \pm 0.015$            &  $3.033 \pm 0.014$      &  $3.033 \pm 0.014$      \\[+1mm]
  $A_L$                           & $1$                          & $1.030 \pm 0.054$           & $1.037 \pm 0.038$            &  $1.051 \pm 0.049$      &  $1.053 \pm 0.034$      \\[+1mm] 
 \hline \\[-1mm] 
  $100\theta_\textrm{rec}$        & $1.0429\pm 0.0091$           & $1.04184 \pm 0.00026$       & $1.04185 \pm 0.00026$        &  $1.04191 \pm 0.00024$  &  $1.04191 \pm 0.00024$  \\[+1mm]
  $\Omega_m$                      & $0.296\pm 0.011$             & $0.3072^{+0.0081}_{-0.0092}$& $0.3062^{+0.0076}_{-0.0091}$ &  $0.3017 \pm 0.0051$    &  $0.3015 \pm 0.0051$    \\[+1mm]
  $\sigma_8$                      & $0.787\pm 0.026$             & $0.8036 \pm 0.0076$         & $0.8033 \pm 0.0075$          &  $0.7998 \pm 0.0064$    &  $0.7998 \pm 0.0064$    \\[+1mm]      \hline\\[-1mm]
  $\chi_{\textrm{min}}^2$         & $1460.61$                    & $30548.15$                  & $30557.01$                   & $32011.35$              &  $32018.44$             \\[+1mm]
  $\textrm{DIC}$                  & $1468.20$                    & $30602.82$                  & $30610.15$                   & $32063.57$              &  $32073.18$             \\[+1mm]
  $\Delta\textrm{DIC}$            & $-$                          & $+0.90$                     & $+1.05$                      & $-2.59$                 &  $-0.87$                \\[+1mm]
\end{tabular}
\\[+1mm]
\begin{flushleft}
\end{flushleft}
\end{ruledtabular}
\label{tab:results_flat_LCDM_Alens}
\end{table*}


\begin{figure*}[htbp]
\centering
\mbox{\includegraphics[width=175mm]{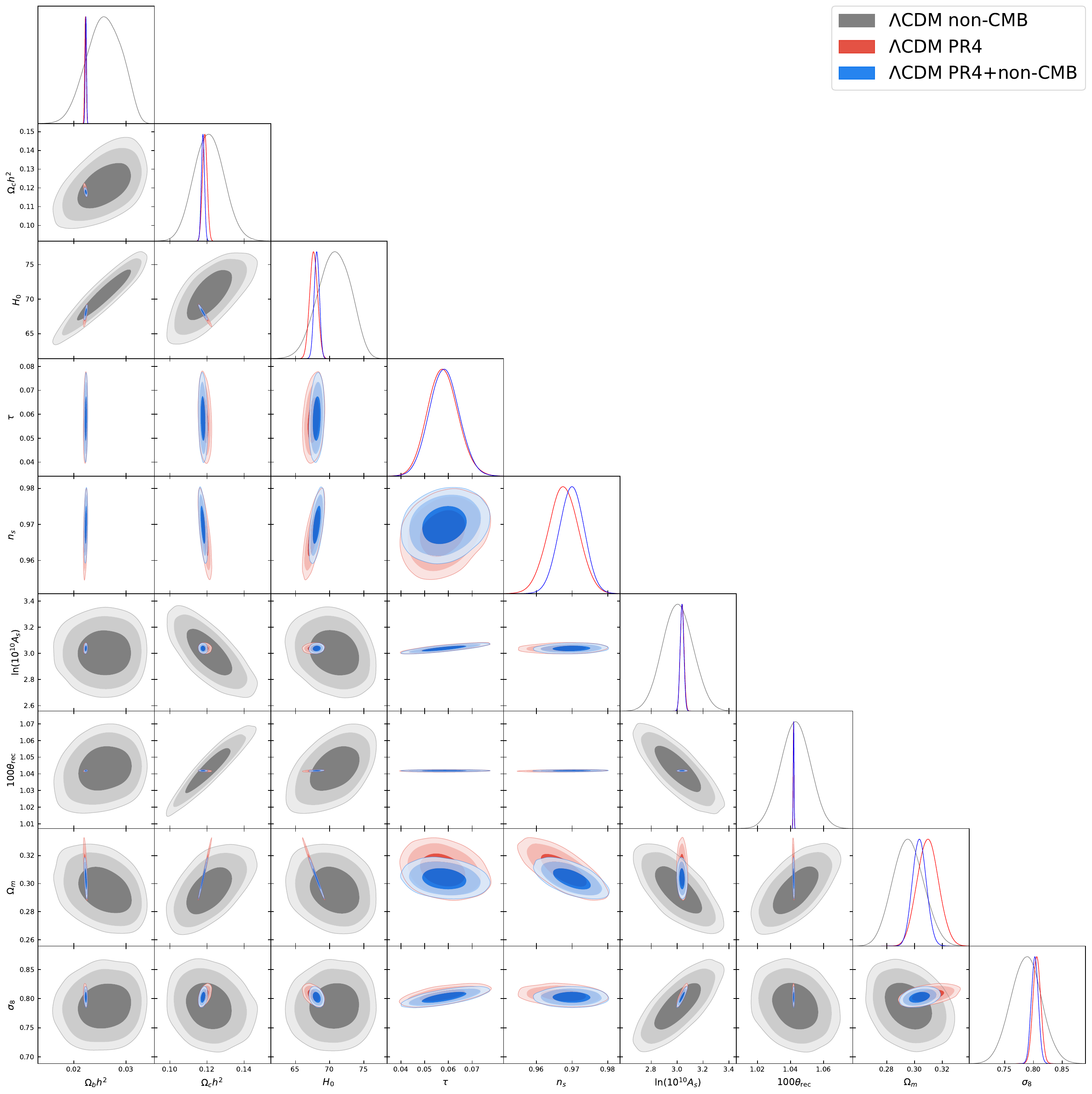}}
        \caption{One-dimensional likelihoods and 1$\sigma$, 2$\sigma$, and $3\sigma$ likelihood confidence contours of $\Lambda$CDM model parameters favored by non-CMB, PR4, and PR4+non-CMB datasets.
}
\label{fig:nonCMB_vs_PR4_LCDM}
\end{figure*}


\begin{figure*}[htbp]
\centering
\mbox{\includegraphics[width=175mm]{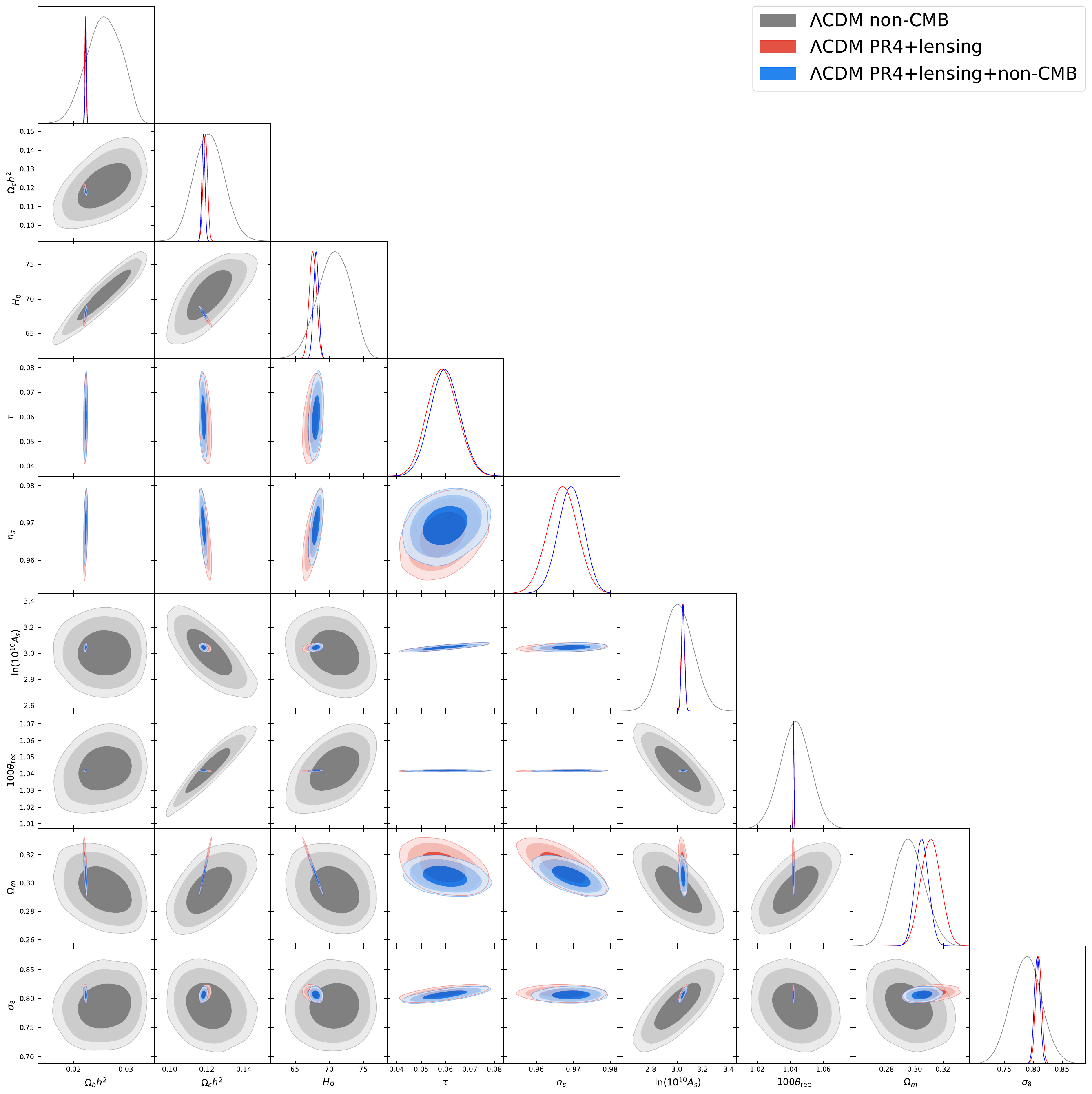}}
        \caption{One-dimensional likelihoods and 1$\sigma$, 2$\sigma$, and $3\sigma$ likelihood confidence contours of $\Lambda$CDM model parameters favored by non-CMB, PR4+lensing, and PR4+lensing+non-CMB datasets. 
}
\label{fig:LCDM}
\end{figure*}


\begin{figure*}[htbp]
\centering
\mbox{\includegraphics[width=175mm]{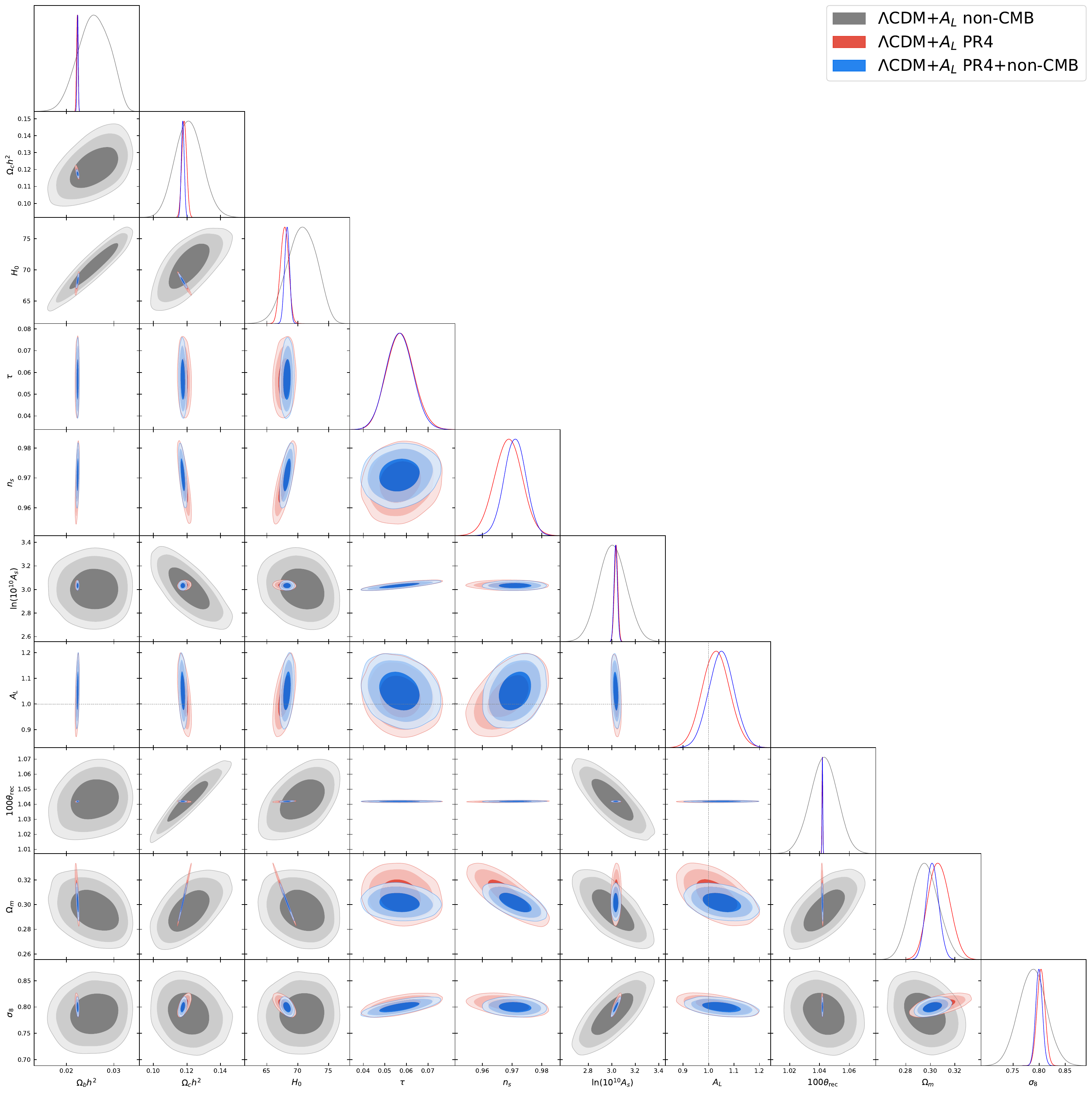}}
        \caption{One-dimensional likelihoods and 1$\sigma$, 2$\sigma$, and $3\sigma$ likelihood confidence contours of $\Lambda$CDM+$A_L$ model parameters favored by non-CMB, PR4, and PR4+non-CMB datasets. 
}
\label{fig:nonCMB_vs_PR4_LCDM_AL}
\end{figure*}


\begin{figure*}[htbp]
\centering
\mbox{\includegraphics[width=175mm]{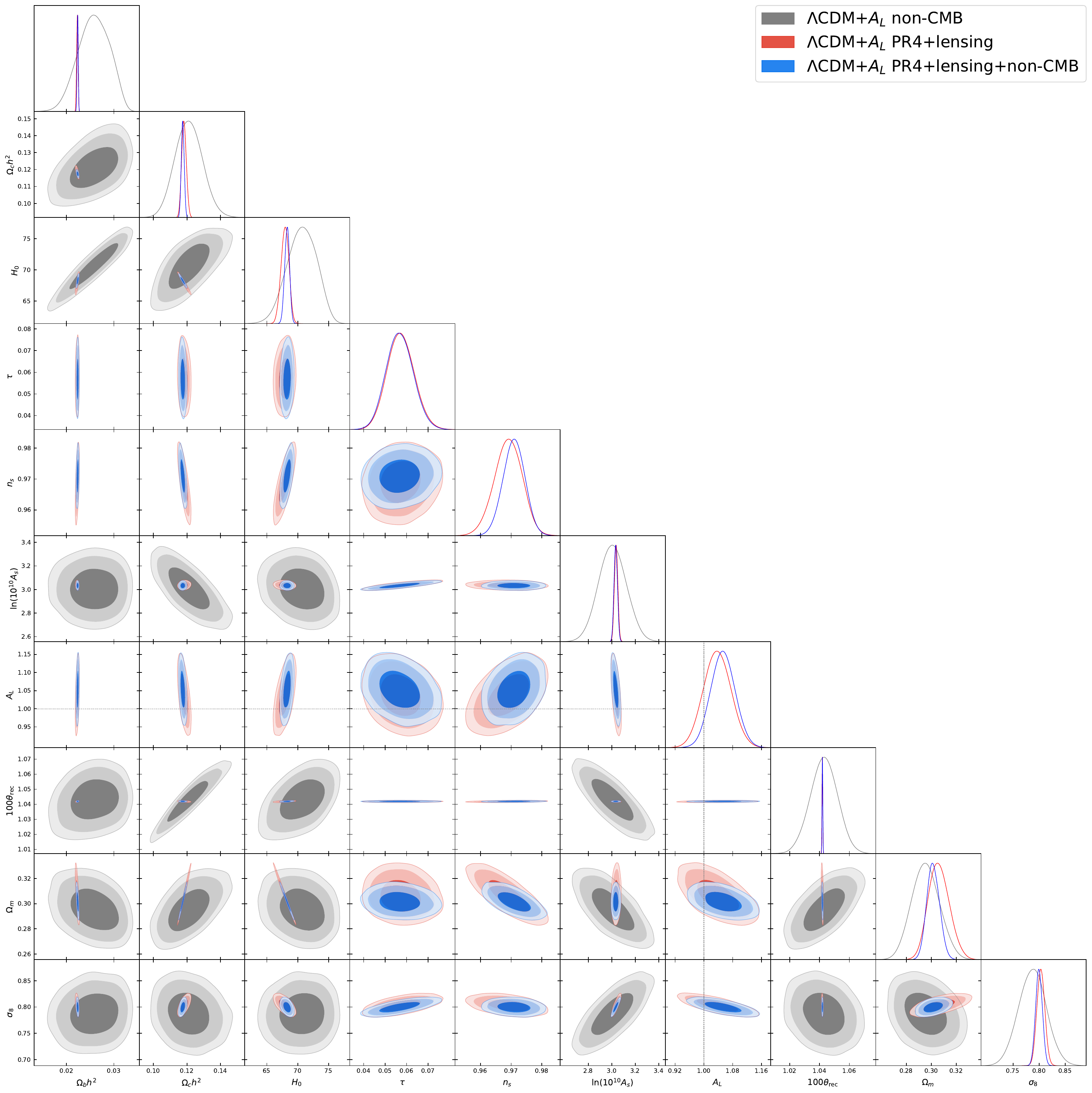}}
        \caption{One-dimensional likelihoods and 1$\sigma$, 2$\sigma$, and $3\sigma$ likelihood confidence contours of $\Lambda$CDM+$A_L$ model parameters favored by non-CMB, PR4+lensing, and PR4+lensing+non-CMB datasets. 
}
\label{fig:LCDM_AL}
\end{figure*}


\section{Results and Discussion}
\label{sec:ResultsandDiscussion}

We present the $\Lambda$CDM$(+A_L)$ models and $w_0$CDM$(+A_L)$ and $w_0w_a$CDM$(+A_L)$ parameterizations cosmological parameter constraints in Tables \ref{tab:results_flat_LCDM} -- \ref{tab:results_flat_w0waCDM_Alens} and in Figs.\ \ref{fig:nonCMB_vs_PR4_LCDM} -- \ref{fig:CPL_AL}.

\subsection{Comparison with the Tristram et al.\ PR4 results \cite{Tristram:2023haj}}
\label{subsec:Tristrametal}

In this subsection we compare our $\Lambda$CDM model PR4 and PR4+lensing data cosmological constraint results, as well as our $\Lambda$CDM$+A_L$ model PR4 data  $A_L$ results, to those reported in \cite{Tristram:2023haj} in their table 5 and Eq.\ (35), except for $\theta_{\rm rec}$ since \cite{Tristram:2023haj} use $\theta_*$ instead of $\theta_{\rm rec}$.

We observe some differences in the mean parameter values from the two analyses, but all are smaller than $1\sigma$. For PR4 data, all differences for our primary parameters are at or below $0.1\sigma$, with the $n_s$ difference of $0.10\sigma$ being the largest. Most derived parameter differences are also small with the $\sigma_8$ difference of $0.12\sigma$ being the largest.

For PR4+lensing data the differences are a little larger. For our primary parameters the largest differences are $0.17\sigma$ for $H_0$ and $0.16\sigma$ for $\Omega_bh^2$ and $n_s$ while for derived parameters the largest differences are $0.32\sigma$ for $\sigma_8$ and $-0.21\sigma$ for $\Omega_m$. 

For PR4 data in the $\Lambda$CDM$+A_L$ model, our value of $A_L=1.030\pm 0.054$ differs from $A_L = 1.039 \pm 0.052$ in \cite{Tristram:2023haj} by $0.12\sigma$.

\subsection{Comparison with the de Cruz Perez et al.\ $\Lambda$CDM$(+A_L)$ PR3 results \cite{deCruzPerez:2024abc}}
\label{subsec:deCruzPerezetalLCDM}

In this subsection we compare our $\Lambda$CDM$(+A_L)$ PR4 \texttt{CLASS}+\texttt{Cobaya} results to the \texttt{CAMB}+\texttt{CosmoMC} results of \cite{deCruzPerez:2024abc}, tables IV and VII, that made use of the P18/PR3 datasets, except for $\theta_{\rm rec}$ since \cite{deCruzPerez:2024abc} use $\theta_{\rm MC}$ instead of $\theta_{\rm rec}$.

Since the non-CMB cosmological parameter constraints of \cite{deCruzPerez:2024abc} were computed assuming different values of $\tau$ and $n_s$ than those used here, as discussed above, we do not compare the two different sets of non-CMB data results. 

Comparing the $\Lambda$CDM model PR3 data results of \cite{deCruzPerez:2024abc} to our PR4 data results, all differences for our primary parameters are $0.76\sigma$ (the $\Omega_c h^2$ difference) or smaller, being $0.55\sigma$ for $\Omega_b h^2$, $-0.45\sigma$ for $H_0$, $-0.43\sigma$ for $n_s$, $-0.34\sigma$ for $\tau$, and $0.24\sigma$ for ${\rm ln} (10^{10} A_s)$. For our derived parameters, the differences are $0.60\sigma$ and $0.59\sigma$ for $\Omega_m$ and $\sigma_8$, respectively.

Comparing the $\Lambda$CDM model PR3+lensing data results of \cite{deCruzPerez:2024abc} to our PR4+lensing data results, all differences for our primary parameters are $0.71\sigma$ (the $\Omega_b h^2$ difference) or smaller, being $0.55\sigma$ for $\Omega_c h^2$, $-0.47\sigma$ for $\tau$, $-0.37\sigma$ for $n_s$, $-0.26\sigma$ for $H_0$, and $-0.054\sigma$ for ${\rm ln} (10^{10} A_s)$. For our derived parameters the differences are $0.42\sigma$ for $\Omega_m$ and $0.28\sigma$ for $\sigma_8$.

Comparing the $\Lambda$CDM model PR3+non-CMB data results of \cite{deCruzPerez:2024abc} to our PR4+non-CMB data results, all differences for our primary parameters are $0.96\sigma$ (the $\Omega_b h^2$ difference) or smaller, being $0.49\sigma$ for $\Omega_c h^2$, $-0.37\sigma$ for $\tau$, $0.19\sigma$ for ${\rm ln} (10^{10} A_s)$, $-0.18\sigma$ for $n_s$, and $-0.055\sigma$ for $H_0$. For our derived parameters the differences are  $0.27\sigma$ for $\Omega_m$ and $0.31\sigma$ for $\sigma_8$.

Comparing the $\Lambda$CDM model PR3+lensing+non-CMB data results of \cite{deCruzPerez:2024abc} to our PR4+lensing+non-CMB data results, all differences for our primary parameters are $1.0\sigma$ (the $\Omega_b h^2$ difference) or smaller, being $0.38\sigma$ for $\Omega_c h^2$, $-0.31\sigma$ for $\tau$, $-0.16\sigma$ for $n_s$, $0.075\sigma$ for $H_0$, and $0$ for ${\rm ln} (10^{10} A_s)$. For our derived parameters the differences are  $0.16\sigma$ for $\Omega_m$ and $\sigma_8$.

Comparing the $\Lambda$CDM$+A_L$ model PR3 data results of \cite{deCruzPerez:2024abc} to our PR4 data results, all differences for our primary parameters are $1.8\sigma$ (the $A_L$ difference) or smaller, being $1.3\sigma$ for $\Omega_b h^2$, $-0.73\sigma$ for $\tau$, $0.48\sigma$ for $H_0$, $0.35\sigma$ for $n_s$, $-0.26\sigma$ for ${\rm ln} (10^{10} A_s)$, and $-0.019\sigma$ for $\Omega_c h^2$. For our derived parameters the differences are $-0.34\sigma$ for $\sigma_8$ and $-0.33\sigma$ for $\Omega_m$.

Comparing the $\Lambda$CDM$+A_L$ model PR3+lensing data results of \cite{deCruzPerez:2024abc} to our PR4+lensing data results, all differences for our primary parameters are $0.94\sigma$ (the $\Omega_b h^2$ difference) or smaller, being $-0.80\sigma$ for $\tau$, $0.64\sigma$ for $A_L$, $-0.38\sigma$ for ${\rm ln} (10^{10} A_s)$, $0.23\sigma$ for $H_0$, $0.074\sigma$ for $n_s$, and $0$ for $\Omega_c h^2$. For our derived parameters the differences are $-0.32\sigma$ for $\sigma_8$ and $-0.11\sigma$ for $\Omega_m$.

Comparing the $\Lambda$CDM$+A_L$ model PR3+non-CMB data results of \cite{deCruzPerez:2024abc} to our PR4+non-CMB data results, all differences for our primary parameters are $1.9\sigma$ (the $A_L$ difference) or smaller, being $1.5\sigma$ for $\Omega_b h^2$, $-0.84\sigma$ for $\tau$, $0.52\sigma$ for $H_0$, $-0.36\sigma$ for ${\rm ln} (10^{10} A_s)$, $0.23\sigma$ for $n_s$, and $-0.008\sigma$ for $\Omega_c h^2$. For our derived parameters the differences are $-0.34\sigma$ for $\sigma_8$ and $-0.27\sigma$ for $\Omega_m$.

Comparing the $\Lambda$CDM$+A_L$ model PR3+lensing+non-CMB data results of \cite{deCruzPerez:2024abc} to our PR4+lensing+non-CMB data results, all differences for our primary parameters are $1.1\sigma$ (the $\Omega_b h^2$ difference) or smaller, being $-0.86\sigma$ for $\tau$, $0.70\sigma$ for $A_L$, $-0.39$ for ${\rm ln} (10^{10} A_s)$, $0.33\sigma$ for $H_0$, $0.072\sigma$ for $\Omega_c h^2$, and $0.039\sigma$ for $n_s$. For our derived parameters the differences are $-0.38\sigma$ for $\sigma_8$ and $-0.14\sigma$ for $\Omega_m$.

In summary, in the $\Lambda$CDM case, the biggest differences between the PR3-based results of \cite{deCruzPerez:2024abc} and the PR4-based results here is the $\Omega_bh^2$ difference of $1.0\sigma$ for the PR3+lensing+non-CMB vs.\ PR4+lensing+non-CMB comparison, with all other parameter differences being less than $1.0\sigma$.

In the $\Lambda$CDM$+A_L$ case, the biggest differences between the PR3-based results of \cite{deCruzPerez:2024abc} and the PR4-based results here are for $A_L$, which are: $1.8\sigma$ (PR3 vs.\ PR4) and $1.9\sigma$ (PR3+non-CMB vs.\ PR4+non-CMB). The only other differences of $1.0\sigma$ or larger are for $\Omega_bh^2$: $1.3\sigma$ (PR3 vs.\ PR4), $1.5\sigma$ (PR3+non-CMB vs.\ PR4+non-CMB), and $1.1\sigma$ (PR3+lensing+non-CMB vs.\ PR4+lensing+non-CMB). 

In the $\Lambda$CDM$(+A_L)$ cases, these differences arise because the PR4 data combinations give smaller $A_L$ values (closer to unity) and lower $\Omega_b h^2$ values compared to the corresponding PR3 data values.

We now discuss the ratios of the de Cruz Perez et al.\ $\Lambda$CDM$(+A_L)$ parameter error bars to those computed here. In the case of parameters with asymmetric upper and lower error bars, we use the averaged error bar when computing this ratio.

The ratios of our PR4 data $\Lambda$CDM model error bars to those of the PR3 data error bars of \cite{deCruzPerez:2024abc} are less than unity. For our primary parameters the ratios are 0.81 ($\tau$), 0.86 ($\Omega_c h^2$), 0.87 ($\Omega_b h^2$), 0.88 (${\rm ln} (10^{10} A_s)$), 0.92 ($H_0$), and 0.98 ($n_s$), while for our derived parameters the ratios are 0.90 ($\Omega_m$) and 0.91 ($\sigma_8$).

The ratios of our PR4+lensing data $\Lambda$CDM model error bars to those of the PR3+lensing data error bars of \cite{deCruzPerez:2024abc} are less than unity in all cases except for $\Omega_bh^2$. For our primary parameters the ratios are 0.85 ($\tau$), 0.86 (${\rm ln} (10^{10} A_s)$), 0.92 ($\Omega_c h^2$), 0.95 ($H_0$), 0.98 ($n_s$), and 1.0 ($\Omega_b h^2$), while for our derived parameters the ratios are 0.86 ($\sigma_8$) and 0.95 ($\Omega_m$).

The ratios of our PR4+non-CMB data $\Lambda$CDM model error bars to those of the PR3+non-CMB data error bars of \cite{deCruzPerez:2024abc} are less than unity. For our primary parameters the ratios are 0.81 ($\tau$), 0.88 (${\rm ln} (10^{10} A_s)$), 0.92 ($\Omega_b h^2$), 0.95 ($\Omega_c h^2$), and 0.97 ($n_s$ and $H_0$), while for our derived parameters the ratios are 0.91 ($\sigma_8$) and 0.98 ($\Omega_m$).

The ratios of our PR4+lensing+non-CMB data $\Lambda$CDM model error bars to those of the PR3+lensing+non-CMB data error bars of \cite{deCruzPerez:2024abc} are less than unity. For our primary parameters the ratios are 0.85 ($\tau$), 0.86 (${\rm ln} (10^{10} A_s)$), 0.92 ($\Omega_b h^2$), 0.98 ($\Omega_c h^2$), and 0.97 ($n_s$ and $H_0$), while for our derived parameters the ratios are 0.86 ($\sigma_8$) and 0.98 ($\Omega_m$).


\begin{table*}
	\caption{Mean and 68\% (or 95\% indicated between parentheses when the value is provided) confidence limits of $w_0$CDM parameterization parameters
        from non-CMB, PR4, PR4+lensing, PR4+non-CMB, and PR4+lensing+non-CMB data.
        $H_0$ has units of km s$^{-1}$ Mpc$^{-1}$. We also include the values of $\chi^2_{\text{min}}$, DIC, and $\Delta$DIC the difference with respect to the $\Lambda$CDM model.
}
\begin{ruledtabular}
\begin{tabular}{lccccc}
  Parameter                     &  Non-CMB                      & PR4                   &  PR4+lensing               &  PR4+non-CMB            & PR4+lensing+non-CMB     \\[+1mm]
 \hline \\[-1mm]
  $\Omega_b h^2$                & $0.0288^{+0.0036}_{-0.0015}$  & $0.02227\pm 0.00014$  & $0.02227 \pm 0.00013$      &  $0.02234 \pm 0.00012$  &  $0.02233 \pm 0.00012$  \\[+1mm]
  $\Omega_c h^2$                & $0.1003^{+0.0079}_{-0.0110}$  & $0.1185\pm 0.0012$    & $0.1186 \pm 0.0011$        &  $0.11742 \pm 0.00093$  &  $0.11774 \pm 0.00092$  \\[+1mm]
  $H_0$                         & $68.6^{+2.0}_{-1.8}$          & $84^{+10}_{-6} (>65)$ & $85^{+10}_{-7} (>67)$      &  $67.66 \pm 0.64$       &  $67.68 \pm 0.64$       \\[+1mm]
  $\tau$                        & $0.0576$                      & $0.0576\pm 0.0062$    & $0.0577 \pm 0.0061$        &  $0.0592 \pm 0.0062$    &  $0.0606 \pm 0.0063$    \\[+1mm]
  $n_s$                         & $0.9683$                      & $0.9683\pm 0.0040$    & $0.9682 \pm 0.0040$        &  $0.9709 \pm 0.0037$    &  $0.9702 \pm 0.0037$    \\[+1mm]
  $\ln(10^{10} A_s)$            & $3.50^{+0.22}_{-0.19}$        & $3.038\pm 0.014$      & $3.038 \pm 0.012$          &  $3.039 \pm 0.014$      &  $3.048 \pm 0.012$      \\[+1mm]
  $w_0$                         & $-0.868^{+0.044}_{-0.038}$    & $-1.49^{ +0.18}_{-0.36} (> -1.90)$  & $-1.51^{+0.18}_{-0.32} (>-1.90)$  &  $-0.980 \pm 0.023$     &  $-0.985 \pm 0.024$     \\[+1mm]  
 \hline \\[-1mm] 
  $100\theta_\textrm{rec}$      & $1.021\pm 0.011$              & $1.04184\pm 0.00027$  & $1.04183 \pm 0.00026$      &  $1.04192 \pm 0.00024$  &  $1.04190 \pm 0.00024$  \\[+1mm]
  $\Omega_m$                    & $0.275^{+0.010}_{-0.012}$     & $0.208^{+0.018}_{-0.064}$  & $0.203^{+0.018}_{-0.058}$  &  $0.3068 \pm 0.0062$    &  $0.3073 \pm 0.0063$    \\[+1mm]
  $\sigma_8$                    & $0.820\pm 0.028$              &  $0.941^{+0.098}_{-0.053}$ & $0.948^{+0.087}_{-0.050}$  &  $0.7961 \pm 0.0095$    &  $0.8021 \pm 0.0084$    \\[+1mm]
\hline\\[-1mm]
  $\chi_{\textrm{min}}^2$       & $1449.90$                     & $30549.51$            & $30557.20$                 & $32010.94$              &  $32020.83$             \\[+1mm]
  $\textrm{DIC}$                & $1459.28$                     & $30599.39$            & $30608.72$                 & $32064.93$              &  $32075.25$             \\[+1mm]
  $\Delta\textrm{DIC}$          & $-8.92$                       & $-2.53$               & $-0.38$                    & $+1.36$                 &  $+1.20$                \\[+1mm]
\end{tabular}
\\[+1mm]
\begin{flushleft}
\end{flushleft}
\end{ruledtabular}
\label{tab:results_flat_XCDM}
\end{table*}

The ratios of our PR4 data $\Lambda$CDM$+A_L$ model error bars to those of the PR3 data error bars of \cite{deCruzPerez:2024abc} are less than unity. For our primary parameters the ratios are 0.77 ($\tau$), 0.81 ($A_L$), 0.88 (${\rm ln} (10^{10} A_s)$), 0.92 ($H_0$ and $n_s$), 0.93 ($\Omega_c h^2$), and 0.94 ($\Omega_b h^2$), while for our derived parameters the ratios are 0.86 ($\sigma_8$) and 0.93 ($\Omega_m$).

The ratios of our PR4+lensing data $\Lambda$CDM$+A_L$ model error bars to those of the PR3+lensing data error bars of \cite{deCruzPerez:2024abc} are less than unity. For our primary parameters the ratios are 0.72 ($\tau$), 0.83 (${\rm ln} (10^{10} A_s)$), 0.88 ($\Omega_b h^2$), 0.91 ($H_0$), and  0.93 ($n_s$, $A_L$, and $\Omega_c h^2$),  while for our derived parameters the ratios are 0.84 ($\sigma_8$) and 0.92 ($\Omega_m$).


\begin{table*}
	\caption{Mean and 68\% (or 95\% indicated between parentheses when the value is provided) confidence limits of $w_0$CDM+$A_L$ parameterization parameters
        from non-CMB, PR4, PR4+lensing, PR4+non-CMB, and PR4+lensing+non-CMB data.
        $H_0$ has units of km s$^{-1}$ Mpc$^{-1}$. We also include the values of $\chi^2_{\text{min}}$, DIC, and $\Delta$DIC the difference with respect to the $\Lambda$CDM model.
}
\begin{ruledtabular}
\begin{tabular}{lccccc}
  Parameter                     &  Non-CMB                     & PR4                          &  PR4+lensing                 &  PR4+non-CMB            & PR4+lensing+non-CMB     \\[+1mm]
 \hline \\[-1mm]
  $\Omega_b h^2$                & $0.0288^{+0.0036}_{-0.0015}$ & $0.02226\pm 0.00015$         & $0.02227\pm 0.00015$         &  $0.02240 \pm 0.00012$  &  $0.02240 \pm 0.00012$  \\[+1mm]
  $\Omega_c h^2$                & $0.1003^{+0.0079}_{-0.0110}$ & $0.1186\pm 0.0014$           & $0.1185^{+0.0013}_{-0.0015}$ &  $0.11693 \pm 0.00099$  &  $0.11691 \pm 0.00098$  \\[+1mm]
  $H_0$                         & $68.6^{+2.0}_{-1.8}$         & $83^{+10}_{-7} (>61)$        & $83^{+10}_{-7} (>62)$        &  $67.67 \pm 0.64$       &  $67.68 \pm 0.64$       \\[+1mm]
  $\tau$                        & $0.0576$                     & $0.0574^{+0.0057}_{-0.0066}$ & $0.0575\pm 0.0064$           &  $0.0577 \pm 0.0063$    &  $0.0577 \pm 0.0062$    \\[+1mm]
  $n_s$                         & $0.9683$                     & $0.9680\pm 0.0045$           & $0.9683\pm 0.0045$           &  $0.9725 \pm 0.0038$    &  $0.9725 \pm 0.0037$    \\[+1mm]
  $\ln(10^{10} A_s)$            & $3.50^{+0.22}_{-0.19}$       & $3.037\pm 0.015$             & $3.038\pm 0.015$             &  $3.034 \pm 0.015$      &  $3.034 \pm 0.014$      \\[+1mm]
  $w_0$                         & $-0.868^{+0.044}_{-0.038}$   & $-1.45^{+0.21}_{-0.41} (> -1.92)$     & $-1.46^{+0.20}_{-0.39} (> -1.91)$    &  $-0.973 \pm 0.024$     &  $-0.973 \pm 0.024$   \\[+1mm]  
  $A_L$                         & $1$                          & $1.002^{+0.052}_{-0.059}$    & $1.006^{+0.037}_{-0.045}$    &  $1.064 \pm 0.051$      &  $1.064 \pm 0.035$     \\[+1mm]  
 \hline \\[-1mm]
  $100\theta_\textrm{rec}$      & $1.021\pm 0.011$             & $1.04182\pm 0.00024$         & $1.04183\pm 0.00027$         &  $1.04196 \pm 0.00024$  &  $1.04195 \pm 0.00024$  \\[+1mm]
  $\Omega_m$                    & $0.275^{+0.010}_{-0.012}$    & $0.218^{+0.021}_{-0.074}$    & $0.214^{+0.019}_{-0.070}$    &  $0.3058 \pm 0.0062$    &  $0.3056 \pm 0.0062$    \\[+1mm]
  $\sigma_8$                    & $0.820\pm 0.028$             & $0.930^{+0.110}_{-0.060}$    & $0.933^{+0.110}_{-0.058}$    &  $0.790 \pm 0.010$      &  $0.791 \pm 0.010$    \\[+1mm]
\hline\\[-1mm]
  $\chi_{\textrm{min}}^2$       & $1449.90$                    & $30549.63$                   & $30555.25$                   & $32010.19$              &  $32018.26$              \\[+1mm]
  $\textrm{DIC}$                & $1459.28$                    & $30600.95$                   & $30611.75$                   & $32063.37$              &  $32072.15$              \\[+1mm]
  $\Delta\textrm{DIC}$          & $-8.92$                      & $-0.97$                      & $+2.65$                      & $-2.79$                 &  $-1.90$                \\[+1mm]
\end{tabular}
\\[+1mm]
\begin{flushleft}
\end{flushleft}
\end{ruledtabular}
\label{tab:results_flat_XCDM_Alens}
\end{table*}

The ratios of our PR4+non-CMB data $\Lambda$CDM$+A_L$ model error bars to those of the PR3+non-CMB data error bars of \cite{deCruzPerez:2024abc} are less than unity. For our primary parameters the ratios are 0.74 ($\tau$), 0.82 ($A_L$ and ${\rm ln} (10^{10} A_s)$), 0.86 ($\Omega_b h^2$), and  0.95 ($\Omega_c h^2$, $n_s$, and $H_0$),  while for our derived parameters the ratios are  0.85 ($\sigma_8$) and 0.94 ($\Omega_m$).

The ratios of our PR4+lensing+non-CMB data $\Lambda$CDM$+A_L$ model error bars to those of the PR3+lensing+non-CMB data error bars of \cite{deCruzPerez:2024abc} are less than unity. For our primary parameters the ratios are 0.72 ($\tau$), 0.78 (${\rm ln} (10^{10} A_s)$), 0.86 ($\Omega_b h^2$), 0.92 ($n_s$), 0.93 ($H_0$), 0.96 ($\Omega_c h^2$), and 0.97 ($A_L$), while for our derived parameters the ratios are  0.85 ($\sigma_8$) and 0.96 ($\Omega_m$).

In the $\Lambda$CDM$(+A_L)$ models, the use of PR4 data leads to more restrictive constraints, particularly for the primary parameters $\tau$ and ${\rm ln} (10^{10} A_s)$, across most data combinations. As noted by Tristram et al.\ \cite{Tristram:2023haj}, replacing PR3 with PR4 data reduces  parameter uncertainties by about 10–-20\%. This improvement is expected, since PR4 data include more refined information than PR3 data, and therefore provide stronger constraints.

The $\Delta$DIC values for the DIC differences between $\Lambda$CDM$+A_L$ and $\Lambda$CDM models, listed here in Table \ref{tab:results_flat_LCDM_Alens} for the PR4 datasets and in table VII of \cite{deCruzPerez:2024abc} for the PR3 datasets are significantly different in some cases. These are $+0.90$ (PR4 data, and weakly against $\Lambda$CDM$+A_L$) versus $-5.52$ (PR3 data, and positively for $\Lambda$CDM$+A_L$); $+1.05$ (PR4+lensing data, and weakly against $\Lambda$CDM$+A_L$) versus $-0.92$ (PR3+lensing data, and weakly for $\Lambda$CDM$+A_L$); $-2.59$ (PR4+non-CMB data, and positively for $\Lambda$CDM$+A_L$) versus $-8.47$ (PR3+non-CMB data, and strongly for $\Lambda$CDM$+A_L$); and $-0.87$ (PR4+lensing+non-CMB data, and weakly for $\Lambda$CDM$+A_L$) versus $-4.01$ (PR3+lensing+non-CMB data, and positively for $\Lambda$CDM$+A_L$).

The biggest difference in $\Delta$DIC values, larger than $5$, is $6.42$ for PR4 versus PR3 and $5.88$ for PR4+non-CMB vs.\ PR3+non-CMB.

Our results indicate that PR4 data yield a weaker preference for $A_L >1$ compared to PR3 data, thus indicating a reduced significance of the CMB lensing anomaly. This behavior is consistent with recent analyses based on PR4 data. We note that the PR4 likelihood exhibits a higher level of numerical noise, which may impact the precise determination of $\chi^2$ minima and should be taken into account when interpreting DIC-based comparisons.

\subsection{Comparison with the de Cruz Perez et al.\ $w_0$CDM$(+A_L)$ PR3 results \cite{deCruzPerez:2024abc}}
\label{subsec:deCruzPerezetalXCDM}

In this subsection we compare our $w_0$CDM$(+A_L)$ PR4 \texttt{CLASS}+\texttt{Cobaya} results to the \texttt{CAMB}+\texttt{CosmoMC} results of \cite{deCruzPerez:2024abc}, table XI, that made use of the P18/PR3 datasets, except for $\theta_{\rm rec}$ since \cite{deCruzPerez:2024abc} use $\theta_{\rm MC}$ instead of $\theta_{\rm rec}$. In our comparisons below, we ignore the few primary parameter cases that do not show a $2\sigma$ detection.

Since the non-CMB cosmological parameter constraints of \cite{deCruzPerez:2024abc} were computed assuming different values of $\tau$ and $n_s$ than those used here, see discussion above, we do not compare the two different sets of non-CMB data results.

\begin{figure*}[htbp]
\centering
\mbox{\includegraphics[width=175mm]{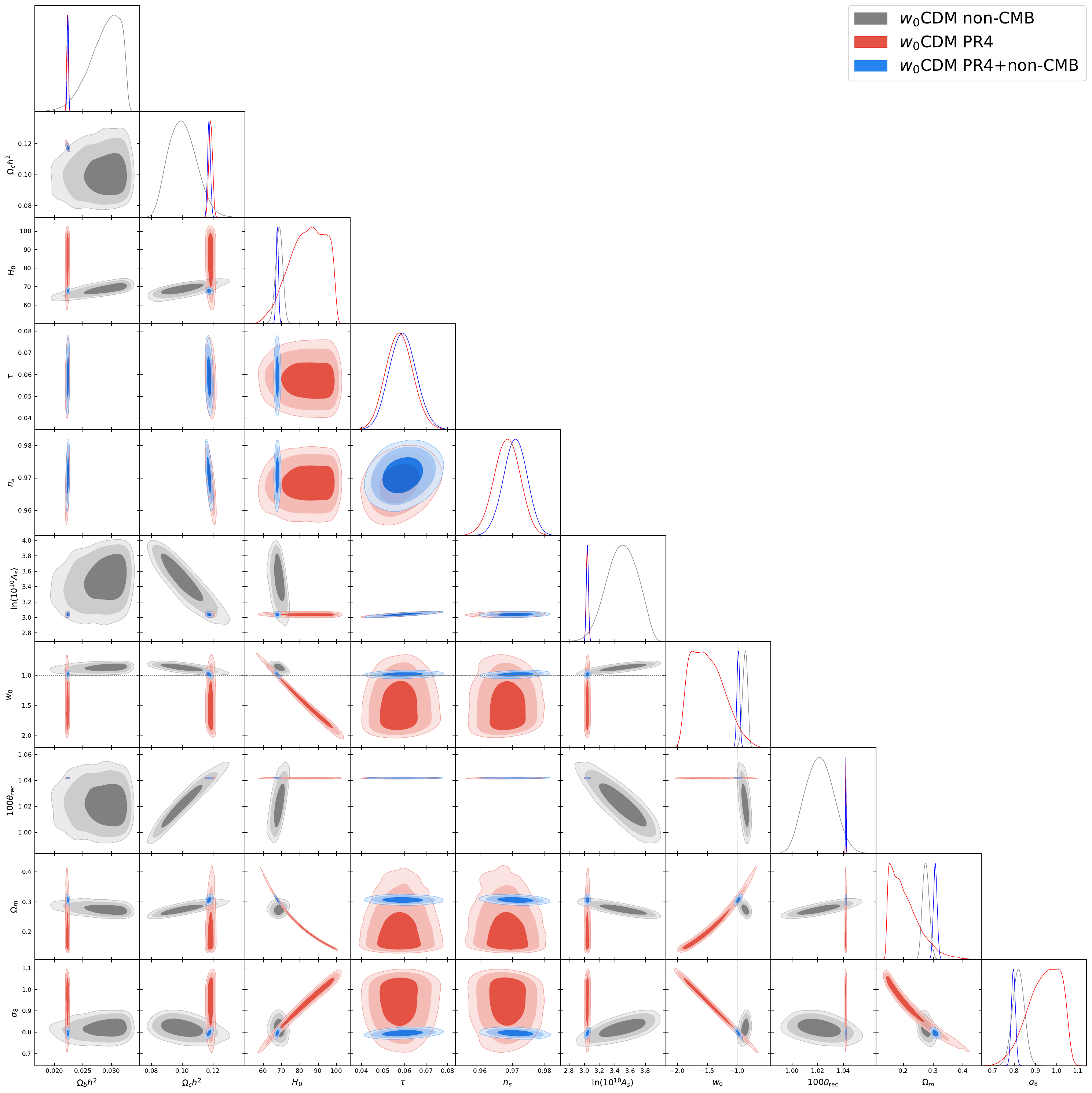}}
        \caption{One-dimensional likelihoods and 1$\sigma$, 2$\sigma$, and $3\sigma$ likelihood confidence contours of $w_0$CDM parameterization parameters favored by non-CMB, PR4, and PR4+non-CMB datasets. 
}
\label{fig:nonCMB_vs_PR4_XCDM}
\end{figure*}


\begin{figure*}[htbp]
\centering
\mbox{\includegraphics[width=175mm]{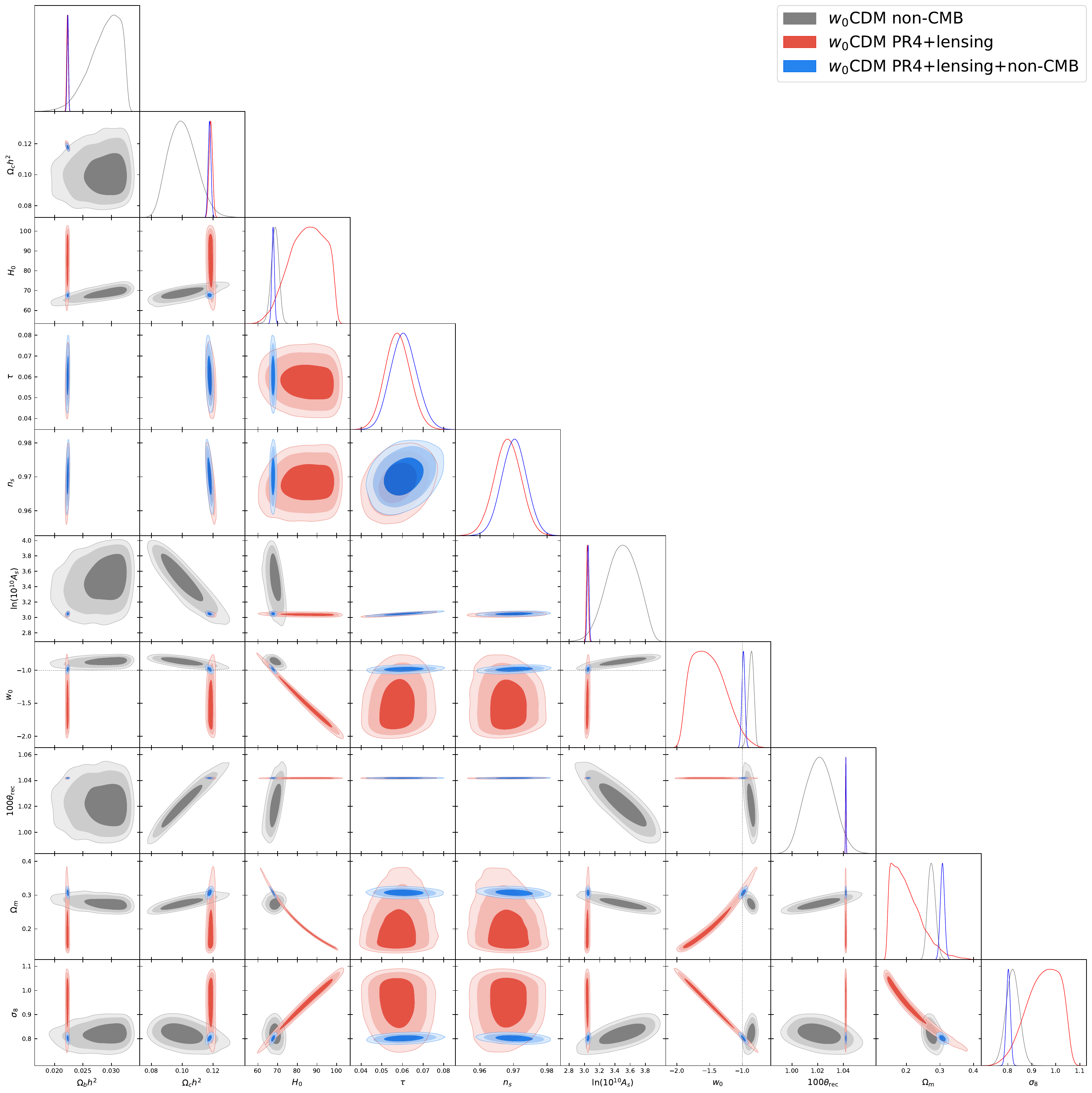}}
        \caption{One-dimensional likelihoods and 1$\sigma$, 2$\sigma$, and $3\sigma$ likelihood confidence contours of $w_0$CDM parameterization parameters favored by non-CMB, PR4+lensing, and PR4+lensing+non-CMB datasets. 
}
\label{fig:XCDM}
\end{figure*}


\begin{figure*}[htbp]
\centering
\mbox{\includegraphics[width=175mm]{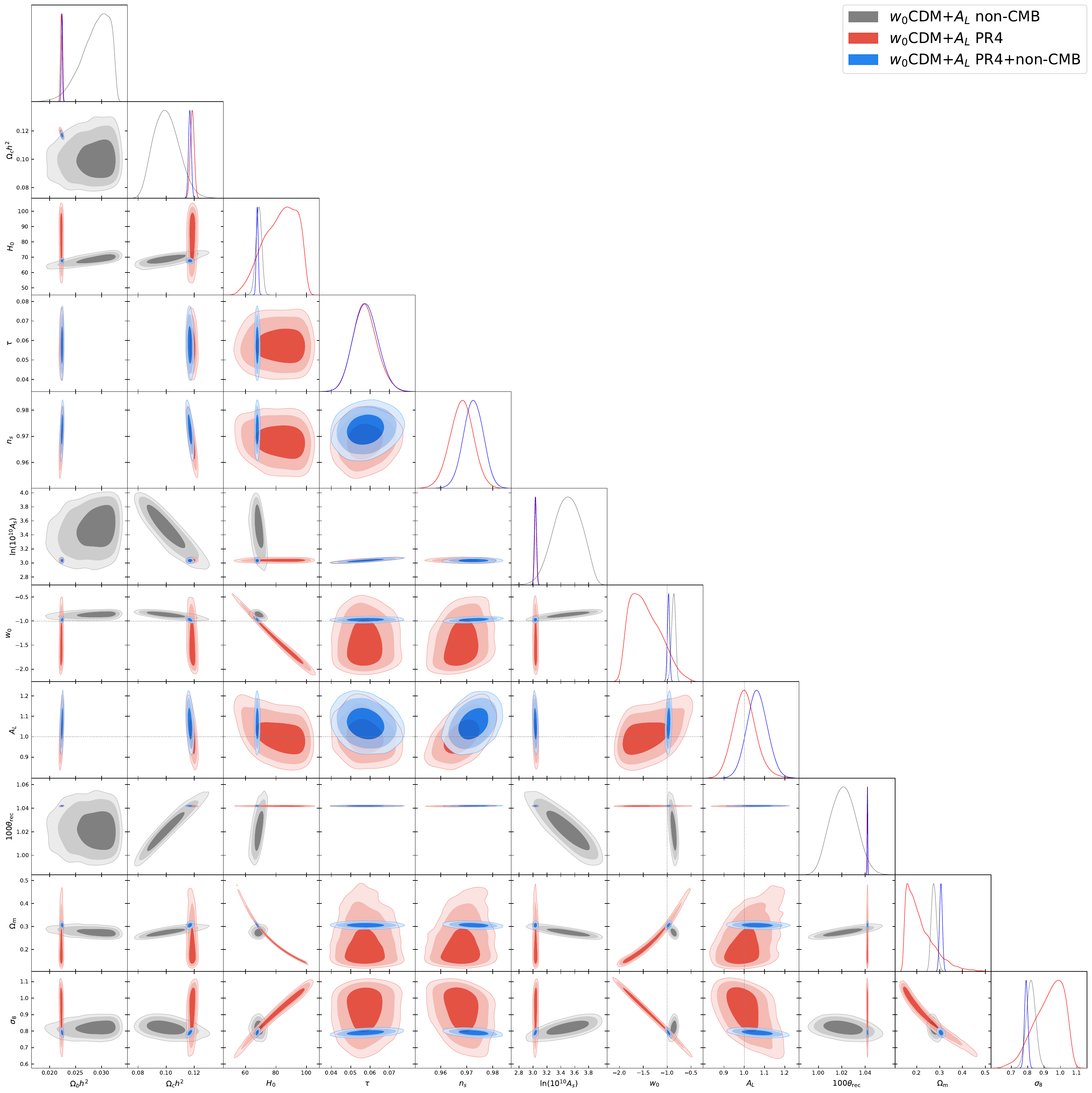}}
        \caption{One-dimensional likelihoods and 1$\sigma$, 2$\sigma$, and $3\sigma$ likelihood confidence contours of $w_0$CDM+$A_L$ parameterization parameters favored by non-CMB, PR4, and PR4+non-CMB datasets. 
}
\label{fig:nonCMB_vs_PR4_XCDM_AL}
\end{figure*}


\begin{figure*}[htbp]
\centering
\mbox{\includegraphics[width=175mm]{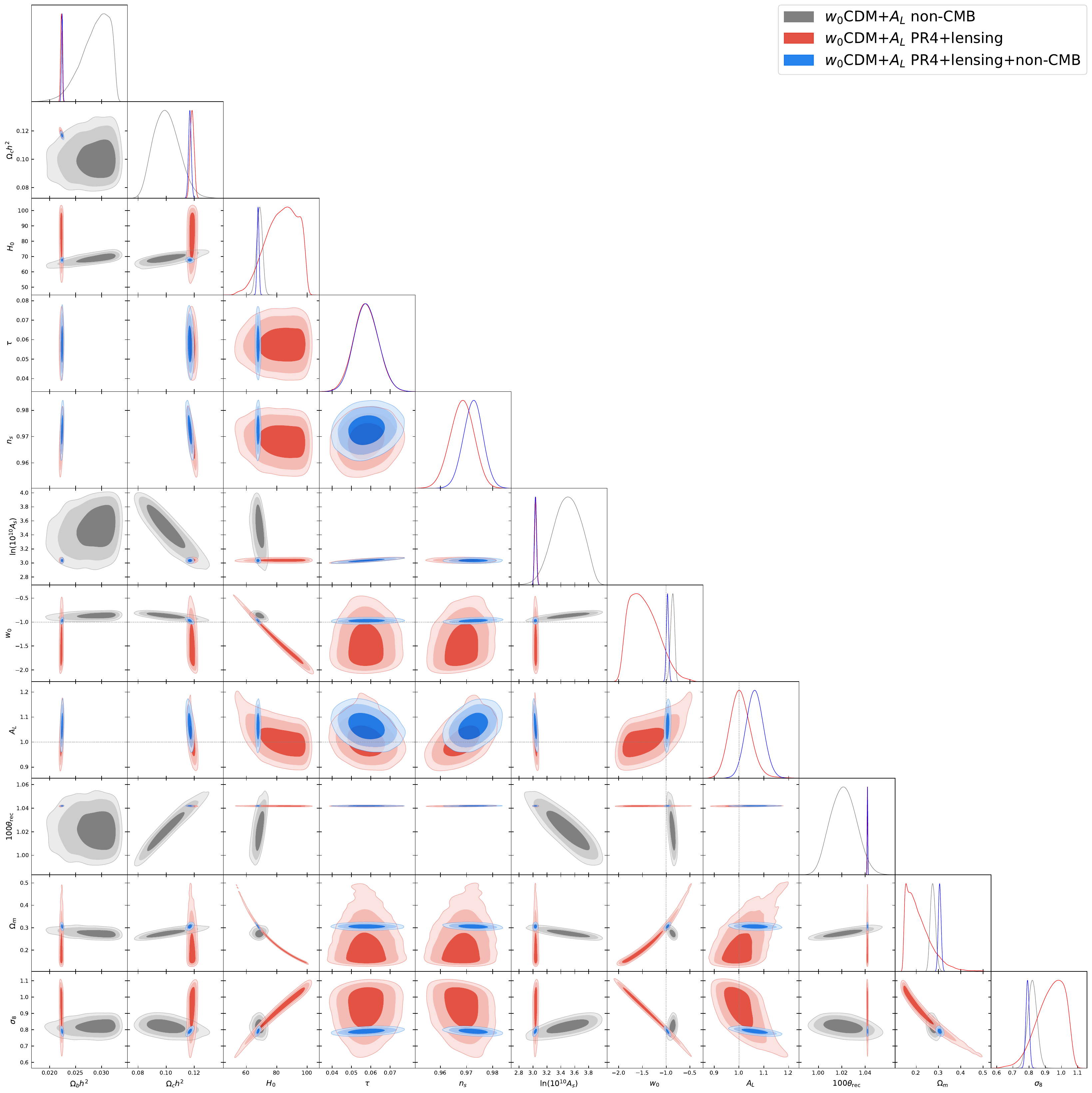}}
        \caption{One-dimensional likelihoods and 1$\sigma$, 2$\sigma$, and $3\sigma$ likelihood confidence contours of $w_0$CDM+$A_L$ parameterization parameters favored by non-Cmb, PR4+lensing, and PR4+lensing+non-CMB datasets. 
}
\label{fig:XCDM_AL}
\end{figure*}


Comparing the $w_0$CDM parameterization PR3 data results of \cite{deCruzPerez:2024abc} to our PR4 data results, all differences for our primary parameters are $0.81\sigma$ (the $\Omega_c h^2$ difference) or smaller, being $0.63\sigma$ for $\Omega_b h^2$, $-0.49\sigma$ for $n_s$, $-0.39\sigma$ for $\tau$, and $0.24\sigma$ for ${\rm ln} (10^{10} A_s)$. For our derived parameters the differences are $0.27\sigma$ for $\sigma_8$ and $-0.14\sigma$ for $\Omega_m$.

Comparing the $w_0$CDM parameterization PR3+lensing data results of \cite{deCruzPerez:2024abc} to our PR4+lensing data results, all differences for our primary parameters are $0.81\sigma$ (the $\Omega_b h^2$ difference) or smaller, being $-0.55\sigma$ for $\tau$, $0.43\sigma$ for $\Omega_c h^2$, $-0.26\sigma$ for $n_s$, and 0 for ${\rm ln} (10^{10} A_s)$. For our derived parameters the differences are $0.11\sigma$ for $\sigma_8$ and $-0.040\sigma$ for $\Omega_m$.

Comparing the $w_0$CDM parameterization PR3+non-CMB data results of \cite{deCruzPerez:2024abc} to our PR4+non-CMB data results, all differences for our primary parameters are $0.92\sigma$ (the $\Omega_b h^2$ difference) or smaller, being $0.50\sigma$ for $\Omega_c h^2$, $-0.34\sigma$ for $\tau$, $-0.24\sigma$ for $n_s$, $0.19\sigma$ for ${\rm ln} (10^{10} A_s)$, $-0.18\sigma$ for $w_0$, and $0.13\sigma$ for $H_0$. For our derived parameters the differences are $0.36\sigma$ for $\sigma_8$ and $0.079\sigma$ for $\Omega_m$.

Comparing the $w_0$CDM parameterization PR3+lensing+non-CMB data results of \cite{deCruzPerez:2024abc} to our PR4+lensing+non-CMB data results, all differences for our primary parameters are $0.92\sigma$ (the $\Omega_b h^2$ difference) or smaller, being $0.42\sigma$ for $\Omega_c h^2$, $-0.30\sigma$ for $\tau$, $-0.23\sigma$ for $n_s$, $-0.15\sigma$ for $w_0$, $0.14\sigma$ for $H_0$, and $0$ for ${\rm ln} (10^{10} A_s)$. For our derived parameters the differences are $0.21\sigma$ for $\sigma_8$ and $0.045\sigma$ for $\Omega_m$.

Comparing the $w_0$CDM$+A_L$ parameterization PR3 data results of \cite{deCruzPerez:2024abc} to our PR4 data results, all differences for our primary parameters are $1.6\sigma$ (the $A_L$ difference) or smaller, being $1.4\sigma$ for $\Omega_b h^2$, $-0.75\sigma$ for $\tau$, $0.39\sigma$ for $n_s$, $-0.34\sigma$ for ${\rm ln} (10^{10} A_s)$, and $-0.24\sigma$ for $\Omega_c h^2$. For our derived parameters the differences are $-0.52\sigma$ for $\sigma_8$ and $0.46\sigma$ for $\Omega_m$.

Comparing the $w_0$CDM$+A_L$ parameterization PR3+lensing data results of \cite{deCruzPerez:2024abc} to our PR4+lensing data results, all differences for our primary parameters are $1.0\sigma$ (the $\Omega_b h^2$ difference) or smaller, being $-0.80\sigma$ for $\tau$, $0.69\sigma$ for $A_L$, $-0.38\sigma$ for ${\rm ln} (10^{10} A_s)$, $0.12\sigma$ for $n_s$, and $-0.047\sigma$ for $\Omega_c h^2$. For our derived parameters the differences are $-0.35\sigma$ for $\sigma_8$ and $0.33\sigma$ for $\Omega_m$.

Comparing the $w_0$CDM$+A_L$ parameterization PR3+non-CMB data results of \cite{deCruzPerez:2024abc} to our PR4+non-CMB data results, all differences for our primary parameters are $1.9\sigma$ (the $A_L$ difference) or smaller, being $1.7\sigma$ for $\Omega_b h^2$, $-0.73\sigma$ for $\tau$, $0.38\sigma$ for $n_s$, $-0.31\sigma$ for ${\rm ln} (10^{10} A_s)$, $0.27\sigma$ for $w_0$, $-0.22\sigma$ for $\Omega_c h^2$, and $0.18\sigma$ for $H_0$. For our derived parameters the differences are $-0.40\sigma$ for $\sigma_8$ and $-0.17\sigma$ for $\Omega_m$.

Comparing the $w_0$CDM$+A_L$ parameterization PR3+lensing+non-CMB data results of \cite{deCruzPerez:2024abc} to our PR4+lensing+non-CMB data results, all differences for our primary parameters are $1.2\sigma$ (the $\Omega_b h^2$ difference) or smaller, being $-0.78\sigma$ for $\tau$, $0.73\sigma$ for $A_L$, $-0.36\sigma$ for ${\rm ln} (10^{10} A_s)$, $0.15\sigma$ for $n_s$ and $w_0$, $0.12\sigma$ for $H_0$, and $-0.075\sigma$ for $\Omega_c h^2$. For our derived parameters the differences are $-0.40\sigma$ for $\sigma_8$ and $-0.068\sigma$ for $\Omega_m$.

In summary, in the $w_0$CDM case, all parameter differences are less than $1.0\sigma$. In the $w_0$CDM$+A_L$ case, the most significant differences between the PR3-based results of \cite{deCruzPerez:2024abc} and the PR4-based results presented here arise for $A_L$: $1.6\sigma$ (PR3 vs.\ PR4) and $1.9\sigma$ (PR3+non-CMB vs.\ PR4+non-CMB). Some differences in $\Omega_b h^2$ are nearly as large: $1.4\sigma$ (PR3 vs.\ PR4), $1.0\sigma$ (PR3+lensing vs.\ PR4+lensing), $1.7\sigma$ (PR3+non-CMB vs.\ PR4+non-CMB), and $1.2\sigma$ (PR3+lensing+non-CMB vs.\ PR4+lensing+non-CMB). All other parameter differences are below $1.0\sigma$. In the $w_0$CDM$+A_L$ case, these differences are due to smaller $A_L$ (closer to unity) and $\Omega_b h^2$ values from the various PR4 data combinations relative to their corresponding PR3 values.

We now discuss the ratios of the \citeauthor{deCruzPerez:2024abc} $w_0$CDM$(+A_L)$ parameter error bars to those computed here. For parameters with asymmetric upper and lower error bars, we use the average of the two when computing this ratio.

The ratios of our PR4 data $w_0$CDM parameterization error bars to those of the PR3 data error bars of \cite{deCruzPerez:2024abc} are less than unity except for $\sigma_8$. For our primary parameters the ratios are 0.79 ($\tau$), 0.86 ($\Omega_c h^2$), 0.88 (${\rm ln} (10^{10} A_s)$), 0.91 ($n_s$), and 0.93 ($\Omega_b h^2$), while for our derived parameters the ratios are 0.89 ($\Omega_m$) and 1.1 ($\sigma_8$).

The ratios of our PR4+lensing data $w_0$CDM parameterization error bars to those of the PR3+lensing data error bars of \cite{deCruzPerez:2024abc} are less than unity. For our primary parameters the ratios are 0.80 (${\rm ln} (10^{10} A_s)$), 0.82 ($\tau$), 0.87 ($\Omega_b h^2$), 0.92 ($\Omega_c h^2$), and 0.98 ($n_s$), while for our derived parameters the ratios are 0.79 ($\Omega_m$) and 0.96 ($\sigma_8$).

The ratios of our PR4+non-CMB data $w_0$CDM parameterization error bars to those of the PR3+non-CMB data error bars of \cite{deCruzPerez:2024abc} are less than unity in all cases except $H_0$. For our primary parameters the ratios are 0.79 ($\tau$), 0.86 ($\Omega_b h^2$), 0.88 (${\rm ln} (10^{10} A_s)$), 0.93 ($\Omega_c h^2$), 0.95 ($n_s$), 0.96 ($w_0$), and 1.0 ($H_0$), while for our derived parameters the ratios are 0.95 ($\sigma_8$) and 0.98 ($\Omega_m$).

The ratios of our PR4+lensing+non-CMB data $w_0$CDM parameterization error bars to those of the PR3+lensing+non-CMB data error bars of \cite{deCruzPerez:2024abc} are less than unity except for $H_0$, $w_0$, and $\Omega_m$. For our primary parameters the ratios are 0.80 (${\rm ln} (10^{10} A_s)$), 0.84 ($\tau$), 0.86 ($\Omega_b h^2$), 0.97 ($\Omega_c h^2$ and $n_s$), and 1.0 ($H_0$ and $w_0$), while for our derived parameters the ratios are 0.94 ($\sigma_8$) and 1.0 ($\Omega_m$).

The ratios of our PR4 data $w_0$CDM$+A_L$ parameterization error bars to those of the PR3 data error bars of \cite{deCruzPerez:2024abc} are less than unity. For our primary parameters the ratios are 0.69 ($A_L$), 0.72 ($\tau$), 0.83 (${\rm ln} (10^{10} A_s)$), 0.88 ($\Omega_b h^2$), 0.92 ($n_s$), and 0.93 ($\Omega_c h^2$), while for our derived parameters the ratios are 0.43 ($\Omega_m$) and 0.71 ($\sigma_8$).

The ratios of our PR4+lensing data $w_0$CDM$+A_L$ parameterization error bars to those of the PR3+lensing data error bars of \cite{deCruzPerez:2024abc} are less than unity. For our primary parameters the ratios are 0.76 ($\tau$), 0.83 (${\rm ln} (10^{10} A_s)$), 0.84 ($A_L$), 0.88 ($\Omega_b h^2$), 0.92 ($n_s$), and 0.93 ($\Omega_c h^2$), while for our derived parameters the ratios are 0.54 ($\Omega_m$) and 0.76 ($\sigma_8$).

The ratios of our PR4+non-CMB data $w_0$CDM$+A_L$ parameterization error bars to those of the PR3+non-CMB data error bars of \cite{deCruzPerez:2024abc} are less than unity except for $w_0$, $\Omega_m$, and $H_0$. For our primary parameters the ratios are 0.74 ($\tau$), 0.80 ($\Omega_b h^2$), 0.81 ($A_L$), 0.88 (${\rm ln} (10^{10} A_s)$), 0.90 ($\Omega_c h^2$), 0.93 ($n_s$), and 1.0 ($w_0$ and $H_0$), while for our derived parameters the ratios are 0.91 ($\sigma_8$) and 1.0 ($\Omega_m$).

The ratios of our PR4+lensing+non-CMB data $w_0$CDM$+A_L$ parameterization error bars to those of the PR3+lensing+non-CMB data error bars of \cite{deCruzPerez:2024abc} are less than unity except for $w_0$, $\Omega_m$, and $H_0$. For our primary parameters the ratios are 0.75 ($\tau$), 0.82 (${\rm ln} (10^{10} A_s)$), 0.86 ($\Omega_b h^2$), 0.89 ($\Omega_c h^2$), 0.93 ($n_s$), 0.95 ($A_L$), and 1.0 ($w_0$ and $H_0$), while for our derived parameters the ratios are 0.91 ($\sigma_8$) and 1.0 ($\Omega_m$).

In the $w_0$CDM$(+A_L)$ parameterizations, PR4 data result in more restrictive error bars particularly for the primary $\tau$ parameter, across most datasets. Unlike the $\Lambda$CDM$(+A_L)$ cases, in the $w_0$CDM$(+A_L)$ parameterizations PR4 data do not always provide tighter constraints than PR3 results, with some of the error bar ratios being unity, or even slightly exceeding it. Especially, for the dark energy equation-of-state parameter $w_0$, when comparing results from PR3(+lensing) data combined with non-CMB data, the PR4 error bars are nearly identical, indicating that the updated PR4 dataset did not provide improved constraints on $w_0$.

The $\Delta$DIC values for the DIC differences between the $w_0$CDM parameterization and the $\Lambda$CDM model, listed here in Table \ref{tab:results_flat_XCDM} for the PR4 datasets and in table XI of \cite{deCruzPerez:2024abc} for the PR3 datasets are only mildly different in most cases. These are $-8.92$ (non-CMB data here, and strongly for $w_0$CDM) vs.\ $-9.37$ (non-CMB data of \cite{deCruzPerez:2024abc}, and strongly for $w_0$CDM); $-2.53$ (PR4 data, and positively for $w_0$CDM) vs.\ $-2.26$ (PR3 data, and positively for $w_0$CDM); $-0.38$ (PR4+lensing data, and weakly for $w_0$CDM) vs.\ $-2.24$ (PR3+lensing data, and positively for $w_0$CDM); $+1.36$ (PR4+non-CMB data, and weakly against $w_0$CDM) vs.\ $+1.87$ (PR3+non-CMB data, and weakly against $w_0$CDM); and $+1.20$ (PR4+lensing+non-CMB data, and weakly against $w_0$CDM) vs.\ $+2.10$ (PR3+lensing+non-CMB data, and positively against $w_0$CDM).

The $\Delta$DIC values for the DIC differences between the $w_0$CDM$+A_L$ parameterization and the $\Lambda$CDM model, listed here in Table \ref{tab:results_flat_XCDM_Alens} for the PR4 datasets and in table XI of \cite{deCruzPerez:2024abc} for the PR3 datasets are significantly different in some cases. These are $-0.97$ (PR4 data, and weakly for $w_0$CDM$+A_L$) vs.\ $-4.85$ (PR3 data, and positively for $w_0$CDM$+A_L$); $+2.65$ (PR4+lensing data, and positively against $w_0$CDM$+A_L$) vs.\ $-0.64$ (PR3+lensing data, and weakly for $w_0$CDM$+A_L$); $-2.79$ (PR4+non-CMB data, and positively for $w_0$CDM$+A_L$) vs.\ $-8.83$ (PR3+non-CMB data, and strongly for $w_0$CDM$+A_L$); and $-1.90$ (PR4+lensing+non-CMB data, and weakly for $w_0$CDM$+A_L$) vs.\ $-4.31$ (PR3+lensing+non-CMB data, and positively for $w_0$CDM$+A_L$).

The $\Delta$DIC values for the DIC differences between the $w_0$CDM$+A_L$ and $w_0$CDM parameterizations, computed from Tables \ref{tab:results_flat_XCDM} and  \ref{tab:results_flat_XCDM_Alens} here for the PR4 datasets and computed from table XI of \cite{deCruzPerez:2024abc} for the PR3 datasets are significantly different in some cases. These are $+1.56$ (PR4 data, and weakly against $w_0$CDM$+A_L$) vs.\ $-2.59$ (PR3 data, and positively for $w_0$CDM$+A_L$); $+3.03$ (PR4+lensing data, and positively against $w_0$CDM$+A_L$) vs.\ $+1.60$ (PR3+lensing data, and weakly against $w_0$CDM$+A_L$); $-4.15$ (PR4+non-CMB data, and positively for $w_0$CDM$+A_L$) vs.\ $-10.7$ (PR3+non-CMB data, and very strongly for $w_0$CDM$+A_L$); and $-3.10$ (PR4+lensing+non-CMB data, and positively for $w_0$CDM$+A_L$) vs.\ $-6.41$ (PR3+lensing+non-CMB data, and strongly for $w_0$CDM$+A_L$).

The biggest difference in $\Delta$DIC values, greater than 5, occurs for PR4+non-CMB vs.\ PR3+non-CMB data, being 6.55 for $w_0$CDM$+A_L$ relative to $w_0$CDM model and 6.04 for $w_0$CDM$+A_L$ relative to $\Lambda$CDM model.


\begin{table*}
	\caption{Mean and 68\% (or 95\% indicated between parentheses when the value is provided) confidence limits of $w_0w_a$CDM parameterization parameters
        from non-CMB, PR4, PR4+lensing, PR4+non-CMB, and PR4+lensing+non-CMB data.
        $H_0$ has units of km s$^{-1}$ Mpc$^{-1}$. We also include the values of $\chi^2_{\text{min}}$, DIC, and $\Delta$DIC the difference with respect to the $\Lambda$CDM model.
}
\begin{ruledtabular}
\begin{tabular}{lccccc}
  Parameter                 &  Non-CMB                      & PR4                        &  PR4+lensing              &  PR4+non-CMB            & PR4+lensing+non-CMB     \\[+1mm]
 \hline \\[-1mm]
  $\Omega_b h^2$            & $0.0287^{+0.0037}_{-0.0014}$  & $0.02228\pm 0.00014$       & $0.02228\pm 0.00013$      &  $0.02229 \pm 0.00013$  &  $0.02228 \pm 0.00013$  \\[+1mm]
  $\Omega_c h^2$            & $0.1017^{+0.0078}_{-0.0130}$  & $0.1184\pm 0.0012$         & $0.1185\pm 0.0012$        &  $0.1181 \pm 0.0010$    &  $0.11841 \pm 0.00096$  \\[+1mm]
 $H_0$                      & $68.8^{+2.0}_{-1.8}$          & $84^{+10}_{-6} (>62)$      & $84^{+10}_{-7} (>64)$     &  $67.70 \pm 0.64$       &  $67.70 \pm 0.64$       \\[+1mm]
  $\tau$                    & $0.0573$                      & $0.0573\pm 0.0061$         & $0.0574\pm 0.0062$        &  $0.0573 \pm 0.0062$    &  $0.0583 \pm 0.0062$    \\[+1mm]
  $n_s$                     & $0.9686$                      & $0.9686\pm 0.0041$         & $0.9684\pm 0.0040$        &  $0.9691 \pm 0.0038$    &  $0.9684 \pm 0.0037$    \\[+1mm]
  $\ln(10^{10} A_s)$        & $3.48^{+0.33}_{-0.18} (>2.95)$& $3.037\pm 0.014$           & $3.037\pm 0.012$          &  $3.036 \pm 0.014$      &  $3.042 \pm 0.012$      \\[+1mm]
  $w_0$                     & $-0.872\pm 0.059$             & $-1.25^{+0.41}_{-0.48}$    & $-1.27\pm 0.44$           &  $-0.869 \pm 0.060$     &  $-0.863 \pm 0.060$     \\[+1mm]  
 $w_a$                      & $-0.01^{+0.39}_{-0.24}$       & $-1.04^{+0.76}_{-1.70} (>-2.08)$     & $-1.01^{+0.76}_{-1.80} (>-2.87)$    &  $-0.46^{+0.25}_{-0.22}$     &  $-0.50^{+0.25}_{-0.22}$     \\[+1mm]  
 \hline \\[-1mm]
 $w_0+w_a$                  & $-0.89^{+0.35}_{-0.19}$       & $-2.29^{+0.89}_{-1.20}$    & $-2.28^{+0.90}_{-1.20}$   &  $-1.33^{+0.20}_{-0.17}$&  $-1.37^{+0.19}_{-0.17}$\\[+1mm]
  $100\theta_\textrm{rec}$  & $1.024^{+0.010}_{-0.012}$     & $1.04184\pm 0.00027$       & $1.04184\pm 0.00027$      &  $1.04185 \pm 0.00025$  &  $1.04183 \pm 0.00025$  \\[+1mm]
  $\Omega_m$                & $0.277^{+0.011}_{-0.016}$     & $0.212^{+0.018}_{-0.069}$  & $0.210^{+0.019}_{-0.066}$ &  $0.3079 \pm 0.0063$    &  $0.3084 \pm 0.0063$    \\[+1mm]
  $\sigma_8$                & $0.819\pm 0.030$              & $0.938^{+0.110}_{-0.051}$  & $0.941^{+0.100}_{-0.051}$ &  $0.803 \pm 0.010$      &  $0.8076 \pm 0.0087$    \\[+1mm]
\hline\\[-1mm]
  $\chi_{\textrm{min}}^2$   & $1448.29$                    & $30550.42$                  & $30558.93$                & $32009.94$              &  $32018.68$             \\[+1mm]
  $\textrm{DIC}$            & $1461.74$                    & $30598.47$                  & $30606.98$                & $32060.93$              &  $32070.29$             \\[+1mm]
  $\Delta\textrm{DIC}$      & $-6.46$                      & $-3.45$                     & $-2.12$                   & $-5.23$                 &  $-3.76$                \\[+1mm]
\end{tabular}
\\[+1mm]
\begin{flushleft}
\end{flushleft}
\end{ruledtabular}
\label{tab:results_flat_w0waCDM}
\end{table*}


\begin{table*}
	\caption{Mean and 68\% (or 95\% indicated between parentheses when the value is provided) confidence limits of $w_0w_a$CDM+$A_L$ parameterization parameters
        from non-CMB, PR4, PR4+lensing, PR4+non-CMB, and PR4+lensing+non-CMB data.
        $H_0$ has units of km s$^{-1}$ Mpc$^{-1}$. We also include the values of $\chi^2_{\text{min}}$, DIC, and $\Delta$DIC the difference with respect to the $\Lambda$CDM model.
}
\begin{ruledtabular}
\begin{tabular}{lccccc}
  Parameter                     &  Non-CMB                       & PR4                          &  PR4+lensing                 &  PR4+non-CMB            & PR4+lensing+non-CMB     \\[+1mm]
\hline \\[-1mm]
  $\Omega_b h^2$                & $0.0287^{+0.0037}_{-0.0014}$   & $0.02227\pm 0.00015$         & $0.02227\pm 0.00015$         &  $0.02234^{+0.00015}_{-0.00013}$  &  $0.02234^{+0.00014}_{-0.00012}$  \\[+1mm]
  $\Omega_c h^2$                & $0.1017^{+0.0078}_{-0.0130}$   & $0.1185^{+0.0013}_{-0.0015}$ & $0.1185\pm 0.0014$           &  $0.1177 \pm 0.0011$    &  $0.1177 \pm 0.0011$    \\[+1mm]
  $H_0$                         & $68.8^{+2.0}_{-1.8}$           & $83^{+10}_{-6} (>62)$        & $84^{+10}_{-7} (>64)$        &  $67.70 \pm 0.65$       &  $67.72 \pm 0.63$       \\[+1mm]
  $\tau$                        & $0.0573$                       & $0.0575\pm 0.0063$           & $0.0575^{+0.0059}_{-0.0066}$ &  $0.0567 \pm 0.0063$    &  $0.0567 \pm 0.0062$    \\[+1mm]
  $n_s$                         & $0.9686$                       & $0.9681\pm 0.0046$           & $0.9683^{+0.0048}_{-0.0043}$ &  $0.9703 \pm 0.0041$    &  $0.9704 \pm 0.0040$    \\[+1mm]
  $\ln(10^{10} A_s)$            & $3.48^{+0.33}_{-0.18} (>2.95)$ & $3.038\pm 0.015$             & $3.038\pm 0.015$             &  $3.033 \pm 0.014$      &  $3.033 \pm 0.014$      \\[+1mm]
  $w_0$                         & $-0.872\pm 0.059$              & $-1.25^{+0.42}_{-0.50} $     & $-1.26^{+0.42}_{-0.49}$      &  $-0.877 \pm 0.059$     &  $-0.877 \pm 0.060$     \\[+1mm]  
  $w_a$                         & $-0.01^{+0.39}_{-0.24}$        & $-0.97^{+0.78}_{-1.80} (>-2.87)$  & $-0.99^{+0.80}_{-1.80} (>-2.87)$  &  $-0.41^{+0.25}_{-0.22}$     &  $-0.41^{+0.25}_{-0.22}$  \\[+1mm]  
  $A_L$                         & $1$                            & $0.996\pm 0.053$             & $0.9998^{+0.038}_{-0.043}$   &  $1.039\pm 0.053$       &  $1.042\pm 0.037$       \\[+1mm]  
\hline \\[-1mm]
  $w_0+w_a$                     & $-0.89^{+0.35}_{-0.19}$        & $-2.23^{+0.97}_{-1.20}$      & $-2.25^{+0.93}_{-1.20}$      &  $-1.29^{+0.20}_{-0.17}$&  $-1.29^{+0.20}_{-0.17}$\\[+1mm]
  $100\theta_\textrm{rec}$      & $1.024^{+0.010}_{-0.012}$      & $1.04184\pm 0.00027$         & $1.04183\pm 0.00027$         &  $1.04188 \pm 0.00025$  &  $1.04188 \pm 0.00025$  \\[+1mm]
  $\Omega_m$                    & $0.277^{+0.011}_{-0.016}$      & $0.216^{+0.017}_{-0.074}$    & $0.213^{+0.019}_{-0.069}$    &  $0.3071 \pm 0.0064$    &  $0.3069 \pm 0.0062$    \\[+1mm]
  $\sigma_8$                    & $0.819\pm 0.030$               & $0.934^{+0.110}_{-0.055}$    & $0.938^{+0.110}_{-0.057}$    &  $0.799 \pm 0.011$      &  $0.799 \pm 0.011$      \\[+1mm]
\hline\\[-1mm]
  $\chi_{\textrm{min}}^2$       & $1448.29$                     & $30549.10$                    & $30555.88$                   & $32006.65$              &  $32015.66$             \\[+1mm]
  $\textrm{DIC}$                & $1461.74$                     & $30601.49$                    & $30611.48$                   & $32064.55$              &  $32071.69$             \\[+1mm]
  $\Delta\textrm{DIC}$          & $-6.46$                       & $-0.43$                       & $+2.38$                      & $-1.61$                 &  $-2.36$                \\[+1mm]
\end{tabular}
\\[+1mm]
\begin{flushleft}
\end{flushleft}
\end{ruledtabular}
\label{tab:results_flat_w0waCDM_Alens}
\end{table*}

\begin{figure*}[htbp]
\centering
\mbox{\includegraphics[width=175mm]{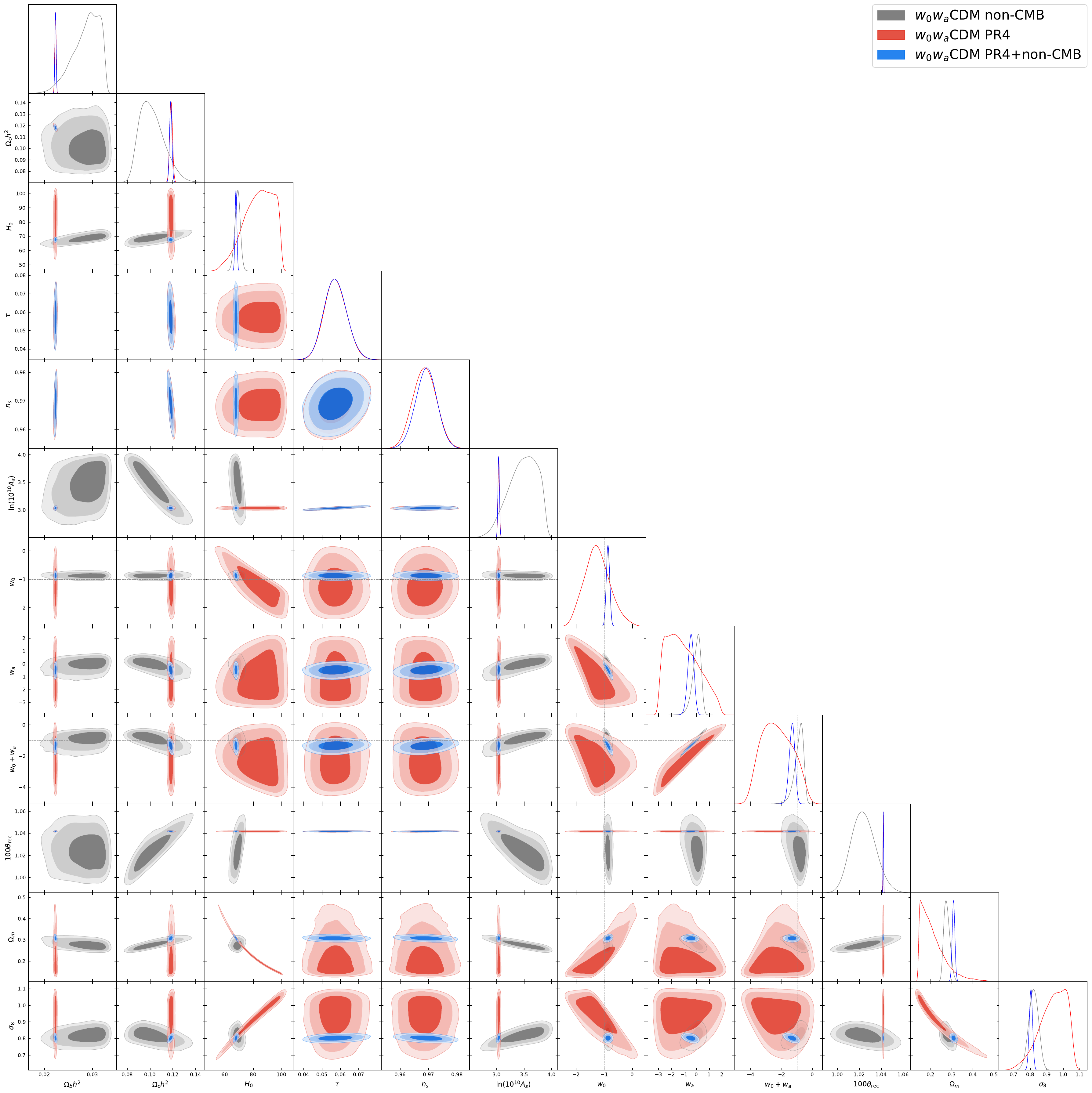}}
        \caption{One-dimensional likelihoods and 1$\sigma$, 2$\sigma$, and $3\sigma$ likelihood confidence contours of $w_0w_a$CDM parameterization parameters favored by non-CMB, PR4, and PR4+non-CMB datasets. 
}
\label{fig:nonCMB_vs_PR4_CPL}
\end{figure*}


\begin{figure*}[htbp]
\centering
\mbox{\includegraphics[width=175mm]{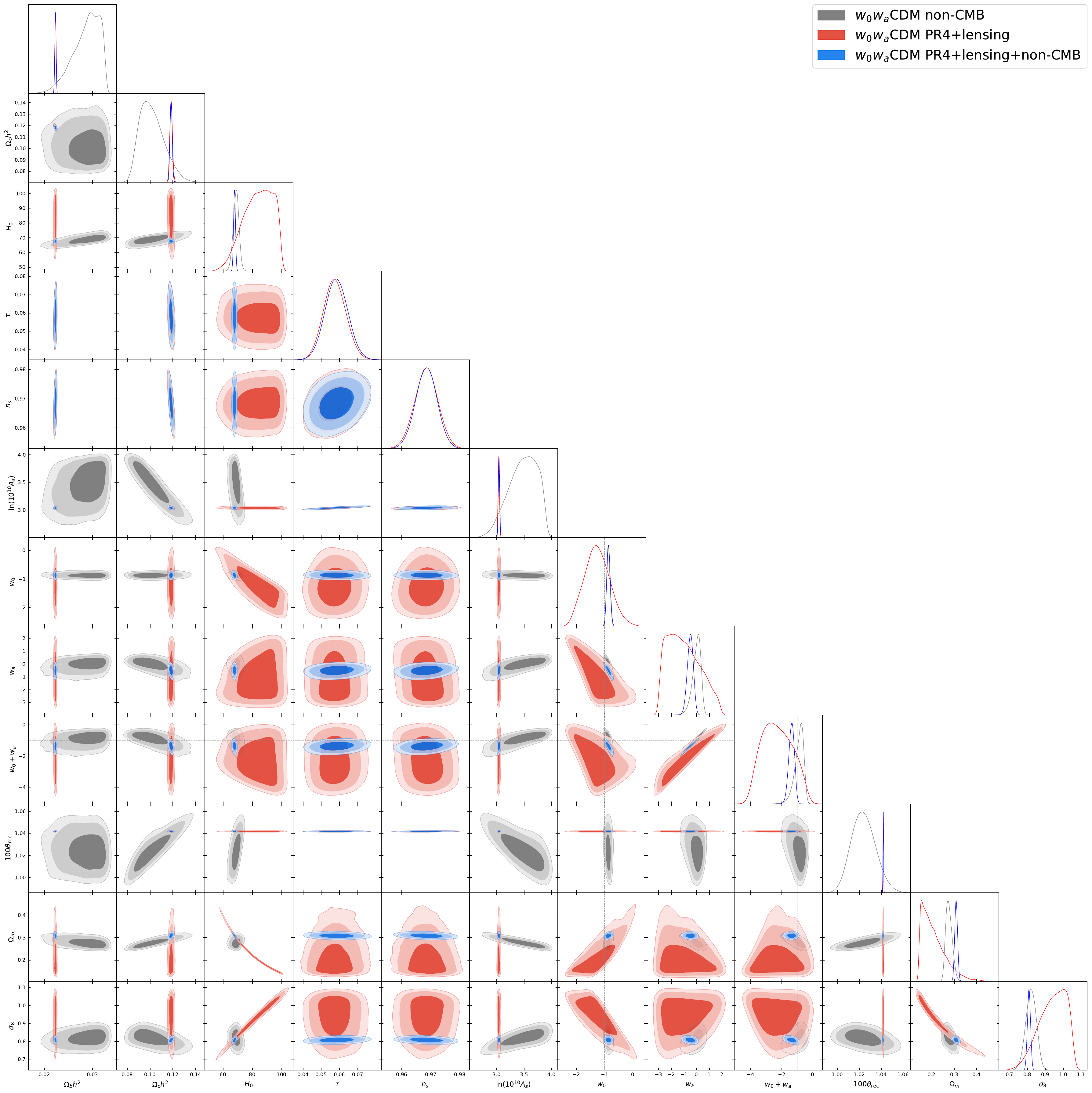}}
        \caption{One-dimensional likelihoods and 1$\sigma$, 2$\sigma$, and $3\sigma$ likelihood confidence contours of $w_0w_a$CDM parameterization parameters favored by non-CMB, PR4+lensing, and PR4+lensing+non-CMB datasets. 
}
\label{fig:CPL}
\end{figure*}


\begin{figure*}[htbp]
\centering
\mbox{\includegraphics[width=175mm]{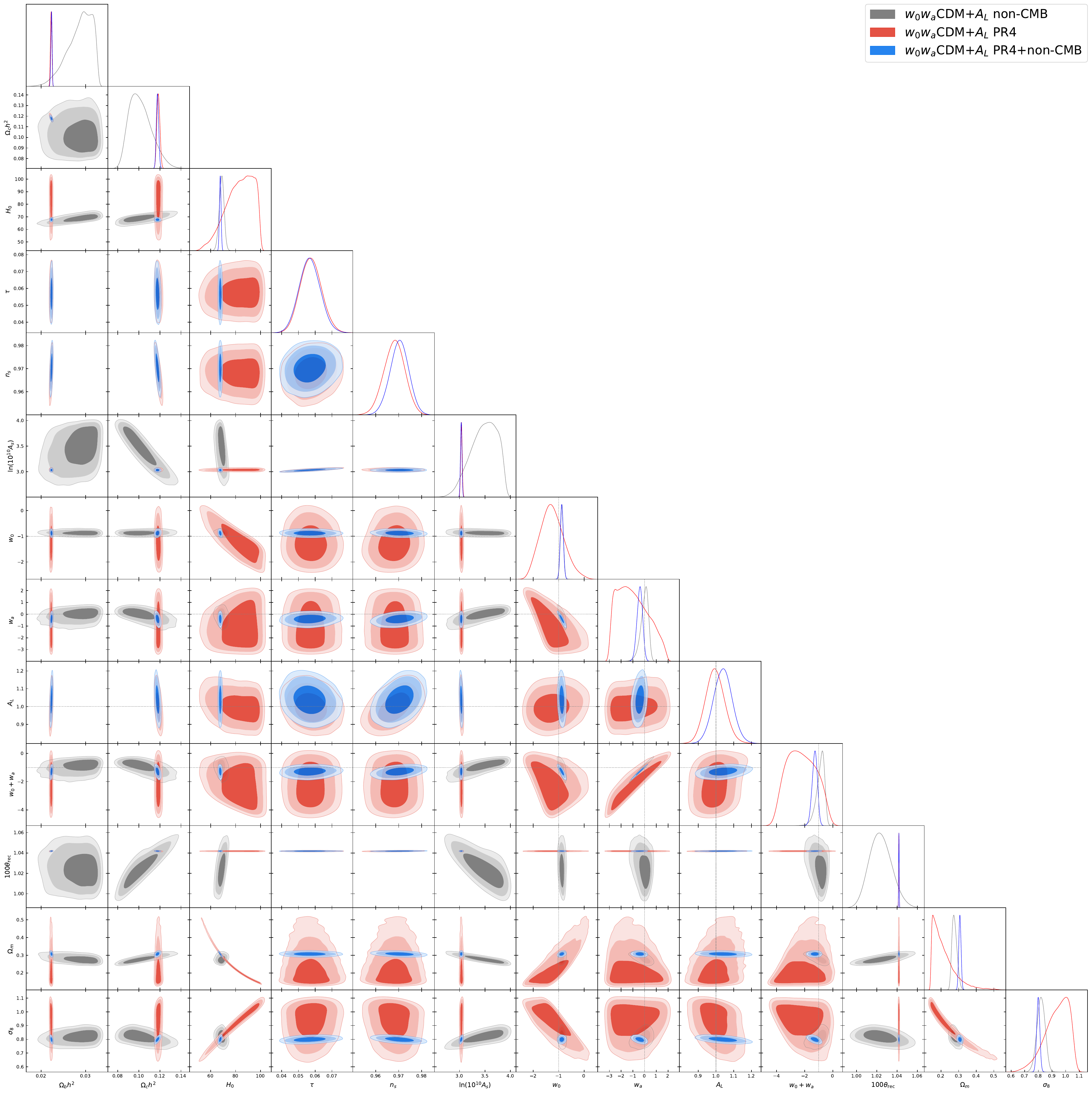}}
        \caption{One-dimensional likelihoods and 1$\sigma$, 2$\sigma$, and $3\sigma$ likelihood confidence contours of $w_0w_a$CDM+$A_L$ parameterization parameters favored by non-CMB, PR4, and PR4+non-CMB datasets. 
}
\label{fig:nonCMB_vs_PR4_CPL_AL}
\end{figure*}


\begin{figure*}[htbp]
\centering
\mbox{\includegraphics[width=175mm]{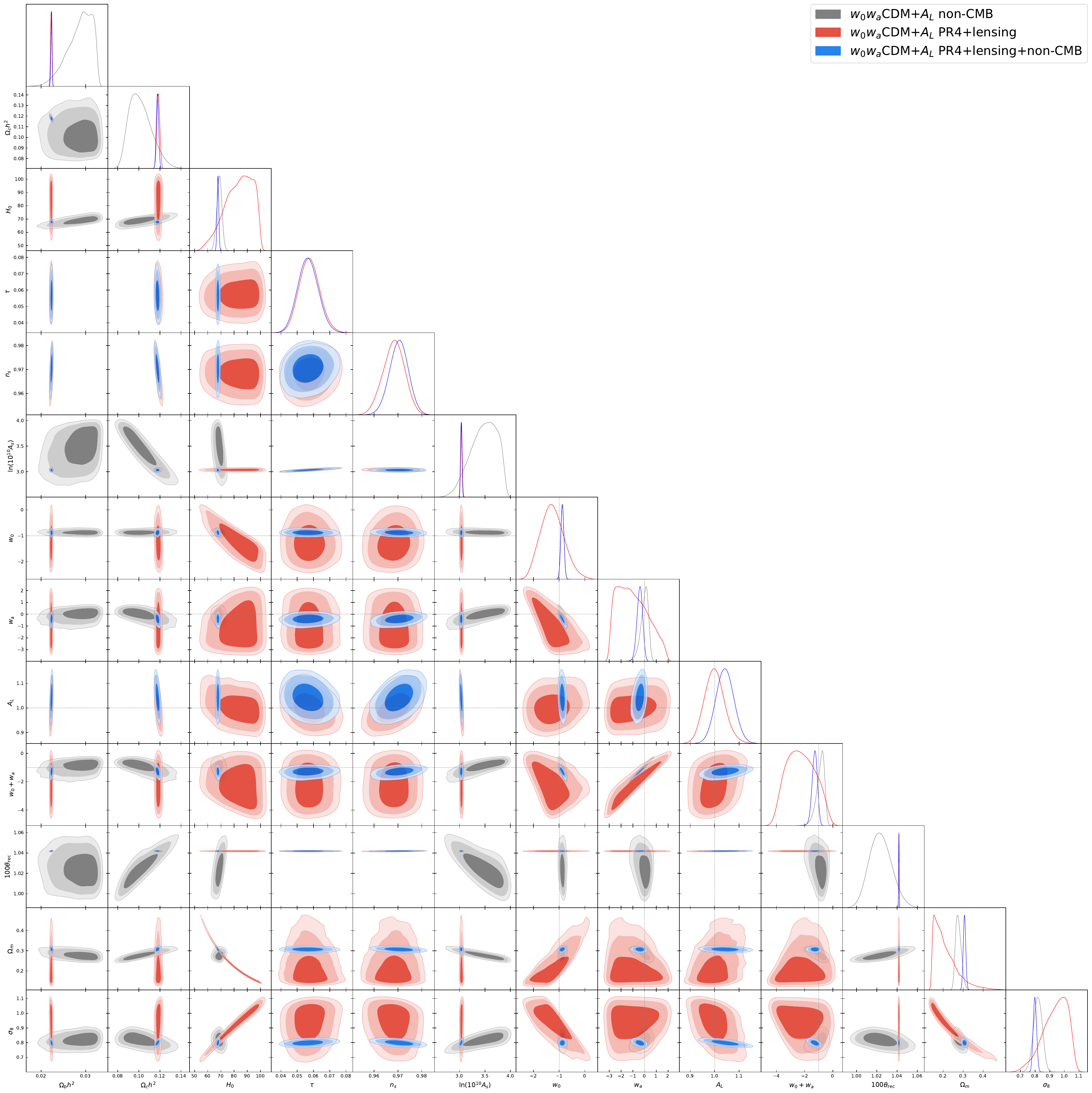}}
        \caption{One-dimensional likelihoods and 1$\sigma$, 2$\sigma$, and $3\sigma$ likelihood confidence contours of $w_0w_a$CDM+$A_L$ parameterization parameters favored by non-CMB, PR4+lensing, and PR4+lensing+non-CMB datasets. 
}
\label{fig:CPL_AL}
\end{figure*}


\subsection{Comparison with the \citeauthor{Park:2024pew} $w_0w_a$CDM$(+A_L)$ PR3 results \cite{Park:2024pew}}
\label{subsec:Parketalw0waCDM}

In this subsection we compare our $w_0w_a$CDM$(+A_L)$ PR4 \texttt{CLASS}+\texttt{Cobaya} results to the \texttt{CAMB}+\texttt{CosmoMC} results of \cite{Park:2024pew}, tables 1 and 2, that made use of the P18/PR3 datasets, except for $\theta_{\rm rec}$ since \cite{Park:2024pew} use $\theta_{\rm MC}$ instead of $\theta_{\rm rec}$. In our comparisons below, we ignore the few primary parameter cases where there is not a $2\sigma$ detection.

Since the non-CMB cosmological parameter constraints of \cite{Park:2024pew} were computed assuming different values of $\tau$ and $n_s$ than those used here, see discussion above, we do not compare the two different sets of non-CMB data results. 

Comparing the $w_0w_a$CDM parameterization PR3 data results of \cite{Park:2024pew} to our PR4 data results, all differences for our primary parameters are $0.81\sigma$ (the $\Omega_c h^2$ difference) or smaller, being $0.58\sigma$ for $\Omega_b h^2$, $-0.54\sigma$ for $n_s$, $-0.33\sigma$ for $\tau$, $0.28\sigma$ for ${\rm ln} (10^{10} A_s)$, and 0 for $w_0$. For our derived parameters the differences are $-0.15\sigma$ for $w_0+w_a$, $0.14\sigma$ for $\sigma_8$, and $0.014\sigma$ for $\Omega_m$.

Comparing the $w_0w_a$CDM parameterization PR3+lensing data results of \cite{Park:2024pew} to our PR4+lensing data results, all differences for our primary parameters are $0.76\sigma$ (the $\Omega_b h^2$ difference) or smaller, being $-0.53\sigma$ for $\tau$, $0.41\sigma$ for $\Omega_c h^2$, $-0.26\sigma$ for $n_s$, $0.054\sigma$ for ${\rm ln} (10^{10} A_s)$, and $0.042\sigma$ for $w_0$. For our derived parameters the differences are $-0.083\sigma$ for $w_0+w_a$, $0.041\sigma$ for $\Omega_m$, and $0.036\sigma$ for $\sigma_8$.

Comparing the $w_0w_a$CDM parameterization PR3+non-CMB data results of \cite{Park:2024pew} to our PR4+non-CMB data results, all differences for our primary parameters are $0.84\sigma$ (the $\Omega_b h^2$ difference) or smaller, being $0.61\sigma$ for $\Omega_c h^2$, $-0.45\sigma$ for $\tau$, $-0.34\sigma$ for $n_s$, $-0.32$ for $w_a$, $0.19\sigma$ for $w_0$, $0.14\sigma$ for ${\rm ln} (10^{10} A_s)$, and $0.12\sigma$ for $H_0$. For our derived parameters the differences are $0.47\sigma$ for $\sigma_8$, $-0.33\sigma$ for $w_0+w_a$, and $0.15\sigma$ for $\Omega_m$.

Comparing the $w_0w_a$CDM parameterization PR3+lensing+non-CMB data results of \cite{Park:2024pew} to our PR4+lensing+non-CMB data results, all differences for our primary parameters are $0.84\sigma$ (the $\Omega_b h^2$ difference) or smaller, being $-0.52\sigma$ for $\tau$, $0.50\sigma$ for $\Omega_c h^2$, $-0.26\sigma$ for $w_a$ and $n_s$, $0.15\sigma$ for $w_0$, $0.11\sigma$ for $H_0$, and $-0.11\sigma$ for ${\rm ln} (10^{10} A_s)$. For our derived parameters the differences are $-0.27\sigma$ for $w_0+w_a$, $0.25\sigma$ for $\sigma_8$, and $0.11\sigma$ for $\Omega_m$.

Comparing the $w_0w_a$CDM$+A_L$ parameterization PR3 data results of \cite{Park:2024pew} to our PR4 data results, all differences for our primary parameters are $1.6\sigma$ (the $A_L$ difference) or smaller, being $1.4\sigma$ for $\Omega_b h^2$, $-0.74\sigma$ for $\tau$, $0.37\sigma$ for $n_s$, $-0.34\sigma$ for ${\rm ln} (10^{10} A_s)$, $0.22\sigma$ for $w_0$, and $-0.19\sigma$ for $\Omega_c h^2$. For our derived parameters the differences are $0.35\sigma$ for $\Omega_m$, $-0.35\sigma$ for $\sigma_8$, and $0.24\sigma$ for $w_0+w_a$.

Comparing the $w_0w_a$CDM$+A_L$ parameterization PR3+lensing data results of \cite{Park:2024pew} to our PR4+lensing data results, all differences for our primary parameters are $1.0\sigma$ (the $\Omega_b h^2$ difference) or smaller, being $-0.74\sigma$ for $\tau$, $0.67\sigma$ for $A_L$, $-0.34\sigma$ for ${\rm ln} (10^{10} A_s)$, $0.15\sigma$ for $w_0$, $0.10\sigma$ for $n_s$, and 0 for $\Omega_c h^2$. For our derived parameters the differences are 
$0.31\sigma$ for $\Omega_m$, $-0.27\sigma$ for $\sigma_8$, and $0.17\sigma$ for $w_0+w_a$.

Comparing the $w_0w_a$CDM$+A_L$ parameterization PR3+non-CMB data results of \cite{Park:2024pew} to our PR4+non-CMB data results, all differences for our primary parameters are $1.8\sigma$ (the $A_L$ difference) or smaller, being $1.4\sigma$ for $\Omega_b h^2$, $-0.78\sigma$ for $\tau$, $0.34\sigma$ for $n_s$, $-0.32\sigma$ for ${\rm ln} (10^{10} A_s)$, $0.19\sigma$ for $H_0$, $0.12\sigma$ for $w_a$, $-0.12\sigma$ for $\Omega_c h^2$, and $-0.036\sigma$ for $w_0$. For our derived parameters the differences are $-0.37\sigma$ for $\sigma_8$, $0.15\sigma$ for $w_0+w_a$ and $-0.14\sigma$ for $\Omega_m$.

Comparing the $w_0w_a$CDM$+A_L$ parameterization PR3+lensing+non-CMB data results of \cite{Park:2024pew} to our PR4+lensing+non-CMB data results, all differences for our primary parameters are $1.1\sigma$ (the $\Omega_b h^2$ difference) or smaller, being $-0.80\sigma$ for $\tau$, $0.66\sigma$ for $A_L$, $-0.35\sigma$ for ${\rm ln} (10^{10} A_s)$, $0.13\sigma$ for $H_0$, $0.12\sigma$ for $n_s$, $0.060\sigma$ for $w_a$, $-0.024\sigma$ for $w_0$, and 0 for $\Omega_c h^2$. For our derived parameters the differences are $-0.31\sigma$ for $\sigma_8$, $-0.079\sigma$ for $\Omega_m$, and $0.076\sigma$ for $w_0+w_a$.

In summary, in the $w_0w_a$CDM case, all parameter differences are less than $1.0\sigma$.

In the $w_0w_a$CDM$+A_L$ case, the biggest differences between the PR3-based results of \cite{Park:2024pew} and the PR4-based results here are  for $A_L$: $1.6\sigma$ (PR3 vs.\ PR4) and $1.8\sigma$ (PR3+non-CMB vs.\ PR4+non-CMB), with some $\Omega_b h^2$ differences being nearly as large: $1.4\sigma$ (PR3 vs.\ PR4), $1.0\sigma$ (PR3+lensing vs.\ PR4+lensing), $1.4\sigma$ (PR3+non-CMB vs.\ PR4+non-CMB), and $1.1\sigma$ (PR3+lensing+non-CMB vs.\ PR4+lensing+non-CMB). All other parameter differences are less than $1.0\sigma$. In the $w_0w_a$CDM$+A_L$ case these differences are due to smaller $A_L$ (closer to unity) and $\Omega_b h^2$ values from the various PR4 data combinations compared to the corresponding PR3 data values.

We now discuss the ratios of the \citeauthor{Park:2024pew} $w_0w_a$CDM$(+A_L)$ parameter error bars to those computed here. In the case of parameters with asymmetric upper and lower error bars, we use the averaged error bar when computing this ratio.

The ratios of our PR4 data $w_0w_a$CDM parameterization error bars to those of the PR3 data error bars of \cite{Park:2024pew} are less than unity except for $\Omega_m$, $\sigma_8$, and $w_0+w_a$. For our primary parameters the ratios are 0.77 ($\tau$), 0.86 ($\Omega_c h^2$), 0.88 (${\rm ln} (10^{10} A_s)$), 0.90 ($w_0$),  0.93 ($\Omega_b h^2$), and 0.95 ($n_s$), while for our derived parameters the ratios are 1.0 ($\Omega_m$ and $\sigma_8$) and 1.1 ($w_0+w_a$).

The ratios of our PR4+lensing data $w_0w_a$CDM parameterization error bars to those of the PR3+lensing data error bars of \cite{Park:2024pew} are less than unity except for $\Omega_c h^2$ and $w_0+w_a$. For our primary parameters the ratios are 0.84 ($\tau$), 0.86 (${\rm ln} (10^{10} A_s)$), 0.87 ($\Omega_b h^2$), 0.88 ($w_0$), 0.98 ($n_s$), and 1.0 ($\Omega_c h^2$), while for our derived parameters the ratios are 0.96 ($\sigma_8$), 0.97 ($\Omega_m$), and 1.1 ($w_0+w_a$).

The ratios of our PR4+non-CMB data $w_0w_a$CDM parameterization error bars to those of the PR3+non-CMB data error bars of \cite{Park:2024pew} are less than unity in all cases except $H_0$ and $\Omega_m$. For our primary parameters the ratios are 0.81 ($\tau$), 0.88 (${\rm ln} (10^{10} A_s)$), 0.91 ($\Omega_c h^2$), 0.93 ($\Omega_b h^2$), 0.94 ($w_a$), 0.95 ($n_s$), 0.98 ($w_0$), and 1.0 ($H_0$), while for our derived parameters the ratios are 0.91 ($\sigma_8$), 0.95 ($w_0+w_a$), and 1.0 ($\Omega_m$).

The ratios of our PR4+lensing+non-CMB data $w_0w_a$CDM parameterization error bars to those of the PR3+lensing+non-CMB data error bars of \cite{Park:2024pew} are less than unity except for $H_0$, $\Omega_m$, and $w_0$. For our primary parameters the ratios are 0.86 (${\rm ln} (10^{10} A_s)$ and $\tau$), 0.93 ($\Omega_b h^2$), 0.95 ($n_s$), 0.96 ($\Omega_c h^2$), 0.98 ($w_a$), and 1.0 ($H_0$ and $w_0$), while for our derived parameters the ratios are 0.96 ($\sigma_8$), 0.97 ($w_0+w_a$), and 1.0 ($\Omega_m$).

The ratios of our PR4 data $w_0w_a$CDM$+A_L$ parameterization error bars to those of the PR3 data error bars of \cite{Park:2024pew} are less than unity. For our primary parameters the ratios are 0.71 ($A_L$), 0.75 ($w_0$), 0.79 ($\tau$), 0.88 (${\rm ln} (10^{10} A_s)$ and $\Omega_b h^2$), 0.93 ($\Omega_c h^2$), and 0.94 ($n_s$), while for our derived parameters the ratios are 0.59 ($\Omega_m$), 0.68 ($\sigma_8$), and 0.94 ($w_0+w_a$).

The ratios of our PR4+lensing data $w_0w_a$CDM$+A_L$ parameterization error bars to those of the PR3+lensing data error bars of \cite{Park:2024pew} are less than unity. For our primary parameters the ratios are 0.74 ($\tau$), 0.78 ($w_0$), 0.85 ($A_L$), 0.88 (${\rm ln} (10^{10} A_s)$ and $\Omega_b h^2$), and 0.93 ($n_s$ and $\Omega_c h^2$), while for our derived parameters the ratios are 0.70 ($\Omega_m$), 0.76 ($\sigma_8$), and 0.93 ($w_0+w_a$) .

The ratios of our PR4+non-CMB data $w_0w_a$CDM$+A_L$ parameterization error bars to those of the PR3+non-CMB data error bars of \cite{Park:2024pew} are less than unity except for $w_0$, $w_0+w_a$, $\Omega_m$, and $H_0$. For our primary parameters the ratios are 0.80 ($\tau$), 0.83 ($A_L$), 0.88 (${\rm ln} (10^{10} A_s)$), 0.92 ($\Omega_c h^2$), 0.93 ($\Omega_b h^2$), 0.95 ($n_s$), 0.98 ($w_a$), and 1.0 ($H_0$), while for our derived parameters the ratios are 0.92 ($\sigma_8$) and 1.0 ($w_0+w_a$ and $\Omega_m$).

The ratios of our PR4+lensing+non-CMB data $w_0w_a$CDM$+A_L$ parameterization error bars to those of the PR3+lensing+non-CMB data error bars of \cite{Park:2024pew} are less than unity except for $w_0$ and $w_0+w_a$. For our primary parameters the ratios are 0.78 ($\tau$), 0.85 (${\rm ln} (10^{10} A_s)$), 0.87 ($\Omega_b h^2$), 0.92 ($\Omega_c h^2$), 0.93 ($n_s$), 0.97 ($A_L$), 0.98 ($w_a$ and $H_0$), and 1.0 ($w_0$), while for our derived parameters the ratios are 0.92 ($\sigma_8$), 0.97 ($\Omega_m$), and 1.0 ($w_0+w_a$).

In the $w_0w_a$CDM$(+A_L)$ parameterizations, PR4 data result in more restrictive error bars particularly for the primary $\tau$ parameter, across most datasets. Unlike the $\Lambda$CDM$(+A_L)$ cases, in the $w_0w_a$CDM$(+A_L)$ parameterizations PR4 data do not necessarily more restrictively constrain some parameters, compared to PR3 data results, with some of the error bar ratios being unity, or even slightly exceeding unity. In particular, for the dark energy equation-of-state parameter $w_0+w_a$, comparing PR3(+lensing) results combined with non-CMB data to those from PR4 shows that the PR4 uncertainties are nearly identical, indicating that the updated PR4 dataset provides no tighter constraints on $w_0+w_a$.
 
The $\Delta$DIC values for the DIC differences between the $w_0w_a$CDM parameterization and the $\Lambda$CDM model, listed here in Table \ref{tab:results_flat_w0waCDM} for the PR4 datasets and in table 1 of \cite{Park:2024pew} for the PR3 datasets are only mildly different in most cases. These are $-6.46$ (non-CMB data here, and strongly for $w_0w_a$CDM) vs.\ $-7.18$ (non-CMB data of \cite{Park:2024pew}, and strongly for $w_0w_a$CDM); $-3.45$ (PR4 data, and positively for $w_0w_a$CDM) vs.\ $-2.48$ (PR3 data, and positively for $w_0w_a$CDM); $-2.12$ (PR4+lensing data, and positively for $w_0w_a$CDM) vs.\ $-2.26$ (PR3+lensing data, and positively for $w_0w_a$CDM); $-5.23$ (PR4+non-CMB data, and positively for $w_0w_a$CDM) vs.\ $-1.85$ (PR3+non-CMB data, and weakly for $w_0w_a$CDM); and $-3.76$ (PR4+lensing+non-CMB data, and positively for $w_0w_a$CDM) vs.\ $-2.45$ (PR3+lensing+non-CMB data, and positively for $w_0w_a$CDM).

The $\Delta$DIC values for the DIC differences between the $w_0w_a$CDM$+A_L$ parameterization and the $\Lambda$CDM model, listed here in Table \ref{tab:results_flat_w0waCDM_Alens} for the PR4 datasets and in table 2 of \cite{Park:2024pew} for the PR3 datasets are significantly different in some cases. These are $-0.43$ (PR4 data, and weakly for $w_0$CDM$+A_L$) vs.\ $-5.33$ (PR3 data, and positively for $w_0$CDM$+A_L$); $+2.38$ (PR4+lensing data, and positively against $w_0$CDM$+A_L$) vs.\ $-0.70$ (PR3+lensing data, and weakly for $w_0$CDM$+A_L$); $-1.61$ (PR4+non-CMB data, and weakly for $w_0$CDM$+A_L$) vs.\ $-9.16$ (PR3+non-CMB data, and strongly for $w_0$CDM$+A_L$); and $-2.36$ (PR4+lensing+non-CMB data, and positively for $w_0$CDM$+A_L$) vs.\ $-4.37$ (PR3+lensing+non-CMB data, and positively for $w_0$CDM$+A_L$).

The $\Delta$DIC values for the DIC differences between the $w_0w_a$CDM$+A_L$ and $w_0w_a$CDM parameterizations, computed from Tables \ref{tab:results_flat_w0waCDM} and  \ref{tab:results_flat_w0waCDM_Alens} here for the PR4 datasets and computed from tables 1 and 2 of \cite{Park:2024pew} for the PR3 datasets are significantly different in some cases. These are $+3.02$ (PR4 data, and positively against $w_0w_a$CDM$+A_L$) vs.\ $-2.85$ (PR3 data, and positively for $w_0w_a$CDM$+A_L$); $+4.50$ (PR4+lensing data, and positively against $w_0w_a$CDM$+A_L$) vs.\ $+1.56$ (PR3+lensing data, and weakly against $w_0w_a$CDM$+A_L$); $+3.62$ (PR4+non-CMB data, and positively against $w_0w_a$CDM$+A_L$) vs.\ $-7.31$ (PR3+non-CMB data, and strongly for $w_0w_a$CDM$+A_L$); and $+1.40$ (PR4+lensing+non-CMB data, and weakly against $w_0w_a$CDM$+A_L$) vs.\ $-1.92$ (PR3+lensing+non-CMB data, and weakly for $w_0w_a$CDM$+A_L$).

The biggest difference in $\Delta$DIC values, larger than 5, is for PR4+non-CMB vs.\ PR3+non-CMB, being 10.93 for $w_0w_a$CDM$+A_L$ compared to $w_0w_a$CDM and 7.55 for $w_0w_a$CDM$+A_L$ compared to $\Lambda$CDM, and 6.07 PR4 vs.\ PR3 for $w_0w_a$CDM$+A_L$ compared to $\Lambda$CDM.

\begin{table}
\caption{Consistency check parameter $\log_{10} \mathcal{I}$ for PR4 vs.\ non-CMB data sets and PR4+lensing vs.\ non-CMB data sets in the $\Lambda\textrm{CDM}$($+A_L$) models, and the $w_0$CDM($+A_L$) and $w_0 w_a$CDM($+A_L$) parameterizations.
}
\begin{ruledtabular}
\begin{tabular}{lcc}
\\[-1mm] 
  Data                           & PR4 vs non-CMB   & PR4+lensing vs non-CMB  \\[+1mm]
\cline{1-3}\\[-1mm]
                                 & \multicolumn{2}{c}{$\Lambda$CDM model}    \\[+1mm]
\cline{2-3}\\[-1mm]
  $\log_{10} \mathcal{I}$        & $0.860$          & $0.705$                \\[+1mm]
\hline \\[-1mm]
                                 & \multicolumn{2}{c}{$\Lambda$CDM+$A_L$ model} \\[+1mm]
\cline{2-3}\\[-1mm]
   $\log_{10} \mathcal{I}$       & $1.617$          & $1.121$                \\[+1mm]
\hline \\[-1mm]
                                 & \multicolumn{2}{c}{$w_0$CDM parameterization} \\[+1mm]
\cline{2-3}\\[-1mm]
   $\log_{10} \mathcal{I}$       & $-1.359$         & $-1.573$              \\[+1mm]
\hline \\[-1mm]
                                 & \multicolumn{2}{c}{$w_0$CDM+$A_L$ parameterization} \\[+1mm]
\cline{2-3}\\[-1mm]
   $\log_{10} \mathcal{I}$       & $-0.682$         & $-0.201$          \\[+1mm]
\hline \\[-1mm]
                                 & \multicolumn{2}{c}{$w_0 w_a$CDM parameterization}  \\[+1mm]
\cline{2-3}\\[-1mm]
   $\log_{10} \mathcal{I}$       & $-0.157$         & $-0.340$         \\[+1mm]
\hline \\[-1mm]
                                 & \multicolumn{2}{c}{$w_0 w_a\Lambda$CDM+$A_L$ parameterization}   \\[+1mm]
\cline{2-3}\\[-1mm]
   $\log_{10} \mathcal{I}$       & $-0.286$         & $0.334$          \\[+1mm]
\end{tabular}
\\[+1mm]
\end{ruledtabular}
\label{tab:consistency}
\end{table}


\subsection{Consistency of PR4(+lensing) data and non-CMB data cosmological constraints} 
\label{subsec:Consistency}

Table \ref{tab:consistency} lists the consistency check parameter $\log_{10} \mathcal{I}$ values for P18 vs.\ non-CMB datasets and for P18+lensing vs.\ non-CMB datasets in all six models/parameterizations considered here: $\Lambda\textrm{CDM}$($+A_L$), $w_0$CDM($+A_L$), and $w_0 w_a$CDM($+A_L$). The corresponding PR3 data values are given in tables X and XIV of \cite{deCruzPerez:2024abc} and table 3 of \cite{Park:2024pew}.

Comparing the PR3 and PR4 results, the PR4 and non-CMB constraints are generally more consistent (or less inconsistent) than the PR3 and non-CMB constraints, with the two exceptions being the $w_0$CDM$+A_L$ and the $w_0w_a$CDM$+A_L$ parameterization results, where the PR4 and non-CMB constraints are more inconsistent than the PR3 and non-CMB constraints. The PR4+lensing and non-CMB constraints exhibit less uniform changes compared to the PR4 case, with the $\Lambda\textrm{CDM}$($+A_L$) PR4 results here being a little less consistent than the PR3 ones, and the $w_0$CDM($+A_L$) and $w_0w_a$CDM($+A_L$) PR4 results here being a little less inconsistent (or more consistent) than the PR3 ones.

The most significant changes in Jeffreys' scale categories occur for the $w_0$CDM parameterization, with the PR4 data cases here being less inconsistent, only {\it strongly} so, compared to the {\it decisive} inconsistency of the PR3 data cases. With this, according to our consistency criterion, PR4 and PR4+lensing data can be jointly utilized with non-CMB data to constrain cosmological parameters, unlike the PR3 cases, \cite{deCruzPerez:2024abc}. Other changes in Jeffreys' scale categories are less significant. Overall, according to our consistency criterion, PR4 and PR4+lensing data can be jointly utilized with non-CMB data to constrain cosmological parameters in all six models/parameterizations we consider in this paper.

From Table \ref{tab:consistency} we see that PR4(+lensing) and non-CMB data constraints are {\it substantially} and {\it strongly} consistent in the $\Lambda$CDM and $\Lambda$CDM$+A_L$ models, respectively. These results are qualitatively consistent with the very good overlap of the red PR4(+lensing) and gray non-CMB data constraint contours in Figs.\ \ref{fig:nonCMB_vs_PR4_LCDM}--\ref{fig:LCDM_AL}.

These $\Lambda$CDM($+A_L$) model results are also qualitatively consistent with the numerical parameter values listed in Tables \ref{tab:results_flat_LCDM} and \ref{tab:results_flat_LCDM_Alens}. Comparing the values determined from non-CMB data with those from PR4 data, we find for $\Lambda$CDM that only the primary parameters $H_0$ ($1.3\sigma$) and $\Omega_b h^2$ ($1.1\sigma$) and the derived parameter $\Omega_m$ ($-1.0\sigma$) differ by $1\sigma$ or more, while for $\Lambda$CDM$+A_L$ only the primary parameters $H_0$ ($1.2\sigma$) and $\Omega_b h^2$ ($1.1\sigma$) differ by $1\sigma$ or more. We note that the PR4 and PR4+lensing data results in these tables are not that different so the differences between the non-CMB data and the PR4+lensing data will also be close to those mentioned above. We also note that the size of the differences between non-CMB and PR4 data primary parameter results do not change significantly  between the $\Lambda$CDM and $\Lambda$CDM$+A_L$ cases.

From Table \ref{tab:consistency} we see for the $w_0$CDM parameterization that the PR4 and PR4+lensing data results are {\it strongly} inconsistent with the non-CMB data results. These findings are qualitatively consistent with the lack of overlap between some of the red PR4(+lensing) and gray non-CMB data 2$\sigma$ constraint contours in Figs.\ \ref{fig:nonCMB_vs_PR4_XCDM} and \ref{fig:XCDM}. These 2$\sigma$ contours do not overlap in the $\Omega_b h^2$---$\Omega_c h^2$, $\Omega_b h^2$---${\rm ln}(10^{10}A_s)$, $\Omega_b h^2$---$100 \theta_{\rm rec}$, and $w_0$---$\Omega_m$ subpanels in these figures, and there is only a very slight overlap between $2\sigma$ contours in the ${\rm ln}(10^{10}A_s$)---$\Omega_m$ subpanel in Fig.\  \ref{fig:nonCMB_vs_PR4_XCDM}. For the $w_0$CDM$+A_L$ parameterization, PR4 and non-CMB data constraints are {\it substantially} inconsistent while PR4+lensing and non-CMB data constraints are mildly, less than {\it substantially}, inconsistent. These results are again qualitatively consistent with the behavior of the $2\sigma$ contours shown in Fig.\ \ref{fig:nonCMB_vs_PR4_XCDM_AL}, where the red PR4 and gray non-CMB contours do not overlap in the $\Omega_b h^2$---$\Omega_c h^2$, $\Omega_b h^2$---${\rm ln}(10^{10}A_s)$, $\Omega_b h^2$---$100 \theta_{\rm rec}$, and $w_0$---$\Omega_m$ subpanels, and there is only a very slight overlap between $2\sigma$ contours in the ${\rm ln}(10^{10}A_s$)---$\Omega_m$ subpanel. Similarly, in Fig.\ \ref{fig:XCDM_AL}, the red PR4+lensing and gray non-CMB data 2$\sigma$ constraint contours fail to overlap in the $\Omega_b h^2$---$\Omega_c h^2$, $\Omega_b h^2$---${\rm ln}(10^{10}A_s)$, and $\Omega_b h^2$---$100 \theta_{\rm rec}$ subpanels, and there is only very slight overlaps between $2\sigma$ contours in the  ${\rm ln}(10^{10}A_s$)---$\Omega_m$ and $w_0$---$\Omega_m$ subpanels.

These $w_0$CDM($+A_L$) parameterization results are also qualitatively consistent with the numerical parameter values listed in Tables \ref{tab:results_flat_XCDM} and \ref{tab:results_flat_XCDM_Alens}. Comparing the differences between the values determined from non-CMB data and from PR4 data, we find (ignoring the $H_0$, for $A_L = 1$ and $A_L$ varying, and $w_0$, for $A_L = 1$, cases where there are not $2\sigma$ detections) for $w_0$CDM that (all) primary parameters $\Omega_b h^2$ ($4.3\sigma$), ln($10^{10} A_s$) ($2.4\sigma$) and $\Omega_c h^2$ ($-2.3\sigma$) and (all) derived parameters $\Omega_m$ ($3.1\sigma$), $\sigma_8$ ($-2.0\sigma$), and $\theta_{\rm rec}$ ($-1.9\sigma$) differ by $1\sigma$ or more, while for $w_0$CDM$+A_L$ (all) primary parameters $\Omega_b h^2$ ($4.3\sigma$), $w_0$ ($2.7\sigma$), ln($10^{10} A_s$) ($2.4\sigma$) and $\Omega_c h^2$ ($-2.3\sigma$) and (all) derived parameters $\Omega_m$ ($2.4\sigma$), $\theta_{\rm rec}$ ($-1.9\sigma$), and $\sigma_8$ ($-1.7\sigma$) differ by $1\sigma$ or more. We note that the PR4 and PR4+lensing data results in these tables are very similar, so the differences between the non-CMB data and PR4+lensing data will not differ much from those listed in the preceding sentence. We also note that the size of the differences between non-CMB and PR4 data primary parameter results does not significantly differ between the $w_0$CDM and $w_0$CDM$+A_L$ cases.

From Table \ref{tab:consistency} we see that PR4(+lensing) and non-CMB data constraints are mildly, less than {\it substantially}, inconsistent for the $w_0w_a$CDM parameterization, as are the PR4 and non-CMB data constraints for the $w_0w_a$CDM$+A_L$ parameterization, while PR4+lensing and non-CMB data constraints are mildly, less than {\it substantially}, consistent for the $w_0w_a$CDM$+A_L$ parameterization. These results are qualitatively consistent with the mostly good overlap of the red PR4(+lensing) and gray non-CMB data constraint contours in Figs.\ \ref{fig:nonCMB_vs_PR4_CPL}---\ref{fig:CPL_AL}, with the exception of the lack of 2$\sigma$ contours overlap in the $\Omega_b h^2$---$100 \theta_{\rm rec}$ subpanels, and there is only very slight overlap between $2\sigma$ contours in the $\Omega_b h^2$---$\Omega_c h^2$ subpanels.

These $w_0w_a$CDM($+A_L$) parameterization results are also qualitatively consistent with the numerical parameter values listed in Tables \ref{tab:results_flat_w0waCDM} and \ref{tab:results_flat_w0waCDM_Alens}. Comparing the differences between the values determined from non-CMB data and from PR4 data, we find (ignoring the $H_0$, ln($10^{10} A_s$), and $w_a$ cases where there are not $2\sigma$ detections) for $w_0w_a$CDM that primary parameters $\Omega_b h^2$ ($4.6\sigma$), $\Omega_c h^2$ ($-2.1\sigma$), and $w_0$ ($0.91\sigma$) and (all) derived parameters $\Omega_m$ ($2.7\sigma$), $\sigma_8$ ($-2.0\sigma$), $\theta_{\rm rec}$ ($-1.8\sigma$), and $w_0+w_a$ ($1.5\sigma$) differ by $1\sigma$ or more, while for $w_0w_a$CDM$+A_L$ primary parameters $\Omega_b h^2$ ($4.6\sigma$), $\Omega_c h^2$ ($-2.1\sigma$), and $w_0$ ($0.89\sigma$) and (all) derived parameters $\Omega_m$ ($2.6\sigma$), $\sigma_8$ and $\theta_{\rm rec}$ ($-1.8\sigma$), and $w_0+w_a$ ($1.4\sigma$) differ by $1\sigma$ or more. We note that the PR4 and PR4+lensing data results in these tables are quite similar, so the differences between the non-CMB data and PR4+lensing data will also be close to those mentioned in the preceding sentence. We also note that the size of the differences between non-CMB and PR4 data primary parameter results does not change significantly between the $w_0w_a$CDM and $w_0w_a$CDM$+A_L$ cases.

\subsection{Comparing differences between non-CMB and PR4 data cosmological parameter values and between non-CMB and PR3 data cosmological parameter values}
\label{subsec:comparing_differences}

In this subsection we compare the differences between non-CMB and PR4 data cosmological parameter values, computed here using \texttt{CLASS}+\texttt{Cobaya}, and between non-CMB and PR3 data cosmological parameter values, computed using \texttt{CAMB}+\texttt{CosmoMC} in \cite{deCruzPerez:2024abc, Park:2024pew}, for the $\Lambda$CDM$(+A_L)$ models and the $w_0$CDM$(+A_L)$ and $w_0w_a$CDM$(+A_L)$ parameterizations. Because the PR4 and PR4+lensing results and the PR3 and PR3+lensing results are very similar, we expect similar results to those below for the non-CMB and PR4+lensing differences and for the non-CMB and PR3+lensing differences. Again, we do not compare the PR4 $\theta_{\rm rec}$ and PR3 $\theta_{\rm MC}$ differences.

While we have already seen in Sec.\ \ref{subsec:Consistency} that PR4 data and non-CMB data constraints are mutually consistent at the chosen significance for the $\Lambda$CDM model and the $w_0$CDM and $w_0w_a$CDM parameterizations, PR3 data and non-CMB data constraints are only mutually consistent for the $\Lambda$CDM model and the $w_0 w_a$CDM parameterization at the chosen significance \cite{deCruzPerez:2024abc, Park:2024pew} but are mutually inconsistent at the chosen significance for the $w_0$CDM parameterization \cite{deCruzPerez:2024abc}. In an unsuccessful attempt to understand the reason for this difference between PR4 and PR3 results we compare here the differences between non-CMB and PR4 data cosmological parameter values and between non-CMB and PR3 data cosmological parameter values.

From Table \ref{tab:results_flat_LCDM} and from table IV of \cite{deCruzPerez:2024abc}, for the $\Lambda$CDM model with $A_L=1$, we find primary cosmological parameter differences, between non-CMB and PR4 (non-CMB and PR3) data, for $\Omega_b h^2$ of $1.1\sigma$ ($0.98\sigma$), for $\Omega_c h^2$ of $0.28\sigma$ ($0.062\sigma$), for $H_0$ of $1.3\sigma$ ($1.4\sigma$), and for ln($10^{10} A_s$) of $-0.24\sigma$ ($-0.40\sigma$), with derived cosmological parameter differences, for $\Omega_m$ of $-1.0\sigma$ ($-1.5\sigma$) and for $\sigma_8$ of $-0.70\sigma$ ($-1.0\sigma$). The non-CMB and PR4 data differences and the non-CMB and PR3 data differences are not significant in the $\Lambda$CDM model.

From Table \ref{tab:results_flat_LCDM_Alens} and from table VII of \cite{deCruzPerez:2024abc}, for the $\Lambda$CDM$+A_L$ model with varying $A_L$, we find primary cosmological parameter differences, between non-CMB and PR4 (non-CMB and PR3) data, for $\Omega_b h^2$ of $1.1\sigma$ ($0.91\sigma$), for $\Omega_c h^2$ of $0.33\sigma$ ($0.33\sigma$), for $H_0$ of $1.2\sigma$ ($0.91\sigma$), and for ln($10^{10} A_s$) of $-0.21\sigma$ ($-0.27\sigma$), with derived cosmological parameter differences, for $\Omega_m$ of $-0.78\sigma$ ($-0.48\sigma$) and for $\sigma_8$ of $-0.61\sigma$ ($-0.57\sigma$). The non-CMB and PR4 data differences and the non-CMB and PR3 data differences are not significant in the $\Lambda$CDM$+A_L$ model.

From Table \ref{tab:results_flat_XCDM} and from table XI of \cite{deCruzPerez:2024abc}, for the $w_0$CDM parameterization with $A_L=1$, we find primary cosmological parameter differences (we ignore primary parameter cases where there is not a $2\sigma$ detection), between non-CMB and PR4 (non-CMB and PR3) data, for $\Omega_b h^2$ of $4.3\sigma$ ($2.1\sigma$), for $\Omega_c h^2$ of $-2.3\sigma$ ($-2.3\sigma$), and for ln($10^{10} A_s$) of $2.4\sigma$ ($2.6\sigma$), with derived cosmological parameter differences, for $\Omega_m$ of $3.1\sigma$ ($1.5\sigma$) and for $\sigma_8$ of $-2.0\sigma$ ($-2.0\sigma$). Unlike the $\Lambda$CDM model, the $w_0$CDM parameterization shows parameter differences between PR3/PR4 CMB data and non-CMB data that generally exceed $2\sigma$. In particular, the $\Omega_b h^2$ and $\Omega_m$ parameters exhibit a significant difference of more than $3\sigma$ for PR4, with PR4 showing more than twice the level of inconsistency compared to PR3.

From Table \ref{tab:results_flat_XCDM_Alens} and from table XI of \cite{deCruzPerez:2024abc}, for the $w_0$CDM$+A_L$ parameterization with varying $A_L$, we find primary cosmological parameter differences (we ignore primary parameter cases where there is not a $2\sigma$ detection), between non-CMB and PR4 (non-CMB and PR3) data, for $\Omega_b h^2$ of $4.3\sigma$ ($2.1\sigma$), for $\Omega_c h^2$ of $-2.3\sigma$ ($-2.1\sigma$), and for ln($10^{10} A_s$) of $2.4\sigma$ ($2.7\sigma$), with derived cosmological parameter differences, for $\Omega_m$ of $2.4\sigma$ ($0\sigma$) and for $\sigma_8$ of $-1.7\sigma$ ($-0.29\sigma$). Similar to the $w_0$CDM parameterization, the $w_0$CDM+$A_L$ parameterization also exhibits parameter differences exceeding $2\sigma$ for the $\Omega_b h^2$, $\Omega_c h^2$, and $\ln(10^{10} A_s)$ parameters between PR3/PR4 CMB data and non-CMB data. In particular, the $\Omega_b h^2$ parameter shows a large $4.3\sigma$ difference for PR4, with PR4 showing a larger overall level of difference compared to PR3.

From Table \ref{tab:results_flat_w0waCDM} and from table 1 of \cite{Park:2024pew}, for the $w_0w_a$CDM parameterization with $A_L=1$, we find primary cosmological parameter differences (we ignore primary parameter cases where there is not a $2\sigma$ detection), between non-CMB and PR4 (non-CMB and PR3) data, for $\Omega_b h^2$ of $4.6\sigma$ ($2.1\sigma$), for $\Omega_c h^2$ of $-2.1\sigma$ ($-3.3\sigma$), and for $w_0$ of $0.91\sigma$ ($0.86\sigma$), with derived cosmological parameter differences, for $w_0+w_a$ of $1.5\sigma$ ($2.3\sigma$), for $\Omega_m$ of $2.7\sigma$ ($2.6\sigma$), and for $\sigma_8$ of $-2.0\sigma$ ($-2.2\sigma$). Similar to the $w_0$CDM parameterization, the $w_0 w_a$CDM parameterization shows parameter differences exceeding $2\sigma$ in the $\Omega_b h^2$ and $\Omega_c h^2$ parameters and the derived parameters $\Omega_m$ and $\sigma_8$. However, the difference in the dark energy equation of state parameter $w_0$ remains below $1 \sigma$. In particular, the $\Omega_b h^2$ parameter exhibits a large $4.6\sigma$ difference in PR4, which is $2.2$ times the value from PR3 and larger than that in the $w_0$CDM parameterization.

From Table \ref{tab:results_flat_w0waCDM_Alens} and from table 2 of \cite{Park:2024pew}, for the $w_0w_a$CDM$+A_L$ parameterization with varying $A_L$, we find primary cosmological parameter differences (we ignore primary parameter cases where there is not a $2\sigma$ detection), between non-CMB and PR4 (non-CMB and PR3) data, for $\Omega_b h^2$ of $4.6\sigma$ ($2.1\sigma$), for $\Omega_c h^2$ of $-2.1\sigma$ ($-3.0\sigma$), and for $w_0$ of $0.89\sigma$ ($0.36\sigma$), with derived cosmological parameter differences, for $w_0+w_a$ of $1.4\sigma$ ($0.80\sigma$), for $\Omega_m$ of $2.6\sigma$ ($0.28\sigma$), and for $\sigma_8$ of $-1.8\sigma$ ($-0.59\sigma$). In the $w_0 w_a$CDM+$A_L$ parameterization, parameter differences exceeding $2\sigma$ appear only in the  $\Omega_b h^2$ and $\Omega_c h^2$ parameters. As in the $w_0 w_a$CDM parameterization, the difference in the dark energy equation of state parameter $w_0$ remains below $1 \sigma$. In particular, the $\Omega_b h^2$ parameter shows a large $4.6\sigma$ difference in PR4, which is $2.2$ times larger than in PR3.

\subsection{Comparing the values of the $w_0+1$, $w_0+w_a+1$, $w_a$, and $A_L-1$ deviations from $0$ for PR4 and PR3 data}
\label{subsec:comparing_deviations}

In this subsection we tabulate and compare the significance of the deviations in the beyond-$\Lambda$CDM-model parameters, $w_0$, $w_0+w_a$, $w_a$, and $A_L$, from their expected values in the $\Lambda$CDM model, for both PR4 and PR3 combination datasets.

For the $w_0$CDM parameterization with $A_L=1$, from Table \ref{tab:results_flat_XCDM} here and table XI of \cite{deCruzPerez:2024abc}, $w_0$ differs from the $\Lambda$CDM value of $-1$, for non-CMB data with PR4 $\tau$ and $n_s$ values by $3.5\sigma$ q [favoring quintessence dynamics] (for non-CMB data with PR3 $\tau$ and $n_s$ values by $4.5\sigma$ q), for PR4 data by $2.7\sigma$ p [favoring phantom dynamics] (for PR3 data by $3.9\sigma$ p), for PR4+lensing data by $2.8\sigma$ p (for PR3+lensing data by $3.4\sigma$ p), for PR4+non-CMB data by $0.87\sigma$ q (for PR3+non-CMB data by $0.58\sigma$ q), and for PR4+lensing+non-CMB data by $0.63\sigma$ q (for PR3+lensing+non-CMB data by $0.43\sigma$ q).

For the $w_0$CDM$+A_L$ parameterization, from Table \ref{tab:results_flat_XCDM_Alens} here and table XI of \cite{deCruzPerez:2024abc}, $w_0$ differs from the $\Lambda$CDM value of $-1$, for PR4 data by $2.1\sigma$ p (for PR3 data by $0.74\sigma$ p), for PR4+lensing data by $2.3\sigma$ p (for PR3+lensing data by $1.3\sigma$ p), for PR4+non-CMB data by $1.1\sigma$ q (for PR3+non-CMB data by $1.5\sigma$ q), and for PR4+lensing+non-CMB data by $1.1\sigma$ q (for PR3+lensing+non-CMB data by $1.3\sigma$ q).\footnote{We note that PR4, PR3, PR4+lensing, and PR3+lensing data do not provide $2\sigma$ determinations of $w_0$ in the $w_0$CDM parameterization.}

For the $w_0w_a$CDM parameterization with $A_L=1$, from Table \ref{tab:results_flat_w0waCDM} here and table 1 of \cite{Park:2024pew}, $w_0$, $w_0+w_a$, and $w_a$ differ respectively from the $\Lambda$CDM values of $-1$, $-1$, and $0$, for non-CMB data with PR4 $\tau$ and $n_s$ values by $2.2\sigma$ q, $0.58\sigma$ q, and $-0.026\sigma$ (for non-CMB data with PR3 $\tau$ and $n_s$ values by $2.3\sigma$ q, $1.5\sigma$ q, and $+0.50\sigma$); for PR4 data by $0.61\sigma$ p, $1.4\sigma$ p, and $-1.4\sigma$ (for PR3 data by $0.58\sigma$ p, $2.0\sigma$ p, and $-1.1\sigma$); for PR4+lensing data by $0.61\sigma$ p, $1.4\sigma$ p, and $-1.3\sigma$ (for PR3+lensing data by $0.55\sigma$ p, $1.8\sigma$ p, and $-0.92\sigma$); for PR4+non-CMB data by $2.2\sigma$ q, $1.7\sigma$ p, and $-1.8\sigma$ (for PR3+non-CMB data by $2.4\sigma$ q, $2.0\sigma$ p, and $-2.1\sigma$); and for PR4+lensing+non-CMB data by $2.3\sigma$ q, $1.9\sigma$ p, and $-2.0\sigma$ (for PR3+lensing+non-CMB data by $2.5\sigma$ q, $2.2\sigma$ p, and $-2.3\sigma$).\footnote{We note that PR4, PR3, PR4+lensing, and PR3+lensing data do not provide $2\sigma$ determinations of $w_a$ in the $w_0w_a$CDM parameterization.}

For the $w_0w_a$CDM$+A_L$ parameterization, from Table \ref{tab:results_flat_w0waCDM_Alens} here and table 2 of \cite{Park:2024pew}, $w_0$, $w_0+w_a$, and $w_a$ differ respectively from the $\Lambda$CDM values of $-1$, $-1$, and $0$, for PR4 data by $0.60\sigma$ p, $1.3\sigma$ p, and $-1.2\sigma$ (for PR3 data by $0.13\sigma$ p, $0.65\sigma$ p, and $-0.62\sigma$); for PR4+lensing data by $0.62\sigma$ p, $1.3\sigma$ p, and $-1.2\sigma$ (for PR3+lensing data by $0.29\sigma$ p, $0.83\sigma$ p, and $-0.69\sigma$); for PR4+non-CMB data by $2.1\sigma$ q, $1.5\sigma$ p, and $-1.6\sigma$ (for PR3+non-CMB data by $2.0\sigma$ q, $1.3\sigma$ p, and $-1.5\sigma$); and for PR4+lensing+non-CMB data by $2.1\sigma$ q, $1.5\sigma$ p, and $-1.6\sigma$ (for PR3+lensing+non-CMB data by $2.0\sigma$ q, $1.4\sigma$ p, and $-1.5\sigma$).\footnote{We note that PR4, PR3, PR4+lensing, and PR3+lensing data do not provide $2\sigma$ determinations of $w_a$ in the $w_0w_a$CDM$+A_L$ parameterization.}

For the $\Lambda$CDM$+A_L$ model, from Table \ref{tab:results_flat_LCDM_Alens} here and table VII of \cite{deCruzPerez:2024abc}, $A_L$ differs from the desired value of $1$, for PR4 data by $0.56\sigma$ (for PR3 data by $2.7\sigma$), for PR4+lensing data by $0.97\sigma$ (for PR3+lensing data by $1.8\sigma$), for PR4+non-CMB data by $1.0\sigma$ (for PR3+non-CMB data by $3.3\sigma$), and for PR4+lensing+non-CMB data by $1.6\sigma$ (for PR3+lensing+non-CMB data by $2.5\sigma$).

For the $w_0$CDM$+A_L$ parameterization, from Table \ref{tab:results_flat_XCDM_Alens} here and table XI of \cite{deCruzPerez:2024abc}, $A_L$ differs from the desired value of $1$, for PR4 data by $0.034\sigma$ (for PR3 data by $1.8\sigma$), for PR4+lensing data by $0.13\sigma$ (for PR3+lensing data by $0.92\sigma$), for PR4+non-CMB data by $1.3\sigma$ (for PR3+non-CMB data by $3.5\sigma$), and for PR4+lensing+non-CMB data by $1.8\sigma$ (for PR3+lensing+non-CMB data by $2.7\sigma$).

For the $w_0w_a$CDM$+A_L$ parameterization, from Table \ref{tab:results_flat_w0waCDM_Alens} here and table 2 of \cite{Park:2024pew}, $A_L$ differs from the desired value of $1$, for PR4 data by $-0.075\sigma$ (for PR3 data by $1.9\sigma$), for PR4+lensing data by $-0.0053\sigma$ (for PR3+lensing data by $0.81\sigma$), for PR4+non-CMB data by $0.74\sigma$ (for PR3+non-CMB data by $3.0\sigma$), and for PR4+lensing+non-CMB data by $1.1\sigma$ (for PR3+lensing+non-CMB data by $2.0\sigma$).

For the combined PR4/PR3+lensing+non-CMB datasets there are not large changes in the deviations from the $\Lambda$CDM model value of $w_0 = -1$, in the $w_0$CDM parameterization, in both the $A_L = 1$ and $A_L$-varying cases. The same is true for the deviations from the $\Lambda$CDM model values of $w_0 = -1$ and $w_0+w_a = -1$, in the $w_0w_a$CDM parameterization, in both the $A_L = 1$ and $A_L$-varying cases.

For the combined PR4/PR3+lensing+non-CMB datasets, the median deviation\footnote{Which corresponds to the $\Lambda$CDM model value among the three models considered.} of $A_L$ from unity has dropped from a significance of $2.5\sigma$ for the PR3+lensing+non-CDM data case to $1.6\sigma$ for the PR4+lensing+non-CMB data case, which is in the expected direction. However, this is much less of a decrease than in the median deviation\footnote{Which corresponds to the $w_0$CDM parameterization value among the three models considered.} of $A_L$ from unity for PR3 data compared to PR4 data, with the median deviation decreasing from $1.8\sigma$ (PR3) to $0.034\sigma$ (PR4).

Comparing the significance of the deviations from $w_0 = -1$ for the $w_0$CDM parameterizations in the $A_L=1$ ($A_L$-varying) case, we find a slight increase (decrease) in significance when going from PR3+lensing+non-CMB data to PR4+lensing+non-CMB data, with the significance going from $0.43\sigma$ q for $A_L=1$ ($1.3\sigma$ q for $A_L$-varying) for PR3+lensing+non-CMB data to $0.63\sigma$ q for $A_L=1$ ($1.1\sigma$ q for $A_L$-varying) for PR4+lensing+non-CMB data.

The opposite happens in the $w_0w_a$CDM parameterizations when we compare the significance of the deviations from $w_0 = -1$ and $w_0+w_a = -1$ in the $A_L=1$ ($A_L$-varying) case, where we find a slight decrease (increase) in significance when going from PR3+lensing+non-CMB data to PR4+lensing+non-CMB data, with the $w_0$ deviation significance going from $2.5\sigma$ q for $A_L=1$ ($2.0\sigma$ q for $A_L$-varying) for PR3+lensing+non-CMB data to $2.3\sigma$ q for $A_L=1$ ($2.1\sigma$ q for $A_L$-varying) for PR4+lensing+non-CMB data, and with the $w_0+w_a$ deviation significance going from $2.2\sigma$ p for $A_L=1$ ($1.4\sigma$ p for $A_L$-varying) for PR3+lensing+non-CMB data to $1.9\sigma$ p for $A_L=1$ ($1.5\sigma$ p for $A_L$-varying) for PR4+lensing+non-CMB data.

\subsection{Comparing parameter values for $A_L=1$ models, for non-CMB, PR4, PR4+lensing, PR4+non-CMB, and PR4+lensing+non-CMB datasets}

In this subsection we compare cosmological parameter values obtained for the $\Lambda$CDM model and the $w_0$CDM and $w_0w_a$CDM parameterizations with $A_L=1$ when non-CMB, PR4, PR4+lensing, PR4+non-CMB, and PR4+lensing+non-CMB data are considered. We take as reference values the ones obtained with PR4 data, therefore the cosmological parameter comparison are done with respect to those values. 

For the $\Lambda$CDM model with $A_L=1$, the results can be found in Table \ref{tab:results_flat_LCDM}. When we compare PR4 and non-CMB cosmological parameter values, we observe the following differences $H_0$ ($1.3\sigma$), $\Omega_b h^2$ ($1.1\sigma$), $\Omega_c h^2$ ($0.28\sigma$), and $\ln(10^{10}A_s)$ ($-0.24\sigma$), and for derived parameters we see $\Omega_m$ ($-1.0\sigma$), $\sigma_8$ ($-0.70\sigma$), and $100\theta_{\rm rec}$ ($0.12\sigma$). These results show some differences between cosmological parameter values determined using PR4 data and non-CMB data, as pointed out in Sec.\ \ref{subsec:Consistency}. 

When comparing PR4 and PR4+lensing cosmological parameter values, we observe the following differences $\ln(10^{10}A_s)$ ($0.33\sigma$), $\Omega_c h^2$ ($0.18\sigma$), $H_0$ ($-0.14\sigma$), $\tau$ ($0.12\sigma$), $\Omega_b h^2$ ($-0.10\sigma$), and $n_s$ ($-0.086\sigma$), and for derived parameters we see $\sigma_8$ ($0.37\sigma$), $\Omega_m$ {($0.14\sigma$), and $100\theta_{\rm rec}$ ($-0.085\sigma$). We conclude that when comparing cosmological parameter values for the PR4 and PR4+lensing datasets within the context of the flat $\Lambda$CDM model there are no significant differences. 

If we look at the cosmological parameter differences between PR4 and PR4+non-CMB data results we observe $\Omega_c h^2$ ($-0.69\sigma$), $H_0$ ($0.68\sigma$), $\tau$ ($0.067\sigma$), $n_s$ ($0.42\sigma$), $\Omega_b h^2$ ($0.34\sigma$), and $\ln(10^{10}A_s)$ ($-0.10\sigma$), and for derived parameters we see $\Omega_m$ ($-0.68\sigma$), $\sigma_8$ ($-0.39\sigma$), and $100\theta_{\rm rec}$ ($0.20\sigma$). While the differences are greater than in the previous PR4 and PR4+lensing case they all remain below the threshold of $1\sigma$, consequently we may claim that there are no appreciable differences when comparing PR4 and PR4+non-CMB data cosmological parameter constraints.

For the last $\Lambda$CDM $A_L=1$ model case, we compare PR4 and PR4+lensing+non-CMB cosmological parameter constraints, getting $H_0$ ($0.54\sigma$), $\Omega_c h^2$ ($-0.52\sigma$), $\ln(10^{10}A_s)$ ($0.38\sigma$), $\Omega_b h^2$ ($0.34\sigma$), $n_s$ ($0.33\sigma$), and $\tau$ ($0.24\sigma$), and finding for derived parameters $100\theta_{\rm rec}$ ($0.14\sigma$), $\Omega_m$ ($-0.54\sigma$), and $\sigma_8$ ($0.072\sigma$). Again no appreciable differences are observed. 

Regarding the $w_0$CDM cosmological parameterization with $A_L=1$, we can find the results in Table \ref{tab:results_flat_XCDM}. When we compare PR4 and non-CMB cosmological parameter constraints we obtain $\Omega_b h^2$ ($4.3\sigma$),  $H_0$ ($-2.4\sigma$), $\ln(10^{10}A_s)$ ($2.4\sigma$), and $\Omega_c h^2$ ($-2.3\sigma$), and for derived parameters we get $\Omega_m$ ($3.1\sigma$), $\sigma_8$ ($-2.0\sigma$), and $100\theta_{\rm rec}$ ($-1.9\sigma$). Some of the differences are very significant and they reflect the already discussed differences (see Sec.\ \ref{subsec:Consistency}) between high-redshift PR4 data and low-redshift non-CMB data. For PR4 data we obtain $w_0 =-1.49^{+0.18}_{-0.36}$, indicating a $2.7\sigma$ preference for phantom dynamical dark energy, whereas for the non-CMB data result we get $w_0=-0.868^{+0.044}_{-0.038}$ indicating a $3.5\sigma$ preference for quintessence behavior. The difference between these values is at the $3.4\sigma$ level of significance. While there are some significant differences between individual PR4 data and non-CMB data results for the $w_0$CDM parameterization with $A_L = 1$, we emphasize that according to our chosen threshold for the consistency test of Sec.\ \ref{subsec:Consistency} these data are mutually consistent for the $w_0$CDM parameterization.

When looking at PR4 and PR4+lensing data results we observe the following differences $H_0$ ($0.082\sigma$), $\Omega_c h^2$ ($0.061\sigma$), $n_s$ ($-0.018$), $\tau$ ($0.011\sigma$), $\Omega_b h^2$ ($0\sigma$), and $\ln(10^{10}A_s)$ ($0\sigma$), and for derived parameters we see $\Omega_m$ ($-0.075\sigma$), $\sigma_8$ ($0.064\sigma$), and $100\theta_{\rm rec}$ ($-0.027\sigma$). For PR4+lensing data we obtain $w_0 =-1.51^{+0.18}_{-0.32}$, which deviates from $w_0=-1$ by $2.8\sigma$ indicating a preference for phantom dark energy dynamics. The difference with respect to the PR4 value is $-0.050\sigma$. In light of these results there is good compatibility between the two sets of cosmological parameter constraints. 

The PR4 and PR4+non-CMB data comparison gives $H_0$ ($-2.7\sigma$)}, $\Omega_c h^2$ ($-0.71\sigma$), $n_s$ ($0.48\sigma$), $\Omega_b h^2$ ($0.38\sigma$), $\tau$ ($0.18\sigma$), and $\ln(10^{10}A_s)$ ($0.051\sigma$), and for derived parameters $\Omega_m$ ($5.2\sigma$), $\sigma_8$ ($-2.7\sigma$), and $100\theta_{\rm rec}$ ($0.22\sigma$). The significant differences that we observe for $H_0$, $\Omega_m$ and $\sigma_8$ is  reminiscent of the PR4 vs.\ non-CMB differences previously discussed. For PR4+non-CMB data we obtain $w_0=-0.980\pm 0.023$, which shows a preference of $0.87\sigma$ for quintessence behavior and is in $2.8\sigma$ tension with the phantom-like PR4 result. 

Comparing the PR4 and PR4+lensing+non-CMB cosmological parameter constraints we get $H_0$ ($-2.7\sigma$), $\ln(10^{10}A_s)$ ($0.54\sigma$), $\Omega_c h^2$ ($-0.50\sigma$), $n_s$ ($0.35\sigma$), $\tau$ ($0.34\sigma$), and $\Omega_b h^2$ ($0.33\sigma$), and for derived parameters we obtain $\Omega_m$ ($5.2\sigma$), $\sigma_8$ ($-2.6\sigma$), and $100\theta_{\rm rec}$ ($0.17\sigma$). We note that the differences are very similar to the previous case. For PR4+lensing+non-CMB data we obtain quintessence-like $w_0=-0.985\pm 0.024$, deviating from $w_0=-1$ by $0.63\sigma$ and showing a difference of $2.8\sigma$ with the phantom-like PR4 result. 

Finally, for the flat $w_0w_a$CDM cosmological parameterization with $A_L=1$, the results are shown in Table \ref{tab:results_flat_w0waCDM}. When we compare PR4 and non-CMB cosmological parameter constraints we observe the following differences $\Omega_b h^2$ ($4.6\sigma$), $\ln(10^{10}A_s)$ ($2.5\sigma$), $H_0$ ($-2.4\sigma$), and $\Omega_c h^2$ ($-2.1\sigma$), and for derived parameters $\Omega_m$ ($2.7\sigma$), $\sigma_8$ ($-2.0\sigma$), and $100\theta_{\rm rec}$ ($-1.8\sigma$). Once again, we observe some significant differences between the two sets of cosmological parameters which we previously discussed in Sec.\ \ref{subsec:Consistency}. When PR4 data are considered, we obtain phantom-like $w_0=-1.25^{+0.41}_{-0.48}$ ($0.61\sigma$ away from $w_0=-1$), $w_a =-1.04^{+0.76}_{-1.70}$ ($1.4\sigma$ away from $w_a =0$), and phantom-like $w_0+w_a=-2.29^{+0.89}_{-1.20}$ ($1.4\sigma$ away from $w_0+w_a=-1$ ). On the other hand, for the non-CMB data case, we get quintessence-like $w_0=-0.872\pm 0.059$ ($2.2\sigma$), $w_a =-0.01^{+0.39}_{-0.24}$ ($-0.026\sigma$), and quintessence-like $w_0+w_a=-0.89^{+0.35}_{-0.19}$ ($-0.58\sigma$). The three pairs of values differ at $0.91\sigma$, $1.3\sigma$, and $1.5\sigma$, respectively. While there are some significant differences between individual PR4 data and non-CMB data results for the $w_0w_a$CDM parameterization with $A_L = 1$, we emphasize that according to our chosen threshold for the consistency test of Sec.\ \ref{subsec:Consistency} these data are mutually consistent for the $w_0w_a$CDM parameterization.

Looking at PR4 and PR4+lensing cosmological parameter constraints we observe no significant differences being their values, with differences being $\Omega_c h^2$ ($0.059\sigma$), $\tau$ ($0.011\sigma$), $n_s$ ($-0.035\sigma$), $\Omega_b h^2$ ($0\sigma$), $H_0$ ($0\sigma$), and $\ln(10^{10}A_s)$ ($0\sigma$), and for derived parameters we see $\Omega_m$ ($-0.028\sigma$), $\sigma_8$ ($0.025\sigma$), and $100\theta_{\rm rec}$ ($0\sigma$). For the PR4+lensing dataset we obtain phantom-like $w_0=-1.27\pm 0.44$ ($0.61\sigma$ away from $w_0=-1$), $w_a =-1.01^{+0.76}_{-1.80}$ ($1.3\sigma$ away from $w_a =0$), and phantom-like $w_0+w_a=-2.28^{+0.90}_{-1.20}$ ($1.4\sigma$ away from $w_0+w_a=-1$), with the differences with respect to the PR4 results being $-0.031\sigma$, $0.015\sigma$, and $0.0067\sigma$, respectively. With these results we confirm that PR4 and PR4+lensing data cosmological parameter constraints are very compatible.  

The PR4 and PR4+non-CMB results comparison yields the following differences $H_0$ ($-2.7\sigma$), $\Omega_c h^2$ ($-0.19\sigma$), $n_s$ ($0.089\sigma$), $\Omega_b h^2$ ($0.052\sigma$), $\ln(10^{10}A_s)$ ($-0.051\sigma$), and $\tau$ ($0\sigma$), and for derived parameters $\Omega_m$ ($5.0\sigma$), $\sigma_8$ ($-2.6\sigma$), and $100\theta_{\rm rec}$ ($0.027\sigma$). Again we find that significant differences affect the $H_0$, $\Omega_m$, and $\sigma_8$ parameters. In the PR4+non-CMB case we obtain quintessence-like $w_0=-0.869\pm 0.060$ ($2.2\sigma$ away from $w_0=-1$), $w_a =-0.46^{+0.25}_{-0.22}$ ($1.8\sigma$ away from $w_a =0$), and phantom-like $w_0+w_a=-1.33^{+0.20}_{-0.17}$ ($1.7\sigma$ away from $w_0+w_a=-1$), still indicating a preference for a dynamical dark energy component. The differences with respect to the PR4 results are $0.92\sigma$, $0.73\sigma$, and $1.1\sigma$, respectively. With these results, we find a certain level of difference between PR4 and PR4+non-CMB data cosmological parameter constraints. 

As for the PR4 and PR4+lensing+non-CMB comparison we obtain $H_0$ ($-2.7\sigma$), $\ln(10^{10}A_s)$ ($0.27\sigma$), $\tau$ ($0.11\sigma$), $n_s$ ($-0.036\sigma$), $\Omega_c h^2$ ($0.0065\sigma$), and $\Omega_b h^2$ ($0\sigma$), and for derived parameters we get $\Omega_m$ ($5.1\sigma$), $\sigma_8$ ($-2.5\sigma$), and $100\theta_{\rm rec}$ ($-0.027\sigma$). Therefore we report results very similar to those in the previous case. When PR4+lensing+non-CMB data are used we obtain quintessence-like $w_0=-0.863\pm 0.060$ ($2.3\sigma$ away from $w_0=-1$), $w_a =-0.50^{+0.25}_{-0.22}$ ($2.0\sigma$ away from $w_a =0$), and phantom-like $w_0+w_a=-1.37^{+0.19}_{-0.17}$ ($1.9\sigma$ away from $w_0+w_a=-1$), still indicating a preference for a dynamical dark energy component. The differences with respect to the PR4 results are $0.93\sigma$, $0.68\sigma$, and $1.0\sigma$, respectively.

\subsection{Comparing parameter values for $A_L$-varying models, for non-CMB, PR4, PR4+lensing, PR4+non-CMB, and PR4+lensing+non-CMB datasets} 

Similarly to the previous subsection, in this one we compare cosmological palrameter values obtained when non-CMB, PR4, PR4+lensing, PR4+non-CMB and PR4+lensing+non-CMB data are considered, but now for the $A_L$-varying cases, $\Lambda$CDM$+A_L$, $w_0$CDM$+A_L$, and $w_0w_a$CDM$+A_L$. Again we take as reference values those obtained with PR4 data, therefore the cosmological parameter comparison are done with respect to those values. 

The results for the $\Lambda$CDM+$A_L$ cosmological model can be found in Table \ref{tab:results_flat_LCDM_Alens}. When we compare the PR4 and the non-CMB cosmological parameter constraints within the context of this model we obtain the following differences $H_0$ ($1.2\sigma$), $\Omega_b h^2$ ($1.1\sigma$), $\Omega_c h^2$ ($0.33\sigma$), and $\ln(10^{10}A_s)$ ($-0.21\sigma$), and for derived parameters we get $\Omega_m$ ($-0.78\sigma$), $\sigma_8$ ($-0.61\sigma$), and $100\theta_{\rm rec}$ ($0.12\sigma$). In the analogous comparison but for the case with $A_L=1$, we observed mild differences, and the same is found now. This means that in this case the inclusion of the $A_L$ parameter in the analysis does not have a big impact. When PR4 data are used, we get $A_L=1.030\pm 0.054$, which deviates from $A_L=1$ by $0.56\sigma$. 

The PR4 and PR4+lensing cosmological parameter comparison yields the following differences $H_0$ ($0.080\sigma$), $n_s$ ($0.064\sigma$), $\Omega_c h^2$ ($-0.052\sigma$), $\ln(10^{10}A_s)$ ($0.047\sigma$), $\Omega_b h^2$ ($0.045\sigma$), and $\tau$ ($0.011\sigma$), and for derived parameters $\Omega_m$ ($-0.084\sigma$), $\sigma_8$ ($-0.028\sigma$), and $100\theta_{\rm rec}$ ($0.027\sigma$), showing good agreement. In the PR4 case we obtain $A_L=1.030\pm 0.054$ ($0.56\sigma$ in favor of $A_L>1$) and when PR4+lensing data are analyzed we get $A_L=1.037\pm 0.038$ ($0.97\sigma$ in favor of $A_L>1$). Therefore the inclusion of the CMB lensing data enhances the preference for $A_L>1$. 

When comparing constraints obtained with PR4 and PR4+non-CMB data, we do not observe significant differences, with their values being $\Omega_c h^2$ ($-0.53\sigma$), $H_0$ ($0.52\sigma$), $n_s$ ($0.41\sigma$), $\Omega_b h^2$ ($0.38\sigma$), $\ln(10^{10}A_s)$ ($-0.15\sigma$), and $\tau$ ($-0.045\sigma$), and for derived parameters we find $\Omega_m$ ($-0.52\sigma$), $\sigma_8$ ($-0.38\sigma$), and $100\theta_{\rm rec}$ {($0.20\sigma$). When PR4+non-CMB data are considered we obtain $A_L=1.051\pm 0.049$ ($1.0\sigma$ in favor of $A_L>1$), and differing by $0.29\sigma$ with the one obtained from PR4 data. Once again the differences in the cosmological parameter constraints do not differ significantly from the ones obtained in the corresponding case with $A_L=1$.  

Finally for the $\Lambda$CDM+$A_L$ models, we compare the constraints obtained with PR4 and PR4+lensing+non-CMB data, obtaining very similar results to the previous comparison $\Omega_c h^2$ ($-0.55\sigma$), $H_0$ ($0.54\sigma$), $n_s$ ($0.40\sigma$), $\Omega_b h^2$ ($0.38\sigma$), $\ln(10^{10}A_s)$ ($-0.15\sigma$), and $\tau$ ($-0.045\sigma$), and for derived parameters $\Omega_m$ ($-0.54\sigma$), $\sigma_8$ ($-0.38\sigma$), and $100\theta_{\rm rec}$ ($0.20\sigma$). Using PR4+lensing+non-CMB data gives the value $A_L=1.053\pm 0.034$ ($1.6\sigma$ in favor of $A_L>1$), differing by $0.36\sigma$ from the one obtained with PR4 data. Therefore the simultaneous consideration of PR4, lensing, and non-CMB data strengthens the signal in favor of $A_L>1$.  

The results for the $w_0$CDM+$A_L$ parameterization are shown in Table \ref{tab:results_flat_XCDM_Alens}. Using PR4 data we obtain $w_0=-1.45^{+0.21}_{-0.41}$, indicating a preference for phantom behavior at $2.1\sigma$, and $A_L=1.002^{+0.052}_{-0.059}$, which differs from $A_L=1$ by only $0.03\sigma$.

Comparing PR4 and non-CMB cosmological parameter constraints in the $w_0$CDM$+A_L$ parameterization we obtain $\Omega_b h^2$ ($4.3\sigma$), $\ln(10^{10}A_s)$ ($2.4\sigma$), $\Omega_c h^2$ ($-2.3\sigma$), $H_0$ ($-2.0\sigma$), and for derived parameters we get $\Omega_m$ ($2.4\sigma$), $100\theta_{\rm rec}$ ($-1.9\sigma$), and $\sigma_8$ ($-1.7\sigma$). These differences are approximately as large as the ones obtained in the earlier corresponding comparison with $A_L=1$, consequently we may claim that the inclusion of the $A_L$ parameter in the analysis does not significantly reduce the differences between PR4 and non-CMB results. When non-CMB data are considered we obtain $w_0=-0.868^{+0.044}_{-0.038}$, which indicates a preference for a quintessence behavior at $3.5\sigma$ significance, differing from the PR4 value at $2.7\sigma$. While there are some significant differences between individual PR4 data and non-CMB data results for the $w_0$CDM$+A_L$ parameterization, we emphasize that according to our chosen threshold for the consistency test of Sec.\ \ref{subsec:Consistency} these data are mutually consistent for the $w_0$CDM$+A_L$ parameterization.

As for the PR4 and PR4+lensing comparison, we report the following differences $\Omega_c h^2$ ($-0.052\sigma$), $\Omega_b h^2$ ($0.047\sigma$), $n_s$ ($0.047\sigma$), $\ln(10^{10}A_s)$ ($0.047\sigma$), $\tau$ ($0.012\sigma$), and $H_0$ ($0\sigma$), and for derived parameters we get $\Omega_m$ ($-0.052\sigma$), $100\theta_{\rm rec}$ ($0.028\sigma$), and $\sigma_8$ ($-0.024\sigma$). Following the same trend observed in the earlier corresponding comparison with $A_L=1$, the differences are small. For the PR4+lensing data we get $w_0=-1.46^{+0.20}_{-0.39}$ ($2.3\sigma$ in favor of phantom behavior) and $A_L=1.006^{+0.037}_{-0.045}$ ($0.13\sigma$ in favor of $A_L>1$). The differences from the PR4 results are $-0.023\sigma$ and $0.058\sigma$, respectively. 

The comparison of PR4 and PR4+non-CMB cosmological parameter constraints in the $w_0$CDM$+A_L$ parameterization yields $H_0$ ($-2.2\sigma$), $\Omega_c h^2$ ($-0.97\sigma$), $n_s$ ($0.76\sigma$), $\Omega_b h^2$ ($0.73\sigma$), $\ln(10^{10}A_s)$ ($-0.14\sigma$), and $\tau$ ($0.035\sigma$), and for derived parameters $\Omega_m$ ($4.0\sigma$), $\sigma_8$ ($-2.3\sigma$), and $100\theta_{\rm rec}$ ($0.41\sigma$). We observe some reductions in the differences with respect to the earlier case with $A_L=1$, specially for the $\Omega_m$ and $\sigma_8$ parameters, but the differences still remain significant. When PR4+non-CMB data are used we get $w_0=-0.973\pm 0.024$ ($1.1\sigma$ in favor of quintessence behavior) and $A_L=1.064\pm 0.051$ ($1.3\sigma$ in favor of $A_L>1$); the differences with the PR4 data results are $2.3\sigma$ and $0.85\sigma$, respectively. 

In the final case for the $w_0$CDM+$A_L$ parameterization, we compare PR4 and PR4+lensing+non-CMB constraints, obtaining very similar results to the previous case $H_0$ ($-2.2\sigma$), $\Omega_c h^2$ ($-0.99\sigma$), $n_s$ ($0.77\sigma$), $\Omega_b h^2$ ($0.73\sigma$), $\ln(10^{10}A_s)$ ($-0.15\sigma$), and $\tau$ ($0.036\sigma$), and for derived parameters we get $\Omega_m$ ($4.0\sigma$), $\sigma_8$ ($-2.3\sigma$), and $100\theta_{\rm rec}$ ($0.38\sigma$). When PR4+non-CMB data are analyzed we get $w_0=-0.973\pm 0.024$ ($1.1\sigma$ in favor of quintessence behavior) and $A_L=1.064\pm 0.035$ ($1.8\sigma$ in favor of $A_L>1$), which represents an enhancement of the preference for $A_L>1$ with respect to the PR4+non-CMB case. These differ from the PR4 results by $2.3\sigma$ and $0.99\sigma$, respectively. 

Cosmological parameter constraints for the $w_0w_a$CDM+$A_L$ cosmological parameterization can be found in Table \ref{tab:results_flat_w0waCDM_Alens}. For the PR4 case we obtain for the equation of state parameters, $w_0=-1.25^{+0.42}_{-0.50}$, phantom-like and deviating by $0.6\sigma$ from $w_0=-1$ and $w_a=-0.97^{+0.78}_{-1.80}$. The combination $w_0+w_a=-2.23^{+0.97}_{-1.20}$ shows a preference for phantom-like dynamical dark energy at $1.27\sigma$ significance level and the lensing parameter $A_L=0.996\pm 0.053$ ($0.075\sigma$) is in good agreement with the expected value $A_L=1$. 

Comparing PR4 and non-CMB cosmological parameter constraints for the $w_0w_a$CDM$+A_L$ parameterization we obtain $\Omega_b h^2$ ($4.6\sigma$), $\ln(10^{10}A_s)$ ($2.4\sigma$), $H_0$ ($-2.2\sigma$), and $\Omega_c h^2$ ($-2.1\sigma$), and for derived parameters we get $\Omega_m$ ($2.6\sigma$),  $100\theta_{\rm rec}$ {($-1.8\sigma$), and $\sigma_8$ ($-1.8\sigma$). Once again, we observe significant differences, showing that even when the $A_L$ parameter is allowed to vary differences between the PR4 and non-CMB results remain. For non-CMB data we obtain $w_0 = -0.872 \pm 0.059$ (quintessence-like and $2.2\sigma$ away from $w_0 = -1$), $w_a = -0.01^{+0.39}_{-0.24}$ ($0.026\sigma$ away from $w_a = 0$), and $w_0 + w_a = -0.89^{+0.35}_{-0.19}$ (quintessence-like and $-0.58\sigma$ away from $w_0 + w_a = -1$), still indicating a preference for a dynamical dark energy component. The differences with respect to the PR4 results are $0.89\sigma$, $1.2\sigma$, and $1.4\sigma$, respectively. While there are some significant differences between individual PR4 data and non-CMB data results for the $w_0w_a$CDM$+A_L$ parameterization, we emphasize that according to our chosen threshold for the consistency test of Sec.\ \ref{subsec:Consistency} these data are mutually consistent for the $w_0w_a$CDM$+A_L$ parameterization.

In regard to the PR4 and PR4+lensing comparison, the differences are $H_0$ ($0.082\sigma$), $n_s$ ($0.032\sigma$), $\Omega_b h^2$ ($0\sigma$), $\Omega_c h^2$ ($0\sigma$), $\tau$ ($0\sigma$), and $\ln(10^{10}A_s)$ ($0\sigma$), and for derived parameters $\Omega_m$ ($-0.039\sigma$), $\sigma_8$ ($0.032\sigma$), and $100\theta_{\rm rec}$ ($-0.026\sigma$). These differences are small, showing once more the good agreement between PR4 and lensing data. For PR4+lensing data we observe $w_0=-1.26^{+0.42}_{-0.49}$ (phantom-like and $0.62\sigma$ away from $w_0=-1$), $w_a =-0.99^{+0.80}_{-1.80}$ ($1.2\sigma$ away from $w_a =0$), and $w_0+w_a=-2.25^{+0.93}_{-1.20}$ (phantom-like and $1.3\sigma$ away from $w_0+w_a=-1$), showing preference for a dynamical dark energy component. The differences with respect to the PR4 results are $-0.015\sigma$, $-0.010\sigma$, and $-0.013\sigma$, respectively. As for the lensing parameter the value $A_L=0.9998^{+0.038}_{-0.043}$ ($0.0053\sigma$) is in complete agreement with $A_L=1$ and it shows just a $0.056\sigma$ difference from the PR4 value.    

When we compare PR4 and PR4+non-CMB cosmological parameter constraints we obtain $H_0$ ($-2.5\sigma$), $\Omega_c h^2$ ($-0.43\sigma$), $n_s$ ($0.36\sigma$), $\Omega_b h^2$ ($0.35\sigma$),  $\ln(10^{10}A_s)$ {($-0.24\sigma$)}, and $\tau$ ($-0.090\sigma$), and for derived parameters we get $\Omega_m$ ($5.0\sigma$), $\sigma_8$ ($-2.4\sigma$), and $100\theta_{\rm rec}$ ($0.11\sigma$). So, the large differences affecting the $H_0$, $\Omega_m$, and $\sigma_8$ parameters remain. When PR4+non-CMB data are analyzed, we obtain for the non-standard parameters $w_0=-0.877\pm 0.059$ (quintessence-like and $2.1\sigma$ away from $w_0=-1$), $w_a =-0.41^{+0.25}_{-0.22}$ ($1.6\sigma$ away from $w_a =0$), and $w_0+w_a=-1.29^{+0.20}_{-0.17}$ (phantom-like and $1.5\sigma$ away from $w_0+w_a=-1$). The differences with respect to the PR4 results are $0.88\sigma$, $0.69\sigma$, and $0.95\sigma$, respectively. As for the lensing parameter the value $A_L=1.039\pm 0.053$ deviates from $A_L=1$ by $0.74\sigma$ and it shows a $0.57\sigma$ difference with the PR4 value. 

In the last case studied in this subsection we examine the results obtained from comparing PR4 and PR4+lensing+non-CMB cosmological parameter constraints, obtaining very similar results to the previous case, $H_0$ ($-2.5\sigma$), $\Omega_c h^2$ ($-0.43\sigma$), $n_s$ ($0.38\sigma$), $\Omega_b h^2$ ($0.36\sigma$), $\ln(10^{10}A_s)$ ($-0.24\sigma$), and $\tau$ ($-0.091\sigma$), and for derived parameters $\Omega_m$ ($5.0\sigma$), $\sigma_8$ ($-2.4\sigma$), and $100\theta_{\rm rec}$ ($0.11\sigma$). The values obtained for the equation of state parameters are $w_0=-0.877\pm 0.060$ (quintessence-like and $2.1\sigma$ away from $w_0=-1$), $w_a =-0.41^{+0.25}_{-0.22}$ ($1.6\sigma$ away from $w_a =0$), and $w_0+w_a=-1.29^{+0.20}_{-0.17}$ (phantom-like and $1.5\sigma$ away from $w_0+w_a=-1$), with the differences with respect to the PR4 results being $0.88\sigma$, $0.69\sigma$, and $0.95\sigma$, respectively. The value of the lensing parameter is $A_L=1.042\pm 0.037$ indicating a preference for $A_L>1$ at $1.1\sigma$ and differing from the PR4 result at $0.71\sigma$ significance level.

\subsection{Comparing parameter values for $A_L=1$ and $A_L$-varying models/parameterizations for PR4 dataset combinations}
\label{subsec:AL-parameter-variations}

As there is less evidence for $A_L$ deviating from unity for PR4 data combinations (than there is for PR3 data combinations) there are no parameters with $A_L=1$ and $A_L$-varying values that differ by even $1\sigma$.

Except for the $\Lambda$CDM($+A_L$) case, where the $\Lambda$CDM model and the $\Lambda$CDM$+A_L$ model $\sigma_8$ values differ by $0.63\sigma$ for PR4+lensing data, all other differences greater than $0.5\sigma$ occur for the PR4+lensing+non-CMB dataset and are: $0.71\sigma$ for ln$(10^{10}A_s)$ and $0.83\sigma$ for $\sigma_8$ for the $\Lambda$CDM($+A_L$) case; $0.62\sigma$ for $\Omega_c h^2$, $0.76\sigma$ for ln$(10^{10}A_s)$, and $0.85\sigma$ for $\sigma_8$ for the $w_0$CDM($+A_L$) case; and $0.61\sigma$ for $\sigma_8$ for the $w_0w_a$CDM($+A_L$) case.

Comparing $\Delta$DIC values in Table \ref{tab:results_flat_LCDM_Alens} we see in the $\Lambda$CDM$(+A_L)$ case that among the four datasets that include PR4 data, going from the $\Lambda$CDM model to the $\Lambda$CDM$+A_L$ model, i.e., allowing $A_L$ to vary, results in more positive $\Delta$DIC values for datasets that include CMB lensing data. More specifically, $\Delta$DIC increases from $+0.90$ for PR4 data to $+1.05$ for PR4+lensing data and from $-2.59$ for PR4+non-CMB data to $-0.87$ for PR4+lensing+non-CMB data. This is also true for the $w_0$CDM and $w_0w_a$CDM parameterizations, where now we use the $\Delta$DIC values computed between the $A_L$-varying parameterization and the corresponding $A_L=1$ parameterization for the same data set (and not the $\Delta$DIC values listed in the last lines of Tables \ref{tab:results_flat_XCDM} -- \ref{tab:results_flat_w0waCDM_Alens}, which are relative to the corresponding $\Lambda$CDM model, but rather the differences between the $\Delta$DIC values listed in Tables \ref{tab:results_flat_XCDM} and \ref{tab:results_flat_XCDM_Alens} and in Tables \ref{tab:results_flat_w0waCDM} and \ref{tab:results_flat_w0waCDM_Alens}). The same is true for these models/parameterizations when PR3 data are used instead of PR4 data \cite{deCruzPerez:2024abc, Park:2024pew}. This behavior suggests that these lensing data are somewhat in tension with the additional freedom introduced by a varying $A_L$ parameter.

\subsection{Discussion of the PR4+lensing+non-CMB data results}

In this subsection we discuss the results obtained when the largest dataset considered in this work, PR4+lensing+non-CMB, is employed to set constraints on the parameters of the different models/parameterizations under study. Among these, the simplest, and observationally consistent, one is the spatially-flat $\Lambda$CDM model with $A_L = 1$. Within the context of this model and PR4+lensing+non-CMB data, the six primary cosmological parameters take the following values $\Omega_b h^2 = 0.02231\pm 0.00012$, $\Omega_c h^2 = 0.11804\pm 0.00082$, $H_0 = 68.01\pm 0.37$ km s$^{-1}$ Mpc$^{-1}$, $\tau=0.0598\pm 0.0060$, $n_s=0.9693\pm 0.0035$, and $\ln(10^{10}A_s)=3.046\pm 0.012$. As for the derived parameters considered, we have $100\theta_{\text{rec}}=1.04187\pm 0.00024$, $\Omega_m = 0.3048\pm 0.0049$, and $0.8065\pm 0.0049$. 

One of the goals of this subsection is to determine whether the cosmological parameter constraints for the six primary (and three derived) parameters, common to all the models, are model independent. To assess that, we compute the shifts in the cosmological parameter values relative to the $\Lambda$CDM model values, and we check whether they remain below our chosen $1\sigma$ threshold. We also comment on the values obtained for the non-standard primary parameters characterizing the other models/parameterizations considered in this work, namely the lensing consistency parameter $A_L$ and the equation of state parameters $w_0$, $w_a$, and the combination $w_0+w_a$ computed for the $w_0$CDM$(+A_L)$ and $w_0w_a$CDM($+A_L$) parameterizations. 

If we look at Tables \ref{tab:results_flat_LCDM} and \ref{tab:results_flat_LCDM_Alens} we can compare the cosmological parameter constraints obtained with PR4+lensing+non-CMB data for the $\Lambda$CDM and the $\Lambda$CDM+$A_L$ models. In regard to the shift in the values, we get $\Omega_b h^2$ \textcolor{black}{($+0.29\sigma$)}, $\Omega_c h^2$ ($-0.45\sigma$), $H_0$ ($0.47\sigma$), $\tau$ ($-0.35\sigma$), $n_s$ ($0.32\sigma$), and $\ln(10^{10} A_s)$ ($-0.71\sigma$), indicating that all differences relative to the $\Lambda$CDM model values remain below the $1\sigma$ threshold that we have chosen. As for the derived parameters $100\theta_{\rm rec}$, $\Omega_m$, and $\sigma_8$, the differences are $0.12\sigma$, $-0.47\sigma$, and $-0.83\sigma$, respectively. When the lensing consistency parameter $A_L$ is allowed to vary in the $\Lambda$CDM+$A_L$ model, we obtain $A_L=1.053\pm 0.034$, which deviates from the expected value $A_L=1$ by $1.6\sigma$. Regarding how well the two models fit these data, we find $\Delta\text{DIC}=-0.87$, indicating a {\it weak} preference for the $\Lambda$CDM+$A_L$ model over the $\Lambda$CDM model. 

The cosmological parameter constraints for the $w_0$CDM parameterization can be found in Table \ref{tab:results_flat_XCDM}. The shift in the values, when compared with the $\Lambda$CDM model values, are $\Omega_b h^2$ ($0.12\sigma$), $\Omega_c h^2$ ($-0.24\sigma$), $H_0$ ($-0.45\sigma$), $\tau$ ($0.092\sigma$), $n_s$ ($0.18\sigma$), and $\ln(10^{10} A_s)$ ($0.12\sigma$). Again we observe a good agreement for the six primary parameters, since all parameter differences are less than $1\sigma$. The same happens for the derived parameters $100\theta_{\rm rec}$, $\Omega_m$, and $\sigma_8$, with differences $0.09\sigma$, $0.31\sigma$, and $-0.45\sigma$, respectively. Within the context of the $w_0$CDM parameterization, the use of PR4+lensing+non-CMB data shows a slight preference for quintessence-like dynamical dark energy, with $w_0=-0.985\pm 0.024$ and a deviation of $0.63\sigma$ from $w_0=-1$. In this case the DIC criterion, $\Delta\text{DIC}=+1.20$, tells us that the $\Lambda$CDM model is {\it weakly} preferred over the $w_0$CDM parameterization.

For the $w_0$CDM+$A_L$ parameterization, the results for the PR4+lensing+non-CMB data can be seen in Table \ref{tab:results_flat_XCDM_Alens}. Once again, we find that for all the six primary parameters the shifts with respect to the $\Lambda$CDM model remain $<1\sigma$. In particular, the values are $\Omega_b h^2$ \textcolor{black}{($0.53\sigma$)}, $\Omega_c h^2$ ($-0.88\sigma$), $H_0$ ($-0.45\sigma$), $\tau$ ($-0.24\sigma$), $n_s$ ($0.63\sigma$), and $\ln(10^{10} A_s)$ ($-0.65\sigma$). For the derived parameters $100\theta_{\rm rec}$, $\Omega_m$, and $\sigma_8$, the differences are $0.24\sigma$, $0.10\sigma$, and $-1.4\sigma$. The difference in the value of $\sigma_8$ is the only one above $1\sigma$, however, we do not consider this tension significant enough. The equation of state parameter value $w_0=-0.973\pm 0.024$, shows a $1.1\sigma$ preference for quintessence-like dark energy dynamics over the rigid $w_0=-1$ case, and regarding the lensing consistency parameter, $A_L=1.064\pm 0.035$, there is a preference for $A_L>1$ at $1.8\sigma$. Comparing the values of $w_0$ for the $w_0$CDM and the $w_0$CDM+$A_L$ parameterizations we observe a small shift of $0.35\sigma$. As for the performance comparison, with respect to the $\Lambda$CDM model we get $\Delta\text{DIC}=-1.90$, indicting a {\it weak} preference for the dynamical dark energy $w_0$CDM$+A_L$ parameterization over the $\Lambda$CDM model and a {\it positive} preference of $\Delta$DIC = $-3.10$ relative to the $w_0$CDM $A_L=1$ case. Consequently, we conclude that allowing the $A_L$ parameter to vary helps to improve the performance of the $w_0$CDM $A_L=1$ parameterization. 

Looking at Table \ref{tab:results_flat_w0waCDM} we can compare the PR4+lensing+non-CMB data results for the $w_0w_a$CDM dark energy parameterization with the ones obtained for the $\Lambda$CDM model. In regard to the shifts of the six primary parameters we get $\Omega_b h^2$ ($-0.17\sigma$), $\Omega_c h^2$ ($0.29\sigma$), $H_0$ ($-0.42\sigma$), $\tau$ ($-0.17\sigma$), $n_s$ ($-0.18\sigma$), and $\ln(10^{10} A_s)$ ($-0.24\sigma$). The differences in the derived parameters $100\theta_{\rm rec}$, $\Omega_m$, and $\sigma_8$, are $-0.12\sigma$, $0.45\sigma$, and $0.11\sigma$, respectively. In light of these results, we may claim that for this cosmological parameterization the agreement with the $\Lambda$CDM results is good, and all differences stay below $1\sigma$. The dark energy equation of state parameters take values $w_0=-0.863\pm 0.060$ (quintessence-like and $2.3\sigma$ away from $w_0=-1$), $w_a=-0.50^{+0.25}_{-0.22}$ ($2\sigma$ away from $w_a=0$), and $w_0+w_a=-1.37^{+0.19}_{-0.17}$ (phantom-like and $1.9\sigma$ away from $w_0+w_a=-1$). When the DIC criterion is employed we get $\Delta\text{DIC}=-3.76$ indicating {\it positive} evidence in favor of this dynamical dark energy parameterization over the $\Lambda$CDM model. This turns out to be the highest performance encountered, when looking at PR4+lensing+non-CMB data results, among the different models/parameterizations under study. If we compare the values obtained within the $w_0$CDM and the $w_0w_a$CDM parameterization for the equation of state parameter $w_0$ we see that both of them are quintessence-like with $w_0>-1$, and with the difference between them being \textcolor{black}{$1.9\sigma$}. 

For the $w_0w_a$CDM+$A_L$ parameterization the PR4+lensing+non-CMB data results can be found in Table \ref{tab:results_flat_w0waCDM_Alens}. Once again we find good agreement with the $\Lambda$CDM model results, with the shift in the values of the cosmological parameters being $\Omega_b h^2$ ($0.18\sigma$), $\Omega_c h^2$ ($-0.25\sigma$), $H_0$ ($-0.40\sigma$), $\tau$ ($-0.36\sigma$), $n_s$ ($0.21\sigma$), and $\ln(10^{10} A_s)$ ($-0.71\sigma$). As for the derived parameters,  $100\theta_{\rm rec}$, $\Omega_m$, and $\sigma_8$, we do not observe significant differences, with them being $0.03\sigma$, $0.27\sigma$, and $-0.62\sigma$, respectively. In regard to the equation of state parameter values we obtain $w_0=-0.877\pm 0.060$ (quintessence-like and $2.1\sigma$ away from $w_0=-1$), $w_a=-0.41^{+0.25}_{-0.22}$ ($1.6\sigma$ away from $w_a=0$), and $w_0+w_a=-1.29^{+0.20}_{-0.17}$ (phantom-like and $1.5\sigma$ away from $w_0+w_a=-1$). These values do not differ much from the corresponding ones obtained for the $w_0w_a$CDM parameterization with $A_L=1$. In particular, the differences are $w_0$ ($-0.16\sigma$), $w_a$ ($0.27\sigma$), and $w_0+w_a$ ($0.31\sigma$); therefore the variation of the lensing consistency parameter $A_L$ does not significantly change the equation of state parameter values. For the lensing consistency parameter we find $1.042\pm 0.037$ deviating from $A_L=1$ by $1.1\sigma$ with a preference for $A_L>1$. As for the performance, compared to the $\Lambda$CDM model we get $\Delta\text{DIC}=-2.36$, indicating a {\it positive} preference over the $\Lambda$CDM model. However, the addition of the lensing parameter $A_L$ to the list of freely varying parameters does not help to improve the performance when compared to the $w_0w_a$CDM parameterization with $A_L=1$, with a relative $\Delta$DIC = $+1.40$. Comparing the values obtained for the dark energy equation of state parameter $w_0$, for the $w_0$CDM+$A_L$ and $w_0w_a$CDM+$A_L$ parameterizations, we observe that they differ by $1.5\sigma$, which represents a reduction of the differences with respect to the corresponding comparison with $A_L=1$.  

In summary, we find good agreement between the values of the six primary (and three derived, with the one exception of the $\sigma_8$ value for the $w_0$CDM+$A_L$ model) parameters for the six cosmological models under study. Additionally, we observe that in terms of the DIC estimator, only the $w_0w_a$CDM dynamical dark energy parameterization without and with a varying lensing consistency parameter $A_L$ surpasses the performance of the $\Lambda$CDM model. 

While there is good agreement between the six $H_0$ and $\Omega_m$ values for the $\Lambda$CDM($+A_L$) model and $w_0$CDM($+A_L$) and $w_0w_a$CDM($+A_L$) parameterizations, on average the $\Lambda$CDM($+A_L$) model $H_0$ ($\Omega_m$) values are a bit higher (lower) than the $w_0$CDM($+A_L$) and $w_0w_a$CDM($+A_L$) parameterizations ones. Averaging the two $\Lambda$CDM($+A_L$) model (four $w_0$CDM($+A_L$) and $w_0w_a$CDM($+A_L$) parameterizations) central values and using the largest individual error bars, we may summarize these results by $H_0 = 68.14 \pm 0.39$ km s$^{-1}$ Mpc$^{-1}$ and $\Omega_m = 0.3032 \pm 0.0051$ for $\Lambda$CDM($+A_L$) and $H_0 = 67.70 \pm 0.64$ km s$^{-1}$ Mpc$^{-1}$ and $\Omega_m = 0.3071 \pm 0.0063$ for $w_0$CDM($+A_L$) and $w_0w_a$CDM($+A_L$). These $H_0$ values agree with the median statistics result $H_0=68\pm 2.8$ km s$^{-1}$ Mpc$^{-1}$ \cite{Chen:2011ab,Gottetal2001,Calabreseetal2012}, as well as with some local measurements including the value of \cite{Cao:2023eja} $H_0=69.25\pm 2.4$ km s$^{-1}$ Mpc$^{-1}$ from a joint analysis of $H(z)$, BAO, Pantheon+ SNIa, quasar angular size, reverberation-measured \mii\ and \civ\ quasar, and 118 Amati correlation gamma-ray burst data, and the local $H_0=70.39\pm 1.94$ km s$^{-1}$ Mpc$^{-1}$ from JWST TRGB+JAGB and SNIa data \cite{Freedman:2024eph}, but are in mild tension with the local $H_0=73.04\pm 1.04$ km s$^{-1}$ Mpc$^{-1}$ measured using Cepheids and SNIa data \cite{Riess:2021jrx}, also see \cite{Chen:2024gnu}. And the above $\Omega_m$ values agree well with the flat $\Lambda$CDM model value of $\Omega_m = 0.313\pm 0.012$ of \cite{Cao:2023eja} (from the data set listed above that was used to determine $H_0$).


\begin{figure*}[htbp]
\centering
\mbox{\includegraphics[width=85mm]{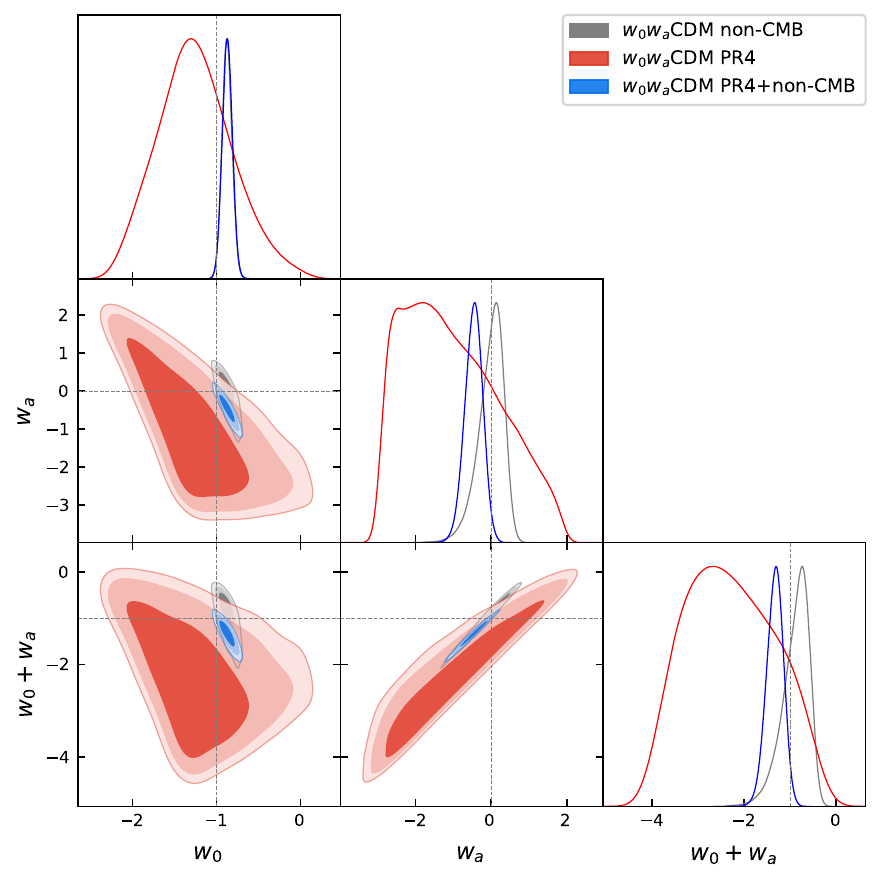}}
\mbox{\includegraphics[width=85mm]{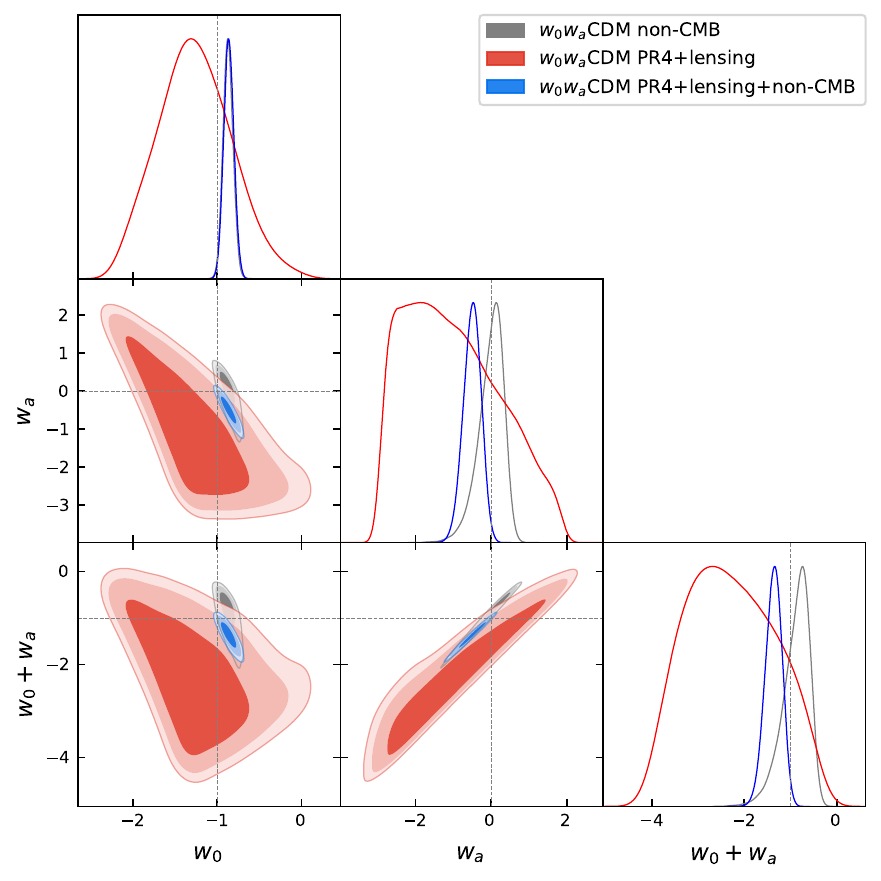}}
        \caption{One-dimensional likelihoods and 1$\sigma$, 2$\sigma$, and $3\sigma$ likelihood confidence contours of $w_0$, $w_a$, and $w_0+w_a$ parameters in the $w_0 w_a$CDM parameterization favored by (left) non-CMB, PR4, and PR4+non-CMB data sets, and (right) non-CMB, PR4+lensing, and PR4+lensing+non-CMB data sets. The horizontal or vertical dotted lines representing $w_0=-1$, $w_a=0$, and $w_0+w_a=-1$ correspond to the values in the standard $\Lambda$CDM model.
}
\label{fig:EoS_CPL}
\end{figure*}

\begin{figure*}[htbp]
\centering
\mbox{\includegraphics[width=85mm]{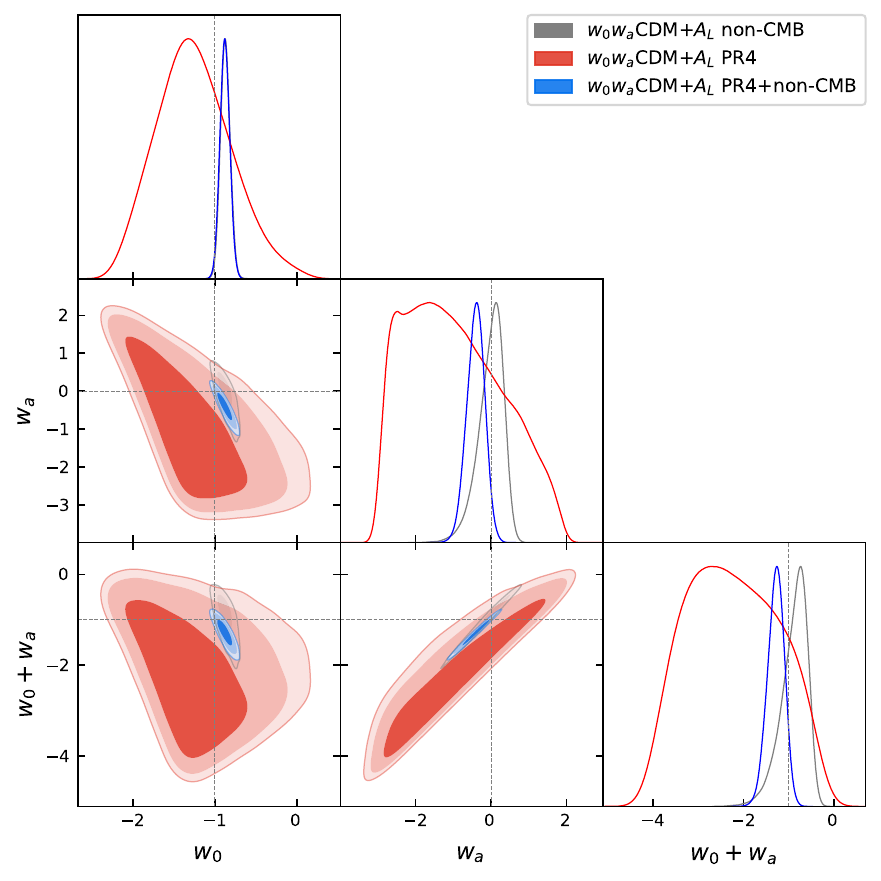}}
\mbox{\includegraphics[width=85mm]{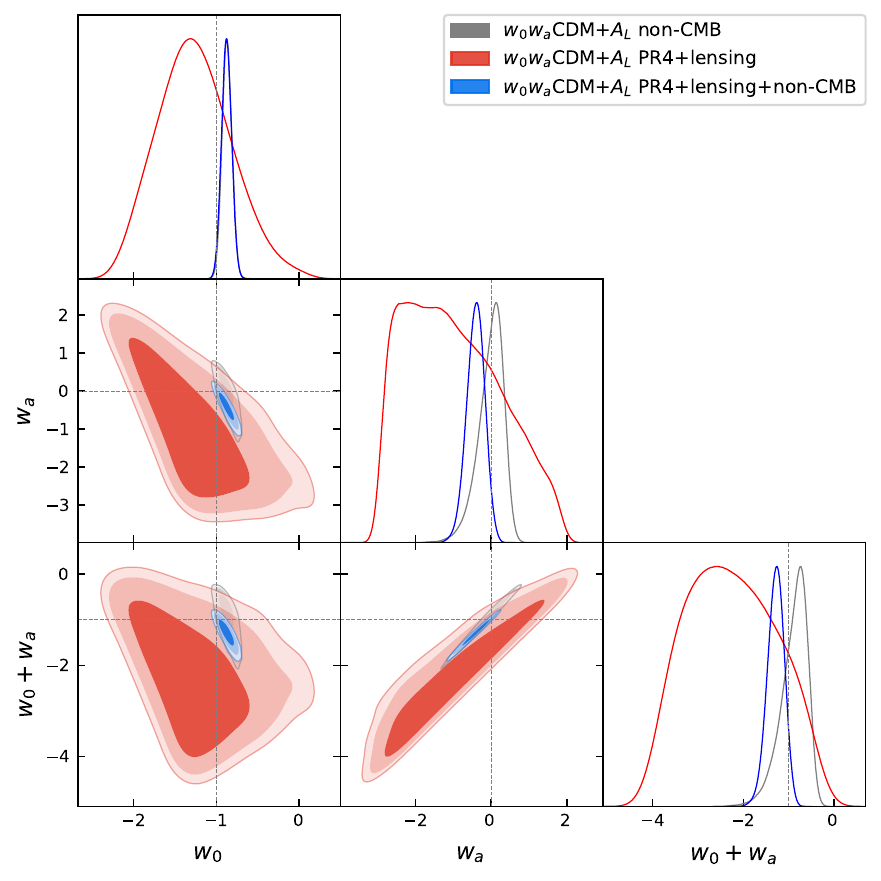}}
        \caption{One-dimensional likelihoods and 1$\sigma$, 2$\sigma$, and $3\sigma$ likelihood confidence contours of $w_0$, $w_a$, and $w_0+w_a$ parameters in the $w_0 w_a$CDM$+A_L$ parameterization favored by (left) non-CMB, PR4, and PR4+non-CMB data sets, and (right) non-CMB, PR4+lensing, and PR4+lensing+non-CMB data sets. The horizontal or vertical dotted lines representing $w_0=-1$, $w_a=0$, and $w_0+w_a=-1$ correspond to the values in the standard $\Lambda$CDM model.
}
\label{fig:EoS_CPL_AL}
\end{figure*}

\begin{figure*}[htbp]
\centering
\mbox{\includegraphics[width=85mm]{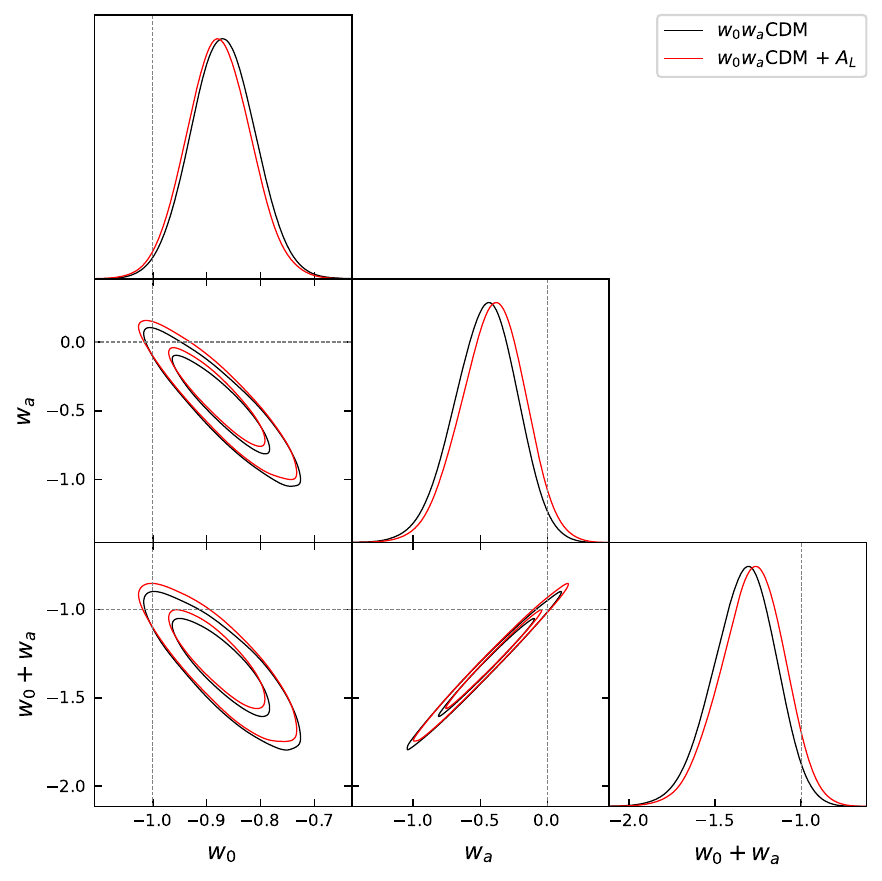}}
\mbox{\includegraphics[width=85mm]{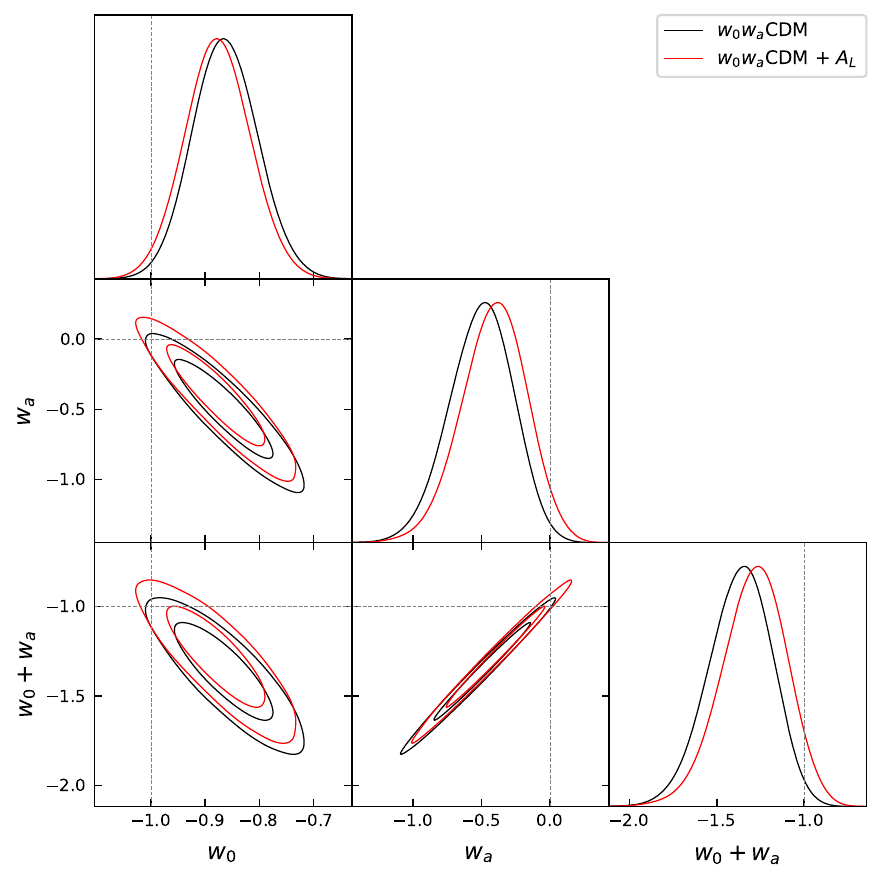}}
        \caption{One-dimensional likelihoods and 1$\sigma$ and 2$\sigma$ likelihood confidence contours of $w_0$, $w_a$, and $w_0+w_a$ parameters in the $w_0 w_a$CDM and $w_0 w_a$CDM+$A_L$ parameterizations, favored by (left) PR4+non-CMB data set, and (right)  PR4+lensing+non-CMB data set. The horizontal or vertical dotted lines representing $w_0=-1$, $w_a=0$, and $w_0+w_a=-1$ correspond to the values in the standard $\Lambda$CDM model.
}
\label{fig:EoS_CPL_and_CPL_AL}
\end{figure*}

\subsection{Discussion of the contour plots for $w_0$, $w_a$ and $w_0+w_a$ for the $w_0w_a$CDM(+$A_L$) parameterizations} 

In this subsection we comment on the contour plots displayed in Figs.\ \ref{fig:EoS_CPL}, \ref{fig:EoS_CPL_AL}, and \ref{fig:EoS_CPL_and_CPL_AL} for the $w_0w_a$CDM and $w_0w_a$CDM+$A_L$ parameterizations, showing the constraints on the equation-of-state parameters $w_0$ and $w_a$ and their combination $w_0+w_a$, obtained with non-CMB, PR4, PR4+lensing, PR4+non-CMB and PR4+lensing+non-CMB datasets. 

In the left panel of Fig.\ \ref{fig:EoS_CPL} we observe that the PR4 and non-CMB contours do not overlap within the $1\sigma$ confidence region, possibly indicating a mild tension between the two datasets as discussed earlier. However, as noted in Sec.\ \ref{subsec:Consistency}, PR4(+lensing) data and non-CMB data constraints are mutually consistent at our chosen level of significance and so these datasets can be used jointly to constrain cosmological parameters. When the datasets are combined, the resulting contours shift away from the PR4 constraints and lie approximately midway between the PR4 and non-CMB regions, although noticeably closer to the non-CMB constraints. This is because the non-CMB dataset has tighter contours, demonstrating its stronger constraining power on the $w_0w_a$CDM parameters compared to PR4 data. The value $w_0+w_a=-1.33^{+0.20}_{-0.17}$ for PR4+non-CMB data indicates that in the limit of high redshifts the equation-of-state parameter exhibits phantom behavior, while the value $w_0 = -0.869\pm0.060$ indicates that at low redshifts the equation-of-state parameter exhibits quintessence behavior.

In the right panel of Fig.\ \ref{fig:EoS_CPL} we see that the constraints obtained after adding CMB lensing data are very similar to the previous results. Although the lensing data contain information from the late-time universe, most of the constraining power still comes from non-CMB data. In this case we find $w_0+w_a=-1.37^{+0.19}_{-0.17}$ for PR4+lensing+non-CMB dataset, indicating a slightly higher preference for high-redshift phantom behavior compared with the previous case, while $w_0 = -0.863\pm0.060$ indicates a slightly higher preference for low-redshift quintessence behavior.

In the left and right panels of Fig.\ \ref{fig:EoS_CPL_AL} we can observe how the constraints change when the lensing parameter $A_L$ is allowed to vary in the analysis. Again we do not observe significant changes between including and excluding the CMB lensing data. In this case the contours from PR4 and non-CMB data move closer to each other, indicating that allowing for extra freedom in the CMB sector partially reduces the difference between the two sets of constraints. However, the overlap between the PR4 and non-CMB contours remains incomplete at the $1\sigma$ level, showing that the tension is reduced but not fully resolved. But again, as shown in Sec.\ \ref{subsec:Consistency}, PR4(+lensing) data and non-CMB data constraints are mutually consistent at our chosen level of significance and so these datasets can be jointly used. The combined PR4+non-CMB constraints continue to lie between those of the individual datasets and the same happens with the PR4+lensing+non-CMB contours. For both PR4+non-CMB and  PR4+lensing+non-CMB datasets, the obtained value is $w_0+w_a=-1.29^{+0.20}_{-0.17}$, which is very similar to the case with $A_L=1$ and also indicates a preference for higher-redshift phantom behavior, with $w_0 = -0.877 \pm 0.059$ (PR4+non-CMB) and $w_0 = -0.877 \pm 0.060$ (PR4+lensing+non-CMB) favoring low-redshift quintessence behavior. 

In Fig.\ \ref{fig:EoS_CPL_and_CPL_AL} we see the comparison of the results obtained for the $w_0w_a$CDM and $w_0w_a$CDM+$A_L$ parameterizations. We see that the constraints obtained by fixing $A_L = 1$ are somewhat similar to those derived when allowing $A_L$ to vary as a free parameter, with larger differences for the PR4+lensing+non-CMB dataset, as discussed next.

\subsection{The effect on the $A_L=1$ $w_0w_a$CDM parameterization evidence for dark energy dynamics when PR3 data are replaced by PR4 data}
\label{subsec:AL-unity-evidence}

For the $w_0w_a$CDM parameterization with $A_L = 1$, we showed in Fig.\ 3 of \cite{Park:2024jns} (also see Fig.\ 5 of \cite{Park:2024pew}) that for PR3+lensing+non-CMB data the $\Lambda$CDM model point at $w_0 = -1$, $w_a = 0$, and $w_0 + w_a = -1$ is slightly more than $2\sigma$ away from the best-fit $w_0w_a$CDM parameterization point, i.e., it lies outside the $2\sigma$ likelihood contours. Here, in the right panel of Fig.\  \ref{fig:EoS_CPL} (and more clearly in the right panel of Fig.\ \ref{fig:EoS_CPL_and_CPL_AL}), we show that for PR4+lensing+non-CMB data the $\Lambda$CDM model point is less than about $1.8\sigma$ away from the best-fit $w_0w_a$CDM parameterization point. 

For the $w_0w_a$CDM parameterization with $A_L = 1$, replacing most of the PR3 data with PR4 data leads to a mild reduction in the significance of dark energy dynamics. This reduction may be due to the weaker evidence in PR4 data for $A_L$ deviating above unity compared with the PR3 case.

\subsection{The effect on the $w_0w_a$CDM$+A_L$ parameterization evidence for dark energy dynamics and the $A_L$ value differing from unity when PR3 data are replaced by PR4 data}
\label{subsec:AL-not-unity-evidence}

For the $w_0w_a$CDM$+A_L$ parameterization with varying $A_L$, we showed in Fig.\ 6 of \cite{Park:2024pew} that for PR3+lensing+non-CMB data the $\Lambda$CDM model point at $w_0 = -1$, $w_a = 0$, and $w_0 + w_a = -1$ is about $1.5\sigma$ away from the best-fit $w_0w_a$CDM$+A_L$ parameterization point, with $A_L$ favored to be $2.0\sigma$ greater than unity (compared to the larger evidence for dark energy dynamics, of greater than $2\sigma$ when $A_L = 1$, see Sec.\ \ref{subsec:AL-unity-evidence}). Here, in the right panel of Fig.\  \ref{fig:EoS_CPL_AL} (and more clearly in the right panel of Fig.\ \ref{fig:EoS_CPL_and_CPL_AL}), we show that for PR4+lensing+non-CMB data the $\Lambda$CDM model point is also less than about $1.5\sigma$ away from the best-fit $w_0w_a$CDM parameterization point (compared to the larger deviation of about $1.8\sigma$ when $A_L = 1$, see \ref{subsec:AL-unity-evidence}), with $A_L$ now only $1.1\sigma$ greater than unity. 

We note that the reduction in significance of the deviation from the $\Lambda$CDM point at $w_0 = -1$ and $w_a = 0$ when $A_L$ is allowed to vary compared to the $A_L = 1$ case is not because the $w_0$ and $w_a$ error bars are larger in the varying $A_L$ case, rather it is because the mean values of $w_0$ and $w_a$ have moved closer to the $\Lambda$CDM $w_0 = -1$ and $w_a = 0$ point.

For the $w_0w_a$CDM$+A_L$ parameterization with varying $A_L$, replacing most of the PR3 data by PR4 data leads to a mild reduction in the significance of dark energy dynamics, from about $1.8\sigma$ for the $A_L = 1$ case to about $1.5\sigma$ for the varying $A_L$ case. This reduction may be due to the fact that PR4 data still show evidence for $A_L$ being greater than unity, although the deviation is weaker than in PR3 data. This suggests that even with PR4 data (instead of PR3 data) there is still mild evidence, $1.1\sigma$, for excess weak lensing smoothing of some Planck CMB anisotropy data over what is expected in the best-fit $w_0w_a$CDM parameterization, and possibly also that this is responsible for some of the evidence for dark energy dynamics in the $w_0w_a$CDM parameterization.

\section{Conclusion}
\label{sec:Conclusion}

In this work we have used various combinations of PR4 CMB data, (PR4) lensing CMB data, and non-CMB data, to test the spatially-flat $\Lambda$CDM model and the spatially-flat $w_0$CDM and $w_0w_a$CDM dynamical dark energy parameterizations, without and with a varying lensing consistency parameter $A_L$. In addition, we have performed a comprehensive comparison between the results obtained using PR3 data and those from PR4 data, allowing us to assess the stability and change of results when moving from PR3 data to PR4 data. 

We have tested the consistency/inconsistency of PR4-based cosmological parameter constraints obtained in the same cosmological model/parameterization but for different data sets. For this test we use the $\log_{10} \mathcal{I}$ estimator that is based on the DIC value. Our results show that mildly significant inconsistencies arise only when analyzing the $w_0$CDM parameterization. In particular, for the PR4 vs.\ non-CMB comparison we obtain $\log_{10} \mathcal{I}=-1.359$ while in the case of the PR4+lensing vs.\ non-CMB analysis we get $\log_{10} \mathcal{I}=-1.573$, with the degree of inconsistency being {\it strong} in both cases. We note that this is below our inconsistency threshold and so these datasets may be used jointly to constrain cosmological parameters. We also find that it is generally easier for the models/parameterizations to jointly accommodate the different data sets when PR4 data are used compared to the PR3 data case. 

When comparing results determined using our largest datasets, PR3+lensing+non-CMB and PR4+lensing+non-CMB, among the $\Lambda$CDM($+A_L$) models and the $w_0$CDM($+A_L$) and $w_0w_a$CDM($+A_L$) parameterizations only $\Omega_b h^2$ differs by $1.0\sigma$ or larger, as follows: $1.0\sigma$ for $\Lambda$CDM, $1.1\sigma$ for $\Lambda$CDM$+A_L$ and $w_0w_a$CDM$+A_L$, and $1.2\sigma$ for $w_0$CDM$+A_L$. The largest difference between $\Omega_b h^2$ values is $1.7\sigma$ for $w_0$CDM+$A_L$ and PR3+non-CMB vs.\ PR4+non-CMB, while the largest differences between $A_L$ values are also for PR3+non-CMB vs.\ PR4+non-CMB and are $1.9\sigma$ (for $\Lambda$CDM$+A_L$ and $w_0$CDM$+A_L$) and $1.8\sigma$ (for $w_0w_a$CDM$+A_L$), with $A_L$ values for PR3 and PR4 differing by $1.8\sigma$ (for $\Lambda$CDM$+A_L$) and $1.6\sigma$ (for $w_0$CDM$+A_L$ and $w_0w_a$CDM$+A_L$).

For the largest datasets we use, PR3/PR4+lensing+non-CMB data, in the $\Lambda$CDM model and the $w_0$CDM and $w_0w_a$CDM parameterizations the PR3-based and PR4-based cosmological constraints differ by $1.0\sigma$ (the $\Omega_b h^2$ difference for the $\Lambda$CDM model) or less. This means that replacing PR3 data by PR4 data does not significantly affect the inferred cosmological parameter values for the $A_L=1$ model/parameterizations.

When the lensing parameter $A_L$ is allowed to vary in the $\Lambda$CDM+$A_L$ model, with the PR3/PR4+lensing+non-CMB data we find a difference of $0.70\sigma$ between the PR3-based result ($A_L=1.087\pm 0.035$, $2.5\sigma$ above unity) and the PR4-based one ($A_L=1.053\pm 0.034$, $1.6\sigma$ above unity). For these data and the $w_0$CDM+$A_L$ parameterization the difference is $0.73\sigma$ between the PR3-based result ($A_L=1.101\pm 0.037$, $2.7\sigma$ above unity) and the PR4-based one ($A_L=1.064\pm 0.035$, $1.8\sigma$ above unity), while for the $w_0w_a$CDM+$A_L$ parameterization the difference is $0.66\sigma$ between the PR3-based result ($A_L=1.078^{+0.036}_{-0.040}$, $2.0\sigma$ above unity) and the PR4-based one ($A_L=1.042\pm 0.037$, $1.1\sigma$ above unity). This indicates that PR4-based results show a weaker preference for anomalous values of the lensing consistency amplitude $A_{L}>1$ compared to the PR3-based values. This is expected on the basis of the findings of Tristram et al.\ \cite{Tristram:2023haj} for a comparison between the $\Lambda$CDM$+A_L$ $A_L$ results for PR3 data and PR4 data where, in our analyses, we find a difference of $1.8\sigma$ between the PR3 data result ($A_L=1.181\pm 0.067$, $2.7\sigma$ above unity) and the PR4 data one ($A_L=1.030\pm 0.054$, $0.56\sigma$ above unity). We emphasize that for PR4 data in the $\Lambda$CDM$+A_L$ model $A_L > 1$ at only $0.56\sigma$ while for PR4+lensing+non-CMB data $A_L > 1$ at $1.6\sigma$ (for $\Lambda$CDM$+A_L$), $1.8\sigma$ (for $w_0$CDM$+A_L$), and $1.1\sigma$ (for $w_0w_a$CDM$+A_L$). We assume that these larger deviations from unity in the PR4+lensing+non-CMB cases are not a consequence of unknown systematics. On the other hand, in the $w_0w_a$CDM$(+A_L)$ parameterizations for the PR3/PR4+lensing+non-CMB data comparison, the results obtained for the equation-of-state parameters $w_0$ and $w_0 + w_a$ remain more stable when moving from PR3-based to PR4-based analyses, however, as discussed next, the smaller differences are important. 

We find that PR4+lensing+non-CMB data in the $w_0w_a$CDM parameterization with $A_L = 1$ favor dynamical dark energy over a cosmological constant at about $1.8\sigma$. This is a slight reduction from the slightly larger than $2\sigma$ favoring of dynamical dark energy over a $\Lambda$ found when PR3+lensing+non-CMB data are used in the $w_0w_a$CDM parameterization \cite{Park:2024jns, Park:2024pew}. It is interesting that replacing PR3 data by PR4 data slightly decreases the evidence for dark energy dynamics. This slight decrease might be the consequence of the decreased evidence for $A_L > 1$ in PR4 data compared to PR3 data.

When we consider PR4+lensing+non-CMB data in the $w_0w_a$CDM$+A_L$ parameterization with a varying lensing consistency parameter $A_L$, we find that dynamical dark energy is still favored over a $\Lambda$, but now at a reduced significance of about $1.5\sigma$, with $A_L > 1$ at $1.1\sigma$. We emphasize again that this reduced significance for dark energy dynamics is a consequence of the $w_0$ and $w_a$ mean values moving closer to the $\Lambda$CDM $w_0 = -1$ and $w_a = 0$ point when $A_L$ is allowed to vary. We had earlier found that PR3+lensing+non-CMB data in the $w_0w_a$CDM$+A_L$ parameterization also favored dark energy dynamics at about $1.5\sigma$, but with $A_L > 1$ at $2.0\sigma$ \cite{Park:2024pew}. So replacing PR3 data with PR4 data does reduce the deviation of $A_L$ above unity but it does not change the slightly reduced evidence for dark energy dynamics. It appears that even for PR4 data part of the evidence for dark energy dynamics in the $w_0w_a$CDM$(+A_L)$ parameterizations might be due to the excess smoothing observed in the Planck CMB spectra (compared to what is expected in the best-fit cosmological model).  

When we analyze the largest dataset considered in this work, PR4+lensing+non-CMB, we observe a good agreement among the six primary cosmological parameters common to all models/parameterizations considered, with all shifts remaining below $1\sigma$. In terms of model comparison, only the $w_0w_a$CDM dark energy parameterization with $A_{L}=1$ is capable of somewhat significantly surpassing the performance of the $\Lambda$CDM model, with $\Delta\textrm{DIC} = -3.76$ and favoring dynamical dark energy at about $1.8\sigma$, while the $w_0w_a$CDM$+A_L$ parameterization has $\Delta\textrm{DIC} = -2.36$ and favors dynamical dark energy at about $1.5\sigma$. Both these parameterizations describe dynamical dark energy that is quintessence-like at low $z$ and phantom-like at high $z$, with $w(z)$ crossing over from $w < -1$ at high $z$ to $w > -1$ at present, an evolutionary behavior that is not easy to accommodate in a simple physically consistent dynamical dark energy model.\footnote{We note that for PR3+lensing+non-CMB data the physically-consistent $\phi$CDM$+A_L$ model, that can only describe quintessence-like dynamical dark energy, has $\Delta\textrm{DIC} = -3.90$, does better than $\Lambda$CDM, and favors dark energy dynamics at $1.7\sigma$ (while for these data $w_0w_a$CDM$+A_L$ has $\Delta\textrm{DIC} = -4.37$, does better than $\Lambda$CDM, and favors dark energy dynamics at about $1.5\sigma$) but the $\phi$CDM model has $\Delta\textrm{DIC} = +1.69$, does not do as well as $\Lambda$CDM, and favors dark energy dynamics at $1.3\sigma$ (while $w_0w_a$CDM has $\Delta\textrm{DIC} = -2.45$, does better than $\Lambda$CDM, and favors dark energy dynamics at a little over $2.0\sigma$) \cite{Park:2024pew, Park:2025fbl}.}    

With newer DESI DR2 BAO data now available, \cite{DESI:2025zgx}, it is of interest to redo our analysis but with DESI DR2 BAO data replacing the BAO data compilation in our non-CMB dataset. While such an analysis has not yet been done, it is possible to use the results of two published analyses to speculate qualitatively on what the results from a complete analysis might indicate. From table V of \cite{DESI:2025zgx}, for the eight parameter $w_0w_a$CDM parameterization with $A_L=1$ and for CMB+lensing+DESI(DR2)+Pantheon+ data we have $w_0 = -0.838 \pm 0.055$ and $w_a = -0.62^{+0.22}_{-0.19}$, where CMB data here are again largely from the PR4 NPIPE pipeline but use the \texttt{CamSpec} likelihood \cite{Rosenberg:2022sdy} instead of the \texttt{HiLLiPoP} likelihood \cite{Tristram:2023haj} and, additionally, PR4 CMB lensing data are now augmented with ACT DR6 CMB lensing data \cite{ACT:2023kun}. On the other hand \cite{RoyChoudhury:2025dhe}, in the twelve parameter $w_0w_a$CDM$+A_L$ parameterization with additional parameters $N_{\rm eff}$ (the number of non-photon radiation species), $\Sigma m_\nu$ (the sum of neutrino masses), and $\alpha_s$ (the running of the scalar spectral index), for CMB+lensing+DESI(DR2)+Pantheon+ data, finds, in table 2, $w_0 = -0.864 \pm 0.056$, $w_a = -0.44^{+0.26}_{-0.22}$, and $A_L = 1.068^{+0.042}_{-0.050}$, for essentially the same CMB data but now using the \texttt{HiLLiPoP} and \texttt{LoLLiPoP} CMB likelihoods. We note that the values \cite{RoyChoudhury:2025dhe} finds for $N_{\rm eff}$, $\Sigma m_\nu$, and $\alpha_s$ are very close to the standard values for these parameters that are assumed in the similar analysis of eight parameter $w_0w_a$CDM in \cite{DESI:2025zgx}, so these additional parameters are unlikely to significantly bias the mean values of $w_0$ and $w_a$ found in the analysis of \cite{RoyChoudhury:2025dhe}. The $w_0w_a$CDM $A_L=1$ parameterization analysis of \cite{DESI:2025zgx} finds that the $\Lambda$CDM $w_0 = -1$ and $w_a = 0$ point is $2.8\sigma$ away from the best-fit $w_0w_a$CDM point, while \cite{RoyChoudhury:2025dhe} find it is only $2.0\sigma$ away from the best-fit $w_0w_a$CDM$+A_L$ point and that $A_L > 1$ at $1.36\sigma$ significance.\footnote{Note that in the $A_{\rm lens}$ row of table 2 of \cite{RoyChoudhury:2025dhe} the three $1\sigma$ lower limits of $-0.54$, $-0.50$, and $-0.52$ should be $-0.054$, $-0.050$, and $-0.052$, S.\ Roy Choudhury, private communication (2026).} It might be significant that the \cite{RoyChoudhury:2025dhe} mean $w_0$ and $w_a$ values are closer to the $w_0 = -1$ and $w_a = 0$ $\Lambda$CDM model values than are the \cite{DESI:2025zgx} mean $w_0$ and $w_a$ values, while the \cite{RoyChoudhury:2025dhe} $w_0$ and $w_a$ error bars are only slightly larger than the \cite{DESI:2025zgx} ones, likely a consequence of the four additional parameters in the parameterization used in \cite{RoyChoudhury:2025dhe}. Ignoring, perhaps not too unreasonably, the differences between the PR4 \texttt{HiLLiPoP} and \texttt{CamSpec} likelihoods, we see that, for CMB+lensing+DESI(DR2)+Pantheon+ data, going from a $w_0w_a$CDM parameterization to a $w_0w_a$CDM$+A_L$ parameterization shows that $A_L > 1$ is now favored at $1.4\sigma$ and also results in a decrease of the evidence for dark dynamics from $2.8\sigma$ to $2.0\sigma$, consistent with what we have found here and earlier \cite{Park:2024pew, Park:2025azv} with different but related data. Of course, this is a qualitative and speculative argument and a proper analysis will be needed to determine believable numerical values.
 
The results presented in this work correspond to the analyses of the standard $\Lambda$CDM model and several popular phenomenological extensions of the standard model. Overall, our findings are encouraging as some of the scenarios we explored show a mild preference over the $\Lambda$CDM model, and a mild preference for dark energy dynamics. Although the statistical significance is not yet sufficient to claim compelling evidence for new physics, and it is important to better understand the connection between larger than unity values of the lensing consistency parameter and the evidence for dark energy dynamics, these indications may point toward the existence of physics beyond the standard cosmological model. While the effect we have discovered is small, it is non-negligible and should be better understood. In this context, it will be important to use better and more data to investigate more physically motivated and theoretically grounded models in order to assess whether the observed trends persist and, if they do, to better understand their fundamental origin.

\acknowledgements
JdCP's research was financially supported by the projects "Plan Complementario de I+D+i en el \'area de Astrof{\'\i}sica" and "Desarrollo de algoritmos de big data y data science aplicados a la física de partículas" funded by the European Union within the framework of the Recovery, Transformation and Resilience Plan - NextGenerationEU and by the Regional Government of Andaluc{\'i}a (References AST22\_00001 and AST22\_8.4\_SR). Also, JdCP acknowledges partial support from MICINN (Spain) project PID2022-138263NB-I0 (AEI/FEDER, UE). C.-G.P.\ was supported by the National Research Foundation of Korea (NRF) grant funded by the Korea government (MSIT) No.\ RS-2026-25473390.

\bibliography{w0waCDM_Alens}

\end{document}